    \documentclass[12pt,a4paper]{article} 
    \usepackage{psfig}  \usepackage{epsfig}  
    \usepackage{graphics} \usepackage{color,array,shadow}
\newcommand{\ra}           {\mbox{$\rightarrow$}}
\newcommand{\bc}           {\begin{center}}
\newcommand{\ec}           {\end{center}}
\newcommand{\bq}           {\begin{eqnarray}}
\newcommand{\eq}           {\end{eqnarray}}
\newcommand{\be}           {\begin{equation}}
\newcommand{\ee}           {\end{equation}}
\newcommand{\bi}           {\begin{itemize}}
\newcommand{\ei}           {\end{itemize}}
\newcommand{\gam}          {\mbox{$\gamma$}}

\newcommand{\pbp}          {\mbox{$\bar{\mbox{p}}\mbox{p}$}}

\newcommand{\pbn}          {\mbox{$\bar{\mbox{p}}\mbox{n}$}}

\newcommand{\kkb}          {\mbox{$\mbox{K}\bar{\mbox{K}}$}}

\newcommand{\p}            {\mbox{$\pi$}}
\newcommand{\pip}          {\mbox{$\pi^+$}}
\newcommand{\pim}          {\mbox{$\pi^-$}}
\newcommand{\piz}          {\mbox{$\pi^0$}}
\newcommand{\rh}           {\mbox{$\rho$}}
\newcommand{\rhz}          {\mbox{$\rho^0$}}
\newcommand{\rhpm}         {\mbox{$\rho^\pm$}}
\newcommand{\rhp}          {\mbox{$\rho^+$}}
\newcommand{\rhm}          {\mbox{$\rho^-$}}
\newcommand{\pipm}         {\mbox{$\pi^{\pm}$}}
\newcommand{\pimp}         {\mbox{$\pi^{\mp}$}}
\newcommand{\etg}          {\mbox{$\eta$}}

\newcommand{\etp}          {\mbox{$\eta^{\prime}$}}

\newcommand{\omg}          {\mbox{$\omega$}}
\newcommand{\kp}           {\mbox{$\mbox{K}^{+}$}}
\newcommand{\km}           {\mbox{$\mbox{K}^{-}$}}

\newcommand{\kn}           {\mbox{$\mbox{K}^{0}$}}
\newcommand{\knb}          {\mbox{$\mbox{$\bar{\mbox{K}}$}^{0}$}}
\newcommand{\ksnb}      {\mbox{$\mbox{$\bar{\mbox{K}}$}^{\ast 0}$}}

\newcommand{\ksp}          {\mbox{$\mbox{K$^{{\ast +}}$}$}}
\newcommand{\ksm}          {\mbox{$\mbox{K$^{{\ast -}}$}$}}
\newcommand{\ksn}          {\mbox{$\mbox{K$^{{\ast 0}}$}$}}
\newcommand{\Piz}          {\mbox{$|\pi^0>$}}
\newcommand{\Pip}          {\mbox{$|\pi^+>$}}
\newcommand{\Pim}          {\mbox{$|\pi^->$}}
\newcommand{\Pipm}         {\mbox{$|\pi^{\pm}>$}}

\newcommand{\nonett}[9]
{
\setlength{\unitlength}{1mm}
\begin{picture}(150.00,90.00)
\put(10.00,45.00){\vector(1,0){70.00}}
\put(45.00,10.00){\vector(0,1){70.00}}
\put(110.00,45.00){\vector(1,0){30.00}}
\put(125.00,30.00){\vector(0,1){30.00}}
\put(82.50,42.50){\makebox(5.00,5.00){$I_3$}}
\put(142.50,42.50){\makebox(5.00,5.00){$I_3$}}
\put(127.50,60.00){\makebox(5.00,5.00){S\quad\ Singlet }}
\put(47.50,80.00){\makebox(5.00,5.00){S\quad\ Octet }}
\put(45.00,45.00){\circle*{2.00}}
\put(70.00,45.00){\circle*{2.00}}
\put(20.00,45.00){\circle*{2.00}}
\put(32.50,70.00){\circle*{2.00}}
\put(57.50,70.00){\circle*{2.00}}
\put(32.50,20.00){\circle*{2.00}}
\put(57.50,20.00){\circle*{2.00}}
\put(125.00,45.00){\circle*{2.00}}
\put(45.00,45.00){\circle{0.00}}
\put(45.00,45.00){\circle{5.00}}
\put(32.50,70.00){\line(1,0){25.00}}
\put(57.50,70.00){\line(1,-2){12.50}}
\put(70.00,45.00){\line(-1,-2){12.50}}
\put(57.50,20.00){\line(-1,0){25.00}}
\put(32.50,20.00){\line(-1,2){12.50}}
\put(20.00,45.00){\line(1,2){12.50}}
\put(60.00,70.00){\makebox(15.00,5.00)[l]{#2}}
\put(15.00,70.00){\makebox(15.00,5.00)[r]{#1}}
\put(15.00,15.00){\makebox(15.00,5.00)[r]{#3}}
\put(6.75,37.50){\makebox(13.25,5.00)[r]{#5}}
\put(60.00,15.00){\makebox(15.00,5.00)[l]{#4}}
\put(70.00,37.50){\makebox(13.75,5.00)[l]{#6}}
\put(47.50,37.50){\makebox(12.50,5.00)[l]{#7}}
\put(47.50,47.50){\makebox(12.50,5.00)[l]{#8}}
\put(127.50,47.50){\makebox(12.50,5.00)[l]{#9}}
\end{picture}
}

\newcommand{\pbarp}{ \bar{\rm{p}}\rm{p} }

\def\lsim{\mathrel{\rlap{\lower4pt\hbox{\hskip1pt$\sim$}}
    \raise1pt\hbox{$<$}}}                
\def\gsim{\mathrel{\rlap{\lower4pt\hbox{\hskip1pt$\sim$}}
    \raise1pt\hbox{$>$}}}                
\begin{document}
{\Large 
\bf\centering
PSI Zuoz Summer School\\
\vskip 4mm
Phenomenology of Gauge Interactions\\
\vskip 4mm
August 13 - 19, 2000
\vskip 20mm
\LARGE
Meson Spectroscopy:\\
\vskip 4mm 
Glueballs, Hybrids and $Q\bar Q$ Mesons\\
\Large
\vfill 
Eberhard Klempt\\
\vskip 2mm
Institut f\"ur Strahlen-- und Kernphysik  \\
\vskip 2mm
der   Universit\"at Bonn \\  
\vskip 2mm
D-53115 BONN \\ 
\vskip 5mm
\normalsize
Electronic mail address: klempt@iskp.uni-bonn.de 
\vspace*{10mm}
}
\newpage
\phantom{rrrr}
\newpage
\vfill
\begin{center}  \begin{Large} \begin{bf}
Meson Spectroscopy:\\ 
Glueballs, Hybrids
and $Q\bar Q$ Mesons\\
  \end{bf}  \end{Large}
  \bigskip
  \bigskip
  \begin{large}
\renewcommand{\thefootnote}{\fnsymbol{footnote}}
\setcounter{footnote}{0}
Eberhard Klempt\,\footnote[1]{\it Electronic mail 
address: klempt@iskp.uni-bonn.de} 
\setcounter{footnote}{0}\\
  \end{large}
\medskip
Institut f\"ur Strahlen-- und Kernphysik  \\
der   Universit\"at Bonn \\  
       D-53115 BONN \\
\end{center}
\bigskip
\begin{center}
{\bf Abstract}
\end{center}
\begin{quotation}
\noindent
Lattice gauge calculations predict the existence of glueballs.
In particular a scalar glueball is firmly expected at a
mass of about 1730 MeV. This prediction has led to an intense study
of scalar isoscalar interactions and to the discovery of 
new meson resonances. 
The number of scalar states observed seems to exceed
the number of states which can be accommodated in the quark
model even when two states, the $a_0(980)$ and $f_0(980)$, 
are interpreted as \kkb\ bound states and are removed
from the list. However, none of these states has a decay 
pattern which is consistent with that of a pure glueball. 
A reasonable interpretation of the number of states 
and of their decay pattern is found only
when mixing of scalar $q\bar q$ states with the scalar glueball 
is taken into account. 
\par
In this paper we scrutinize the evidence for these states
and their production characteristics. The $f_0(1370)$ - a cornerstone 
of all $q\bar q$-glueball mixing scenarios - is shown to be likely of
non-$q\bar q$ nature. The remaining scalar states then do fit into
a nonet classification. If this interpretation should be correct
there would be no room for resonant scalar gluon-gluon interactions,
no room for the scalar glueball. 
\par
We also discuss the status of mesons with exotic quantum numbers, 
of mesons which cannot possibly have $q\bar q$ structure, and argue
that these are, most likely, four-quark states and not
excitations of the gluon string  providing the 
binding between quark and antiquark. 

\end{quotation}
\section{Introduction: why hadron spectroscopy\,?}
SU(3) symmetry considerations had shown to be 
extremely useful in classifying mesons and baryons 
\cite{gell-mann} as composed of quarks and antiquark
or of 3 quarks, respectively. Deep inelastic 
scattering had provided early insight into the 
physics of partons \cite{bjorken}. At that time, the concept 
of particles carrying charges of one or two thirds of the 
electron charge was, however, too far away from
every-days experience, and the reality of quarks was often
not accepted. The breakthrough of the quark model 
in the general perception of physicist was the discovery 
of the J/$\psi$ family of meson resonances 
\cite{Aubert:1974js,Augustin:1974xw} and their
interpretation of $\bar{c}c$ states \cite{Fritzsch:1975tx}.
Attempts to find conventional explanations
for the new narrow states as \omg\omg\ ccompounds
or \pbp\ resonances \cite{durr1,durr2} failed, and the
quark model became the frame of further progress in the 
field. A new theory of strong interactions,
Quantum  Chromo Dynamics or QCD, emerged 
\cite{Weinberg:1973un,Fritzsch:1973pi} which
assigned to quarks a new triple-valued charge 
called color. 
The fact that free quarks were never observed
was understood by the hypothesis that color is
confined \cite{Wilson:1974sk}, even though their 
mutual interactions were supposed to be weak at large 
momentum transfers or at small distances
\cite{Coleman:1973sx}. 
Particles 
which we observe in nature must be color-neutral.
Color-neutral objects can be formed by combining a colored
quark and an antiquark with anticolour to a meson or by 
combining 3 quarks with 3 different colors to a baryon. 
\par
\begin{figure}[ht]
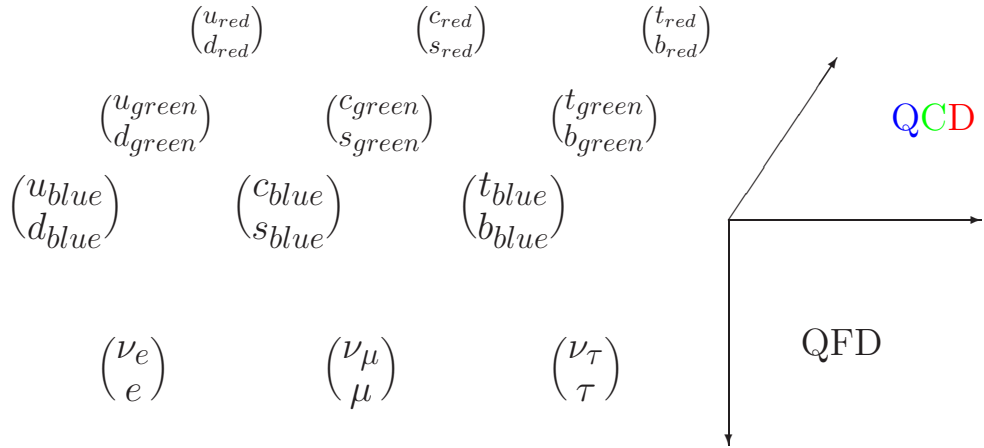

\caption{The Standard Model}
\label{pic:quarklepton}
\end{figure}

According to the Standard Model we have 3 generations
of quarks and leptons which are shown in 
Fig.~\ref{pic:quarklepton}. The unified theory of 
electromagnetic and weak interactions, Quantum Flavor
Dynamics, acts within the 
plane of  Fig.~\ref{pic:quarklepton}; strong interactions
are restricted to the exchange of color via gluons
and act in the third direction. This latter interaction
is of relevance for the further discussion. It is a
renormalisable gauge field theory constructed in 
line with QED. 
Unfortunately, this beautiful theory is of limited use
in hadron spectroscopy; calculations can be carried
out only at large momentum transfer, in the realm of 
perturbative QCD. Only there, observables
can be expressed in a power series in
fine-structure constant $\alpha_s$
of strong interaction. 
\par
Another limit in which QCD can be solved is at
very small energies. The light quarks $u,d$ are nearly
massless, also the strange quark mass is small. In the
limit of vanishing masses there is a new symmetry, chiral
symmetry: massless particles cannot flip spin. 
This symmetry is spontaneously broken and massless 
pseudoscalar bosons, Goldstone bosons must exist
\cite{Goldstone:1961eq}. Finite quark 
masses lead to a breaking of chiral
symmetry in the Lagrangian, the Goldstone bosons
acquire mass and can be identified with the pseudoscalar
mesons. At small
energies, the breaking of chiral symmetry can be
treated perturbatively, and we have access to QCD.
\par
{\large
\setlength{\unitlength}{1.2mm}
\begin{picture}(150.00,90.00)
\put(0,60) {\LARGE${{u}_{blue}}\choose {{d}_{blue}}$}
\put(10,70){\Large${{u}_{green}}\choose {{d}_{\Huge green}}$}
\put(20,80){\large${{u}_{red}}\choose {{d}_{red}}$}
\put(25,60){\LARGE${{c}_{blue}}\choose {{s}_{blue}}$}
\put(35,70){\Large${{c}_{green}}\choose {{s}_{green}}$}
\put(45,80){\large${{c}_{red}}\choose {{s}_{red}}$}
\put(50,60){\LARGE${{t}_{blue}}\choose {{b}_{blue}}$}
\put(60,70){\Large${{t}_{green}}\choose {{b}_{green}}$}
\put(70,80){\large${{t}_{red}}\choose {{b}_{red}}$}
\put(10,42){\LARGE${\Huge\nu_e}\choose {{\Huge e}}$}
\put(35,42){\LARGE${\nu_\mu}\choose {\mu}$}
\put(60,42){\LARGE${\nu_\tau}\choose {\tau}$}
\put(80.00,60.00){\vector(1,0){28.00}}  
\put(80.00,60.00){\vector(0,-1){25.00}} 
\put(80.00,60.00){\vector(2,3){12.00}}  
\put(88,45){QFD}
\put(98,70){{\color{blue} Q}{\color{green} C}{\color{red} D}}
\end{picture}
\vspace*{-55mm}
}
\par
\par
\begin{figure}[h]
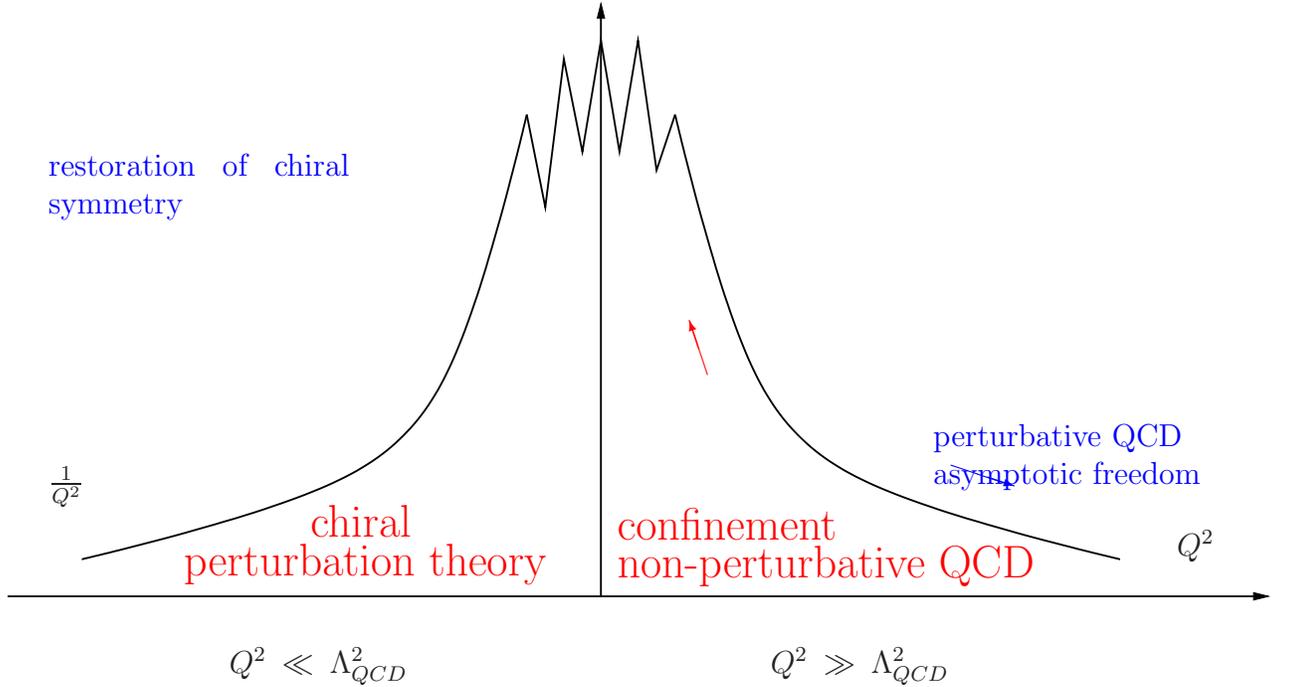

\caption{QCD observables as a function of $Q^2$}
\label{qcd}
\end{figure}

The situation is worse at medium energies, in the 
resonance region. This is the domain of Strong QCD.
The interaction becomes very strong 
and an expansion in the strong-interaction 
fine-structure constant $\alpha_s$ is meaningless. The different
fields in which QCD acts are summarized in 
Fig.~\ref{qcd}. 
\par
To understand what QCD tells us
about the range of $Q^2$ from 0.1 to 4 GeV$^2$, 
'QCD-inspired models' have been developed. These models
may guide us in our attempt
to understand the effective degrees of freedom in
meson and baryon spectroscopy and the effective
interaction between them. E.g. we may ask, 
if we can understand 
spectroscopy by introducing 'constituent' quarks 
with an effective mass. If this should be the case, 
we have to ask what is the residual strong 
interaction when a part
of the interaction has been taken into account
by increasing the current quark mass to an effective one.
\par
A large number of such QCD-inspired models have been 
proposed, like bag models \cite{Chodos:1974pn}, 
quark models with various quark-quark potentials 
\cite{Isgur:1979be}, the Nambu Jona-Lasinio 
model \cite{Nambu:1961tp}, the flux tube model
\cite{Isgur:1983wj}, models based on instanton 
interactions \cite{Blask:1990ez}, QCD sum rules
\cite{Shifman:1979bx}, or lattice QCD. The latter
model claims best reliability; QCD is simulated 
on a lattice, the lattice points are connected
by links representing the gluon fields which adjust
itself to provide a minimum energy in a given 
configuration. Lattice gauge calculations are
believed to reproduce the continuum theory
for a sufficiently large number of lattice points 
at smaller and smaller distances.
\clearpage
\vspace*{-8mm}
\setlength{\unitlength}{1.2mm}
\begin{picture}(150.00,90.00)
\put(-5,15){\epsfig{file=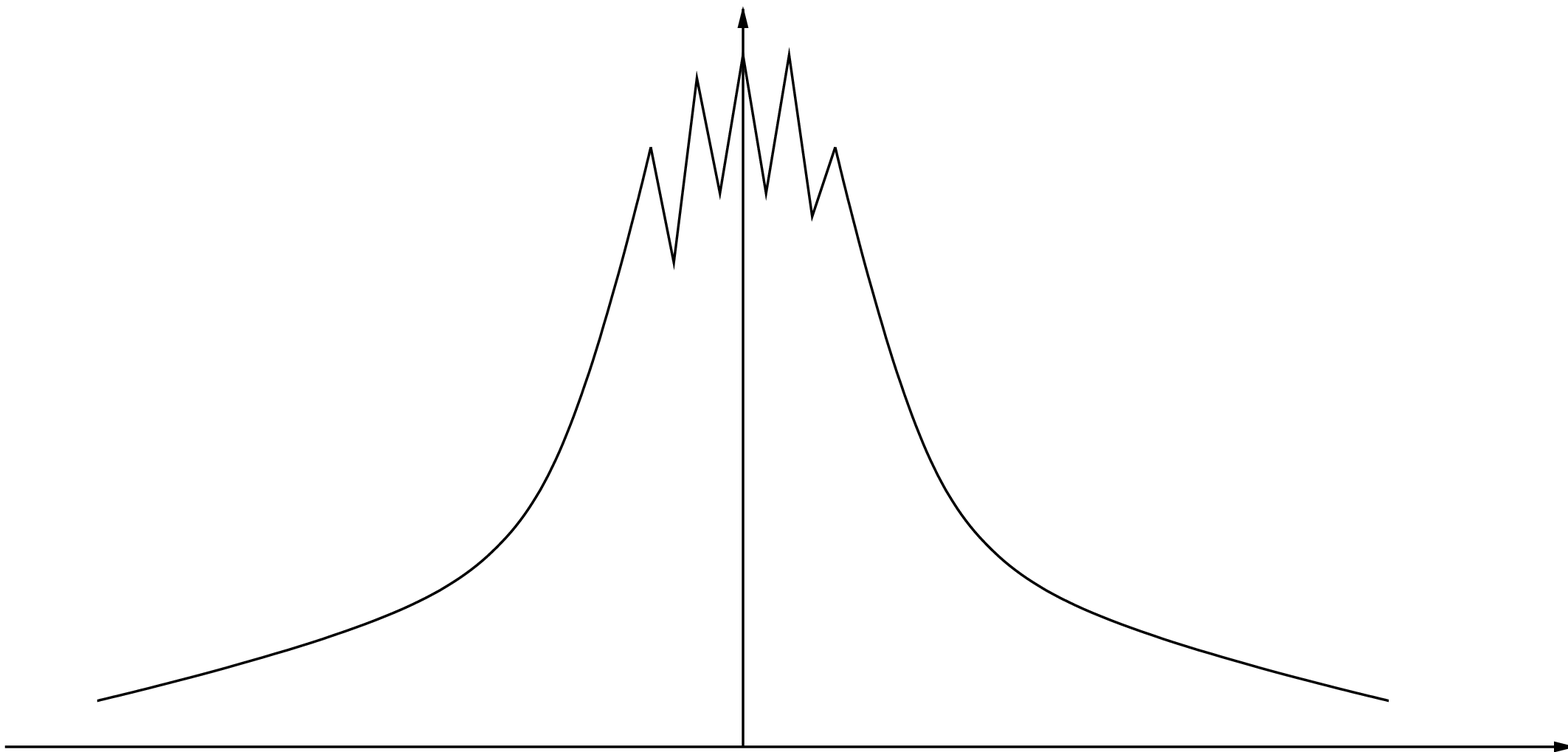,height=8cm}}
\put(0,60){\begin{minipage}[h]{4cm}
               {\color{blue} restoration of chiral symmetry}
          \end{minipage}}
\put(98,30){\begin{minipage}[h]{4cm}
               {\color{blue} perturbative QCD\\
                             asymptotic freedom}
          \end{minipage}}
\put(125,20){$Q^2$}
\put(0,27){$\frac{1}{Q^2}$}
\put(80,7){$Q^2\:\gg\:\Lambda^2_{QCD}$}
\put(20,7){$Q^2\:\ll\:\Lambda^2_{QCD}$}
\put(15,37){\begin{minipage}[h]{6cm}\vspace*{4cm}
               {\color{red} \Large \hspace{15mm} chiral\\ perturbation theory}
          \end{minipage}}
\put(63,20){\begin{minipage}[h]{6cm}
               {\color{red} \Large confinement\\ non-perturbative QCD}
          \end{minipage}}
\put(100,30){{\color{blue} \vector(3,-1){7}}}
\put(73,40){{\color{red}  \vector(-1,3){2}}}
\end{picture}
\par
\vspace{-10mm}
\noindent
\par
\begin{figure}[thb]
\includegraphics[width=0.7\textwidth]{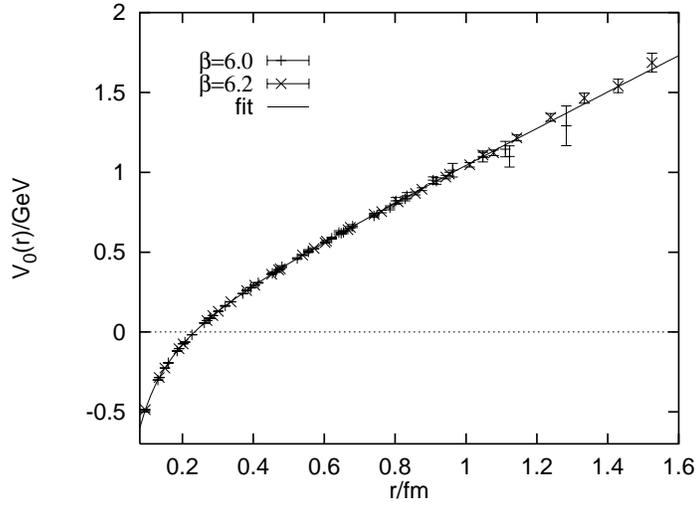}\\
\vspace*{-5mm}
\caption{$b\bar b$ potential from lattice calculations; (from
\protect\cite{Bali:1997am}).}
\label{pot}
\end{figure}
\begin{figure}[bht]
\includegraphics[width=0.9\textwidth]{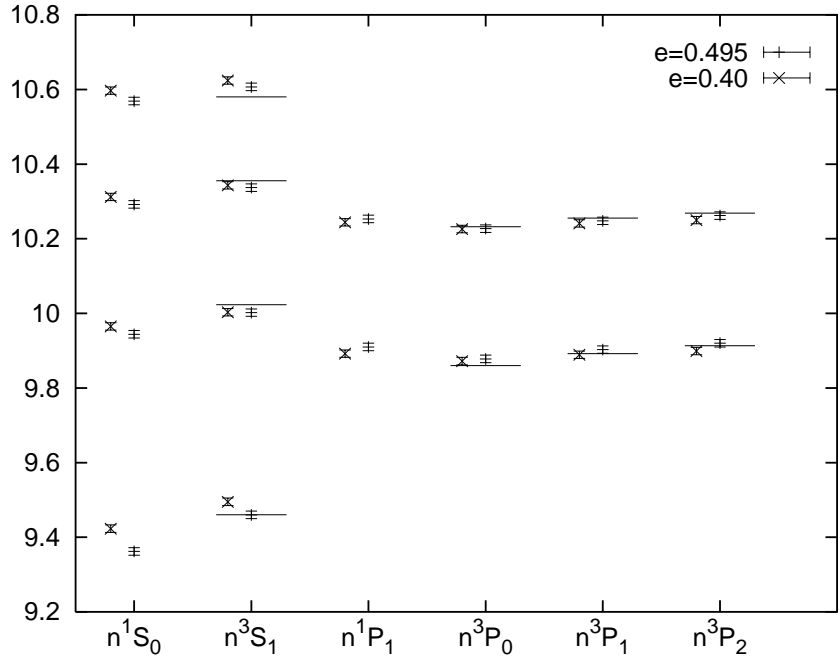}\\
\vspace*{-5mm}
\caption{Comparison of $Y$ states from lattice 
calculations with data (-); (from \protect\cite{Bali:1996bw}).}
\label{y}
\vskip -4mm
\end{figure}

Indeed, QCD on a lattice provides a good estimate
for the potential between a bottom quark and its 
antiquark. The potential is shown in Fig.~\ref{pot}.
It is well described by a sum a of Coulomb-like term
plus a linear (confining) potential: 
\begin{equation}
V(r) = V_0 + \frac{4}{3}\frac{\alpha_s}{r} + cr
\label{alpha}
\end{equation}
At small distances the potential is dominated by the 
Coulomb-like part, with a coupling constant $\alpha_s$. 
The coupling constant is not really a constant; its
value is about 0.12 at the $Z^0$ mass and increases
at low energies. At large distances, in the
confinement region,  the potential  
energy increases linearly; in the lattice calculation
shown in Fig.~\ref{pot} up to 1.6\,fm. The measured masses
of $b\bar b$ states are well reproduced by the lattice 
calculations, see Fig.~\ref{y}.
\par
\begin{figure}[h]
\bc
\epsfig{file=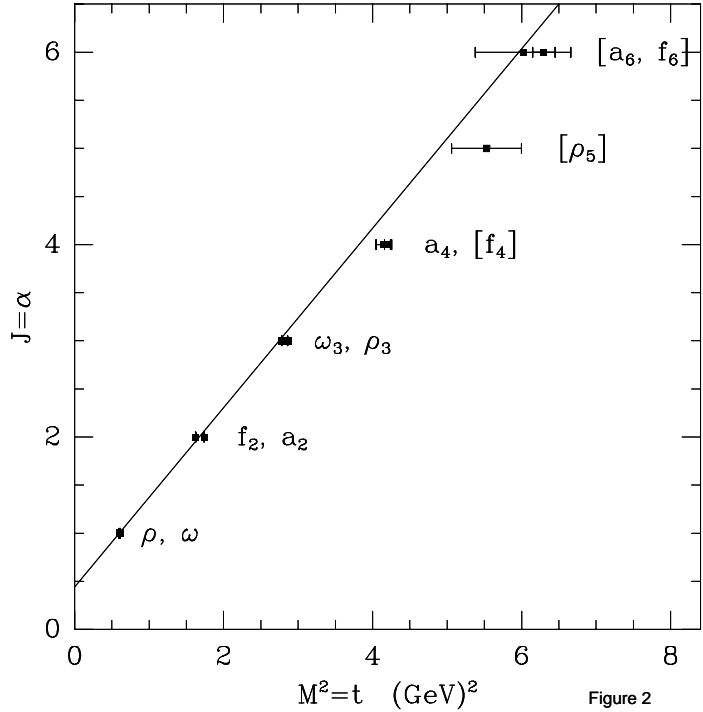,
width=0.7\textwidth,angle=0}
\ec
\vspace{-8mm}
\caption{Spin of meson resonances versus M$^2$. In the
Regge theory, non-integer spins $\alpha$ are allowed in 
exchange processes. Particles have integer spins.}
\label{regge}
  \end{figure}

We may get access to very large quark-antiquark separations
by considering high-spin states. Fig.~\ref{regge} shows
that the spin of $q\bar q$ resonances are linearly 
related to their squared masses. We can understand
this relation assuming that the gluon flux between
the two quarks is concentrated in a rotating flux tube
or a rotating string with a homogeneous mass density. 
The velocity at the ends may be the velocity of light. 
Then the total mass of the string is 
given by
$$
Mc^2 = 2\int_{0}^{r_0}\frac{k dr}{\sqrt{1-v^2/c^2}} = kr_0\pi
$$
and the angular momentum by
$$
J = \frac{2}{\hbar c^2}\int_{0}^{r_0}\frac{k r v dr}{\sqrt{1-v^2/c^2}} = 
\frac{k r_{0}^{2} \pi}{2\hbar c}
$$
$$
{\rm Hence}\qquad\qquad\ J = \frac{1}{2\pi k\hbar c} E^2 + {\rm constant}
$$
From the slope in Fig.~\ref{regge} we find $k = 0.2$\,GeV$^2$
and radii of 
$$
 r_0(\rho ) = 1.2 {\rm fm} \qquad\qquad\ 
 r_0(a_6) = 4 {\rm fm}
$$
\vskip -2mm
\par
The $a_6^{+}(2450)$ is a meson with a mass of 2450 MeV in which
a $u$ quark and a $\bar{d}$ quark are bound by the confining
forces; both quark spins are parallel with the orbital
angular momentum $l=5\hbar$. Quark and antiquark are separated
by 8\,fm\,! 
\par
The type of potential given in (\ref{alpha}) was independently 
suggested to fit the mass spectrum of the J/$\psi$
and $Y$ families of states. This phenomenological approach
yielded a very similar potential; hence we may conclude 
(and there are many other examples) that QCD can be
simulated on a lattice. 
\par
So far, so good. But QCD on the lattice predicts also
one new peculiar type of hadronic matter which does not
comprise constituent quarks, it predicts glueballs.
The search for glueballs has stimulated the field
and has led to extensive searches for these states. 
Many glueball candidates have been suggested in the 
past; also at present there are reasons to believe
that a glueball has been traced in the spectrum of scalar 
mesons. Personally, I do not share this view. Hence
I will discuss the evidence which exists for
the scalar glueball, and why I think that data tell us
that glueballs may not exist at all. 
\par
Apart from glueballs, also bound states of $q\bar q$+gluon
are expected to exist. 
The color 3 of a quark and the $\bar 3$ of an
antiquark may form a color singlet: that is then
a observable state. But colour 3 and $\bar 3$ 
can also couple to 8 which can be colour-neutralized
by one gluon (which is always colour 8). These
states are called hybrids. You may also think
of haybrids as a vibrating color flux tube, as an
excitation of the gluon field. 
\par
The low-mass glueballs have quantum numbers like 
other mesons; hence they would show up in the spectrum
of mesons as additional states. 
Hybrids may have quantum numbers which
are not accessible to normal $q\bar q$ mesons. 
Therefore it seems appropriate to continue with a 
short review of the properties of $q\bar q$ mesons.
\section{The quark model}
\subsection{Mesons and their quantum numbers}
Quarks have spin 1/2 and baryon number 1/3; 
antiquarks have spin 1/2 and baryon number -1/3. 
Quark and antiquark combine to B=0 and to 
spin $S =1$ or $S = 0$ thus  forming conventional mesons. 
The total spin $\vec S$
and the orbital angular momentum $\vec L$
between quark and antiquark couple to the
total angular momentum $\vec J$: $\vec J = \vec L
+ \vec S$.  
 forming conventional mesons.
As $q\bar q$ objects they may have
the following properties:
\subsubsection{Parity $P = (-1)^{L+1}$}
The parity of a meson is due to the orbital 
angular momentum between quark and antiquark $P = (-1)^{L}$
multiplied with their intrinsic parities 
$P_q\cdot P_{\bar q}$ = -1. 
\vspace*{-5mm}
\subsubsection{C-Parity $C = (-1)^{L+S}$}
The total wave function of a meson 
is antisymmetric w.r.t. the
exchange of quark and antiquark. The symmetry
of the wave function 
is given by the product of symmetries 
of the spatial and spin 
wave function, and by the $C$-parity:

\begin{tabular}{lcc}
Spins:  & $(-1)^{S+1}$ &   \\
Space:  & $(-1)^{L}$   & The product is -1, hence \\
Charge: &  C           &    \\
\hline
\end{tabular}
$$
 C = (-1)^{S+L+1} = -1 \quad {\rm and} \quad C = (-1)^{S+L}
$$
\subsubsection{Isospin}
Proton and neutron form an isospin doublet and
so do the {\it up} and the {\it down} quark. We may
construct states of 2 or 3 quarks; the isospin
of the system is determined by adding the quark
isospins using Clebsch-Gordan coefficients.
We could also define $\bar u$ and $\bar d$ as
isospin doublet in a isospin space of 
antiparticles and invent a new table of Clebsch-Gordan 
coefficients. Or we can define iso-doublets in a
way that the doublets of antiparticle transform
under isospin rotations like those of particle
doublets. These are, e.g.:
\bc
\begin{tabular}{cccc}
\ \ $\left(
\begin{array}{l}
p\\
n\\
\end{array}
\right)$\ \
&
$\left(
\begin{array}{l}
u\\
d\\
\end{array}
\right)$\ \
&
$\left(
\begin{array}{l}
\kp\\
\kn\\
\end{array}
\right)$\ \
&
$\left(
\begin{array}{l}
\ksp\\
\ksn\\
\end{array}
\right)$
\end{tabular}
\ec
\bc
\begin{tabular}{cccc}
$\left(
\begin{array}{r}
-\bar n\\
 \bar p\\
\end{array}
\right)$
&
$\left(
\begin{array}{r}
-\bar d\\
 \bar u\\
\end{array}
\right)$ 
&
$\left(
\begin{array}{r}
-\knb\\
\km\\
\end{array}
\right)$
&
$\left(
\begin{array}{r}
\ksnb\\
-\ksm\\
\end{array}
\right)$
\end{tabular}
\ec
We now can contruct $q\bar q$ mesons like pions
or \etg 's.  

\begin{tabular}{lcccr}
$|I=1, I_3=1 >$&=&$-|u\bar{d} >                                $&=&$-|\pip >$\\
$|I=1, I_3=0 >$&=&$\frac{1}{\sqrt{2}}(|u\bar{u}> - |d\bar{d} >)$&=&$|\piz >$\\
$|I=1, I_3=-1>$&=&$  |d\bar{u} >                               $&=&$|\pim >$\\
              && &&\\
$|I=0, I_3=0 >$&=&$\frac{1}{\sqrt{2}}(|u\bar{u}> + |d\bar{d} >)$&=&$|''\etg ''>$
\end{tabular}

The three states -\Pip , \Piz\ and \Pim\ form an iso-triplet, 
\etg\ and \etp\ both form isosinglets.
\subsubsection{ $G$-parity  $P = (-1)^{L+1+I}$}
The C-parity has a defined eigenvalue only
for particles which are their own
antiparticles. The action of C-parity on other states
leads to their antiparticles. We define
\bc
\begin{tabular}{lcllcllcl}
C  \Piz\  &=& + \Piz\ ;&  C \Pip\  &=& \Pim\ ;& C \Pim\  &=& \Pip \\
\end{tabular}
\ec
It is useful to introduce also a
$G$-parity as $C$-parity followed
by a rotation in isospin space by 90$^{\circ}$ 
about the $y$-axis. The rotation 
by 180$^{\circ}$ in isospin space around y-axis
are given by 
$$\bf e^{i\pi I_y}$$ 
We define: \quad\
$\bf G = C\cdot e^{i\pi I_y}$
and introduce {\it Cartesian states}:
$$
\Pipm\ = \frac{1}{\sqrt 2}\mid\pi^x \pm\ i\pi^y> \qquad\ {\rm Then:}
$$
\vspace*{-10mm}
\begin{eqnarray}
G|\pi^{\pm}>& = & C  \frac{1}{\sqrt 2} e^{i\pi I_y} \mid\pi^x \pm\ i\pi^y >
\nonumber \\
         && C  \frac{1}{\sqrt 2}\mid -\pi^x \pm\ i\pi^y> =
         (-1) C \mid\pimp > = - \mid\pipm >\nonumber \\
         &&\nonumber\\
G|\pi^{0}>&  = & C  e^{i\pi I_y}|\pi^{0}> = - C |\pi^{0}> = -|\pi^{0}>
\nonumber 
\end{eqnarray}
\vspace*{-5mm}
Thus we have:
$$
G |\pi > = - |\pi >
$$
\underline{$G$-parity is conserved in strong interactions.}\\ 
For mesons decaying into ${n_\pi}$ pions we have the relation
$$
G = (-1)^I\cdot C = (-1)^{L+S+I} = (-1)^{n_\pi}
$$
\subsection{Meson nonets}
Mesons are characterized by their quantum numbers 
$J^{PC}$ and by their flavor content. In SU(3),
i.e. in the light-quark domain, a set of quantum
numbers $J^{PC}$ leads to a nonet of states. 
Based on SU(3) symmetry, we expect an octet and a
singlet. However, the $s$ quark is heavier than the
$u$ and $d$ quark. The three pairs $u\bar u$, $d\bar d$ and
$s\bar s$ can therefore form mesons which are 
approximately SU(3) eigenstates or - approximately -
mesons composed of $u\bar u$  and $d\bar d$ or $s\bar s$ pairs.
\subsubsection{The pseudoscalar mesons}
The pseudoscalar quantum numbers are  $J^{PC}=0^{-+}$. 
From atomic physics we take the spectroscopic notation
n$^{2s+1}L_J = 1^1S_0$. From the 3 quarks $u,d,s$ and 
their antiquarks the following SU(3) eigenstates
can be constructed:
$$K^0 = d\bar s \qquad\ K^+ = u\bar s $$
$$\pi^- = d\bar u \qquad\ \pi^0 = \frac{1}{\sqrt 2}(u\bar u - d\bar d)
\qquad\ \pi^+ = u\bar d $$
$$K^- = s\bar u \qquad\ K^0 = s\bar d $$
$$\eta_8 = \sqrt{\frac{1}{6}}(u\bar u +d\bar d -2s\bar s ) \qquad\
\eta_1 = \sqrt{\frac{1}{3}}(u\bar u +d\bar d +s\bar s )$$
\par
The 9 states are orthogonal; one of them, the $\eta_1$,
is invariant under rotations in SU(3).
The nonet structure is seen in the well-known
nonet representation:

\vspace*{-15mm}

\begin{center}
\nonett{${K^0}$}{${K^+}$}{${K^-}$}
{$\overline{{K^0}}$}{$\pi^-$}{$\pi^+$}{$\pi^0$}
{$\eta_8$}{$\eta_1$}
\end{center}

\vspace*{-15mm}

The octet state and the singlet state are the SU(3)
eigenstates. They have the same 
quantum numbers and can mix. The mixing angle is
called pseudoscalar mixing angle $\Theta_{PS}$. The 
physical states are given by 
$$|\eta > = \cos\Theta_{PS}|\eta_8> - \sin\Theta_{PS}|\eta_1>  $$
$$|\eta^{\prime}> = \sin\Theta_{PS}|\eta_8> + \cos\Theta_{PS}|\eta_1>$$
We can write
down the flavor wave function for a few angles:

\begin{tabular}{lrccc}
$\Theta_{PS}$ =& 0$^{\circ}$\qquad\     & $|\eta >$        &=&
$\frac{1}{\sqrt 6}\left(u\bar u + d\bar d - 2s\bar s\right)$ \\
              &                 & $|\eta^{\prime}>$&=&
$\frac{1}{\sqrt 3}\left(u\bar u + d\bar d + s\bar s\right)$ \\ 
&&&&\\
$\Theta_{PS}$ =& $-11.1^{\circ}$\qquad\ & $|\eta >$        &=&
$\frac{1}{\sqrt 2}\left(\frac{1}{\sqrt 2}(u\bar u + d\bar d) - s\bar s\right)$\\ 
              &                 & $|\eta^{\prime}>$&=&
$ \frac{1}{\sqrt 2}\left(\frac{1}{\sqrt 2}(u\bar u + d\bar d) + s\bar s\right)$ \\ 
&&&&\\
$\Theta_{PS}$ =& $-19.3^{\circ}$\qquad\ & $|\eta > $       &=&
$ \frac{1}{\sqrt 3}\left(u\bar u + d\bar d - s\bar s\right)$ \\
              &                 & $|\eta^{\prime}>$&=&
$ \frac{1}{\sqrt 6}\left(u\bar u + d\bar d + 2s\bar s\right)$ \\
&&&&\\
$\Theta_{PS}$ =& $35.3^{\circ}$\qquad\  &  $|\eta > $      &=&
$s\bar s$ \\
              &                 &$|\eta^{\prime}> $&=&
$ \frac{1}{\sqrt 2}\left(u\bar u + d\bar d\right)$ \\
\end{tabular}

The large mixing between the $\sqrt{\frac{1}{2}}(u\bar u +d\bar d)$ 
(which we abbreviate as $n\bar n$) and the
$s\bar s$ component in the \etg\ and \etp\ wave functions
has led to speculations that the \etg\ and in particular
the \etp\ may contain a large fraction of glue, that
they are {\it gluish}. 
This requires an extension of the mixing
scheme by introduction of a non-$q\bar q$ or 
{\it inert} component,
with a third state of unknown mass which is dominantly
a glueball. 
\par
\begin{tabular}{cccc}
$|\eta >$ = & ${X_{\eta}}\cdot 
\frac{1}{\sqrt 2}\left(u\bar u + d\bar d\right)$ + 
& ${Y_{\eta}}\cdot\left(s\bar s\right)$ + 
& ${Z_{\eta}}\cdot\left({\rm glue}\right)$ \\
$|\eta^{\prime} >$ = &${X_{\eta^{\prime}}}\cdot 
\frac{1}{\sqrt 2}\left(u\bar u + d\bar d\right)$ + 
& ${Y_{\eta^{\prime}}}\cdot\left(s\bar s\right)$ 
+ & ${Z_{\eta^{\prime}}}\cdot\left({\rm glue}\right)$ \\
&light quark & strange quark 
&inert  \\
&&&$Z_{\eta}=Z_{\eta^{\prime}}\sim 0$\\
\end{tabular}
\par
Indeed, a first systematic survey 
of J/$\psi$ decays into vector and pseudoscalar mesons
suggested that the \etp\ may contain a large glueball 
fraction $(\sim 35\%)$ \cite{Baltrusaitis:1985rz}. 
At that time, the $\iota (1440)$ or $\eta (1440)$ was
believed to have a large glueball fraction and 
mixing of \etg , \etp, and \etg (1440) was held
responsible for the gluish nature of the \etp .
Yet later more precise data excluded this possibility 
\cite{Coffman:1988ve,Jousset:1990ni}. Nevertheless,
this question remained a controversial topic since then.
In a very recent article it was shown that there is
no need to introduce glue into the \etp\ wave
function \cite{Benayoun:2000}.
\par
At present there is no convincing evidence for a
glueball content in the \etp\ wave function; nevertheless
the \etp\ is still suspect of being produced preferentially
in glue-rich processes or in glueball decays.
\par
\subsubsection{Vector and tensor mesons}
On the next page we show - without further comments - 
the nonet of vector and tensor mesons with $J^{PC} = 1^{--}$
and with $2^{++}$, respectively, or of the 
1$^3S_1$ 
and 1$^3P_2$ states.  Both nonets have a
nearly {\it ideal mixing angle} $\Theta_{ideal} = 
35.3^{\circ}$ for which one meson is a purely $s\bar s$ state.
Note that the mass difference between the $s\bar s$
and the $u\bar u+d\bar d$ state is about 250 MeV.
\subsubsection{Other meson nonets}
A meson nonet is fully described by just 4 names. The
pseudoscalar nonet contains 3 pions, four kaon, the \etp\ and the
\etg . 
In Table \ref{nonets} some meson nonets are collected.
The $f_1(1510)$ is chosen as 
$s\bar s$state instead of the $f_1(1420)$ as only the former
has mass and decay modes compatible with values expected 
from SU(3) arguments \cite{Gavillet:1982tv}. The \etg (1295) is
mostly considered to be the radial excitation of the
\etg\ ground state. This assignment is challenged by
its non-observation in radiative J/$\psi$ decays, in
\pbp\ annihilation \cite{Suh:1999kv} and in 2-photon
collisions \cite{Acciarri:2000ev} while the \etg (1440) is
observed in all three reactions.
\par
\begin{table}
\bc
\renewcommand{\arraystretch}{1.3}
\begin{tabular}{|cccc|cccc|cc|}
\hline
L & S & J & n & I=1 & I=1/2 & I=0   & I=0 & $J^{PC}$ & $n^{2s+1}L_J$  \\
\hline
0 & 0 & 0 & 1 & \p  &  K    &\etp   & \etg & $0^{-+}$ & $1^1S_0$\\
0 & 1 & 1 & 1 & \rh & K$^*$ &$\Phi$ & \omg & $1^{--}$ & $1^3S_1$ \\
\hline
1 & 0 & 1 & 1 &$b_1(1235)$&K$_{1B}$ &$h_1(1380)$ &$h_1(1170)$ 
 & $1^{+-}$ & $1^1P_1$\\
1 & 1 & 0 & 1 &$a_0(????)$&K$_{0}^*(1430)$ &$f_0(????)$ &$f_0(????)$ 
 & $0^{++}$ & $1^3P_0$\\
1 & 1 & 1 & 1 &$a_1(1260)$&K$_{1A}$ &$f_1(1510)$ &$f_1(1285)$ 
 & $1^{++}$ & $1^3P_1$\\
1 & 1 & 2 & 1 &$a_2(1320)$&K$_{2}^*(1430)$ &$f_2(1525)$ &$f_2(1270)$ 
 & $2^{++}$ & $1^3P_2$\\
\hline
2 & 0 & 2 & 1 &$\pi_2(1670)$&K$_{2}(1770)$ &$\eta_2(1645)$ &$\eta_2(1870)$ 
 & $2^{-+}$ & $1^1D_2$\\
2 & 1 & 1 & 1 &$\rho(1700)$&K$^*(1680)$ &$\omega(1650)$ &$\Phi(????)$ 
 & $1^{--}$ & $1^3D_1$\\
2 & 1 & 2 & 1 &$\rho_2(????)$&K$_2(1820)$ &$\omega_2(????)$ &$\Phi_2(????)$ 
 & $2^{--}$ & $1^3D_2$\\
2 & 1 & 3 & 1 &$\rho_3(1690)$&K$^*_3(1780)$ &$\omega_3(1670)$ &$\Phi_3(1850)$ 
 & $3^{--}$ & $1^3D_3$\\
\hline
0 & 0 & 0 & 2 &\p (1370) &  K$_0(1460)$    &\etg (????)  & \etg (1440)
 & $0^{-+}$ & $2^1S_0$\\
0 & 1 & 1 & 2 &\rh (1450)& K$^*(1450)$ &$\Phi (1680)$ & \omg (1420)
 & $1^{--}$ & $2^3S_1$\\
\hline
\end{tabular}
\renewcommand{\arraystretch}{1.0}
\ec
\caption{The light mesons. The two mesons K$_{1A}$ and K$_{1B}$ 
mix to form the observed resonances K$_{1}(1280)$ and K$_{1}(1400)$.
The two $\etg_2$ states are from \protect\cite{Adomeit:1996nr}.
In some cases, mesons still need to be identified. 
The scalar mesons resist an unambiguous classification.}
\label{nonets}
\end{table}
\clearpage
%
%
%
%
\hfill {\large \underline{The vector mesons  $J^{PC} = 1^{--}$}}
\vspace*{-10mm}
\begin{center}
\nonett{${K^0}^{\ast}$}{${K^+}^{\ast}$}{${K^-}^{\ast}$}
{$\overline{{K^0}^{\ast}}$}{$\rho^-$}{$\rho^+$}{$\rho^0$}
{$\omega_8$}{$\omega_1$}
\end{center}
\vspace*{-10mm}
\hfill {\large \underline{The tensor mesons  $J^{PC} = 2^{++}$}}
\vspace*{-9mm}
%
%
%
\begin{center}
\nonett{${K_2^0}^\ast(1430)$}{${K_2^+}^\ast(1430)$}{${K_2^-}^\ast(1430)$}
{$\overline{{K_2^0}^\ast}(1430)$}{$a_2^-(1320)$}{$a_2^+(1320)$}
{\hspace*{-12mm}$a_2^0(1320)$}
{\hspace*{-12mm}$f_2(8)$}{$f_2(1)$}
\vspace*{-5mm}
\end{center}
\bc
\renewcommand{\arraystretch}{1.5}
\begin{tabular}{||lccccc||}
\hline
\hline
{$\Theta_{V,T} = 35.3^{\circ}$} &
$|\omega>$&=&$ \frac{1}{\sqrt 2}\left(u\bar u + d\bar d\right)$&
$\sim$&$f_2(1270)$ \\
& $|\Phi > $&$\sim$&$ s\bar s$&=&$f_2(1525)$ \\
\hline
\hline
\end{tabular}
\renewcommand{\arraystretch}{1.0}
\ec
\subsubsection{The Gell-Mann-Okubo mass formula}
You can derive a relation between the masses within a meson
nonet by ascribing to mesons of one nonet a common mass $M_0$
plus the (constituent) masses of the quark and antiquark
it is composed of. The  pion mass is given by
$$
M_{\pi}  = M_0 + 2M_q
$$
where $M_q$ is the mass of the up or down quark, and the Kaon mass by
$$
M_{K}    = M_0 + M_q + M_s
$$
with $M_s$ as strange quark mass. 
The \etg\ contains masses from both the singlet 
and octet
component which we weight according to their fractions:
$$
M_{\eta} = M_8\cos^2\Theta + M_1\sin^2\Theta
$$
$$
M_{\eta^{\prime}} = M_8\sin^2\Theta + M_1\cos^2\Theta
$$
Similarly we determine the singlet and octet masses 
from the flavor decomposition of their wave functions.
$$
M_1 = M_0 + 4/3 M_q + 2/3 M_s
$$
$$
M_8 = M_0 + 2/3 M_q + 4/3 M_s
$$
Thus we arrive at the linear mass formula:
$$
\cos^2\Theta = \frac{3M_{\eta} + M_{\pi} - 4M_{K}}
{4M_{K} - 3M_{\eta^{\prime}} - M_{\pi}}
$$
Often, the linear GMO mass formula is replaced by
the quadratic GMO formula which is given as above but
with M$^2$ values instead of masses. It reads
$$
\cos^2\Theta = \frac{3M^{2}_{\eta} + M_{\pi}^2 - 4M_{K}^2}
{4M_{K}^2 - 3M^{2}_{\eta^{\prime}} - M_{\pi}^2}
$$
\renewcommand{\arraystretch}{1.3}
\bc
\begin{tabular}{|c|c|c|}
\hline
          Nonet members & $\Theta_{\rm linear}$ & $\Theta_{\rm quad} $ \\
$\pi , K , \eta^{\prime} , \eta $ & $-23^{\circ}$ & $-10^{\circ}$ \\
$\rho , K^* , \Phi\ , \omega $    & $36^{\circ}$ & $39^{\circ}$  \\
$a_2(1320) , K^{*}_{2}(1430) , f_2(1525) , f_2(1270)$ & 
$26^{\circ}$ & $29^{\circ}$ \\
$\rho_3(1690) , K^*_3(1780) , \Phi_3(1850)\ , \omega_3(1670)$ & $29^{\circ}$ & $28^{\circ}$  \\
\hline
\end{tabular}
\ec
\renewcommand{\arraystretch}{1.0}
\subsubsection{Meson decays}
The decays of mesons belonging to a given nonet are related
by SU(3) symmetry. The coefficients governing these relations are
called SU(3) isoscalar factors and listed by the Particle Data
Group \cite{pdg}. We show here two simple examples.
\par
A glueball is, by definition, a flavor singlet. It may decay
into two octet mesons. Hence we have the decay $1 \rightarrow\ 8 \times\ 8$.
In the listings we find
$$
(\Lambda ) \rightarrow\ (N\bar K\ \ \Sigma\pi\ \ \Lambda\eta\ \ \Xi K) =
\frac{1}{\sqrt 8}(2 \ \ 3 \ -1 \ -2)^{1/2}
$$
The particles stand for their SU(3) assignment, the $\Lambda$ can be
octet or singlet. The $^{1/2}$ is understood for every coefficient.
Translated into decays of a flavor singlet meson into two pseudoscalar
mesons it reads
$$
({\rm glueball}) \rightarrow\ (K\bar K\ \ \pi\pi\ \ \eta_8\eta_8\ \ \bar KK) =
\frac{1}{\sqrt 8}(2 \ \ 3 \ -1 \ -2)^{1/2}
$$
Hence glueballs have squared couplings to \kkb , \p\p , $\eta_8\eta_8$ of
4\,:\,3\,:\,1\,. The decay into two isosinglet mesons $\eta_1\eta_1$ has
an independent coupling and is not restricted by these SU(3) relations. The
decay into $\eta_1\eta_8$ is forbidden: a singlet cannot decay into
a singlet and an octet meson. This selection rule holds even for
any pseudoscalar mixing angle: the two mesons \etg\ and \etp\ have 
orthogonal SU(3) flavor states and a flavor singlet cannot dissociate
into two states which are orthogonal. 
\par
As second example we choose decays of vector mesons into two
pseudoscalar mesons. We compare the two decays K$^*$\ra K\p\ and
$\rho$\ra\p\p . These are decays of octet particles into 
two octet particles, $8 \rightarrow\ 8 \times 8$. Two
octets can couple to an octet with symmetry or antisymmetry
w.r.t. their exchange. The two pions in $\rho$ decay must be 
antisymmetric, hence we have to use the isoscalar factors for 
$8_2 \rightarrow\ 8 \times 8$. 
$$
(K^*) \rightarrow\ (K\pi\ \ K\eta\ \ \pi K\ \ \eta K) =
\frac{1}{\sqrt{12}}(3 \ \ 3 \ \ 3 \ -3)^{1/2}
$$
$$
(\rho ) \rightarrow\ (K\bar K\ \ \pi\pi\ \ \eta\pi\ \ \pi\eta\ \ \bar KK) =
\frac{1}{\sqrt{12}}(2 \ \ 8 \ \ 0 \ \ 0 \ -2)^{1/2}
$$
Hence we derive K$^*$\ra K\p +\p K $\propto\ 6, \ \ \rho$\ra\p\p\ 
$\propto\ 8$, \ \ or
$$
\frac{\Gamma_{K^*\ra K\p +\p K}}{\Gamma_{\rho\ra\p\p}} = \frac{6}{8}
\left(\frac{0.291}{0.358}\right)^3 = 0.40
$$
The latter factor is the ratio of the decay
momenta $q$ to the 3rd power. 
The transition probability is proportional to $q$; for
low momenta, the centrifugal barrier scales with $q^{2l}$
where $l$ is the angular momentum.
\par
From data we know that the width ratio is 0.34. So the relations
are o.k. at the level of $\sim 20\%$. This is a typical level of
SU(3) breaking effects. We have neglected many things: mesons have
a size; the $\rho$ and K$^*$ sizes are different; the angular 
barrier factor should include Blatt-Weisskopf corrections. An
application of SU(3) to vector and tensor mesons can be found
in \cite{Peters:1995jv}.
\subsubsection{Scalar mesons}
Of particular interest is the spectrum of scalar
mesons since the lowest-mass glueball is expected to
have quantum numbers $J^{PC} = 0^{++}$
or 1$^3P_0$ (see Fig.~\ref{glueballs}). Unfortunately,
the information on scalar mesons is not unique. A few
years ago, only little was known about scalar mesons.
The  experimental situation has improved in the meanwhile
but the discussion is still controversial. Below you
find the ground state nonet of scalar mesons as most
physicists in the field would agree upon. Clearly,
the situation is unsatisfactory.
There is a number of candidates to fill in the question
marks but there is no general agreement which meson
should go where.
\vspace*{-1cm}

\begin{center}
\nonett{${K_0^0}^\ast(1430)$}{${K_0^+}^\ast(1430)$}{${K_0^-}^\ast(1430)$}
{$\overline{{K_0^0}^\ast}(1430)$}{$a_0^-(????)$}{$a_0^+(????)$}
{\hspace*{-12mm}$a_0^0(????)$}
{\hspace*{-12mm}$f_0(????)$}{$f_0(????)$}
\end{center}

\vspace*{-1cm}

\par
\subsection{Beyond the quark model}
Mesons which are composed of a constituent quark and antiquark
are referred to as conventional mesons. In addition,  other 
forms of hadronic matter are supposed to exist. These are
\underline{glueballs}, excitations of the QCD vacuum or
hadrons without any constituent quark,
\underline{hybrids}, hadrons in which the gluonic string
mediating the color flux between quark and antiquark is 
excited, or \underline{multiquark states} with 2 or 3 $q\bar q$ pairs
as constituent particles. If two (three) $q\bar q$ pairs are clustered into
two separate mesons (nucleons), we speak of mesonic molecules
or of quasi-nuclear states. 
\par
In the next sections 
of this manuscript I will concentrate on new information 
on scalar mesons and will attempt my own interpretation. 
It has the disadvantage that very few 
physicists working in the field will agree to this view,
it has the advantage that it is what I believe to be
close to the 'truth'. Physicists who disagree with the view
presented here may check their favored interpretation 
against the experimental findings and the consequences
drawn here. 
\par
An excellent and unbiased modern review of meson spectroscopy can 
be found in \cite{Godfrey:1999pd}. There are many open questions
at present concerning glueballs and hybrids; these are
discussed in detail in this report. 
\par
\section{Glueballs and their ground state}
Glueballs and hybrids reflect new degrees of freedom brought into hadron 
spectroscopy by QCD and are therefore of prime interest. Indeed, the 
main motivation of current experiments on meson spectroscopy is the 
quest to search for glueballs and hybrids, to establish their 
non-$q\bar q$ character and to determine their properties: masses, total 
and partial widths, and their mixing with ordinary $q\bar q$ states 
having the same quantum numbers. There are indeed strong candidates, 
both for glueballs and for hybrids. The number of scalar states with 
$I^G(J^{PC})=0^+(0^{++})$ seems to be too large to be accommodated within the 
quark model. On the other hand, none of the states has decay properties 
as expected for a pure glueball. Mixing scenarios have hence been 
proposed in which the pattern of observed states is understood as 
quarkonia mixing with a primordial glueball intruding 
into the $q\bar q$ world. The discussion of the scalar mesons 
is the content of this section.

\subsection{Where to find glueballs and how}
\subsubsection{Glueballs and their masses}
\par
\begin{figure}[h!]        
\begin{center}        
\epsfig{file=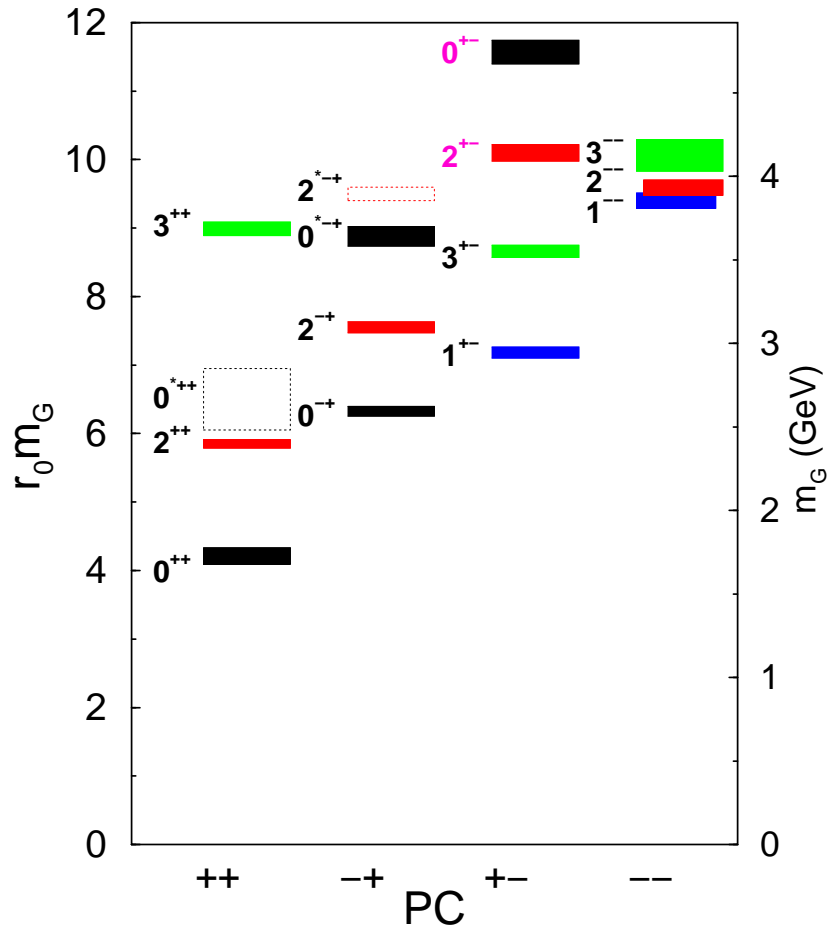,width=0.8\textwidth,angle=0
}
\caption{The glueball spectrum from an anisotropic lattice 
study \protect\cite{lqcd}}
\label{glueballs}
\end{center}  
\vspace*{-5mm}
\end{figure}

The most trusted predictions for the glueball mass spectrum
are based on lattice gauge calculations. They have become
increasingly precise with the advent of high-speed computers
and new and efficient codes. Fig.~\ref{glueballs} shows a recent
calculation of the glueball mass spectrum.
The ground state is a scalar state, at about 1730 MeV; followed
by a tensor and pseudoscalar glueball with masses of 2300 and 
2350 MeV, respectively. The uncertainty of these calculations is
estimated to be of the order of 100 MeV. These masses are supported 
by other models, bag models~\cite{Chodos:1974pn},  
flux tubes \cite{Isgur:1983wj}, or QCD sum 
rules \cite{narison}. In publications
on bag model calculations, rather low glueball masses were 
quoted since the \etg (1440) was identified as pseudoscalar glueball
and was used as a mass scale. 
\par
The low-lying glueballs all have quantum
numbers which allow mixing with conventional mesons.
Hence you should
expect mixing between glueballs and $q\bar q$ mesons.
The glueball strength is then diluted over
two or more physical states. This is very difficult to 
establish (unless you impose the existence of a glueball
right from the beginning), and even more difficult to rule out. 
\par
Glueballs are compact objects. The size of the
ground state is predicted to be smaller than that of 
$q\bar q$ mesons, $\sim 0.5$fm or smaller. 
\subsubsection{Hints for glueball hunters}
 \begin{minipage}[t][4cm][t]{.6\textwidth}
Glueballs are supposed to be produced preferentially
in gluon-rich processes like, e.g., radiative
J/$\Psi$ decays. The J/$\Psi$ is narrow: the OZI
rule suppresses decays of the $c\bar c$ system into light
quarks and the $D\bar D$ threshold is far above
the mass of the J/$\Psi$.  
In most decays the J/$\Psi$ undergoes a transition
into 3 gluons which then convert into hadrons. But the
J/$\Psi$ can also decay into 2 gluons and a photon. The
2 gluons can interact and must form glueballs - if they exist.
\end{minipage}\hfill
 \begin{minipage}[t][4cm][t]{.4\textwidth}
\vspace*{-1cm}
  \begin{center}
  \begin{tabular}{c}
    \hspace{1cm}
    \epsfig{file=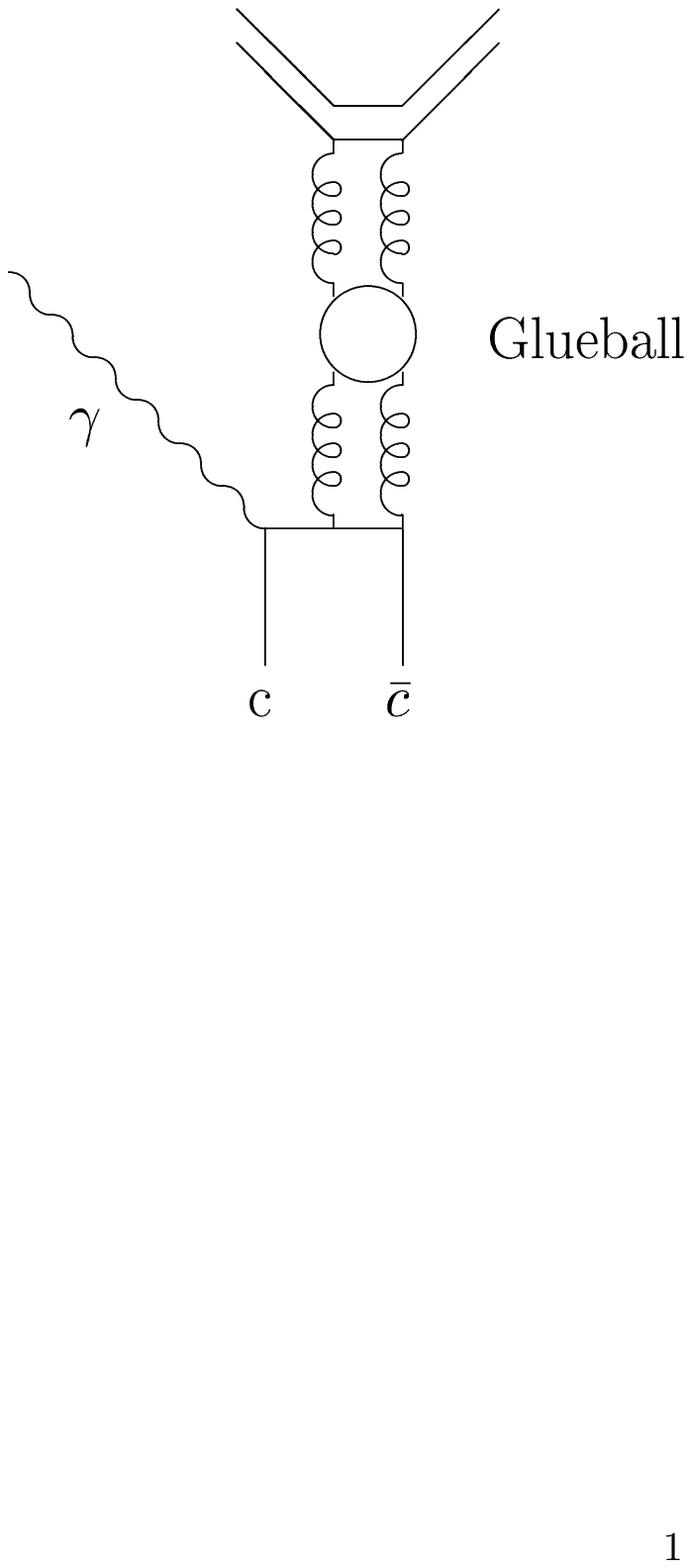,height=6cm,clip=,bbllx=120,bblly=360,bburx=340,bbury=600}
  \end{tabular}
  \end{center}
\end{minipage}
\vskip 1.5cm
\noindent
\begin{minipage}[t][4cm][t]{.6\textwidth}
Central production is another process in which glueball should be
produced abundantly. In central production two hadrons pass by
each other 'nearly untouched' 
and are scattered diffractively in forward direction. No valence 
quarks are exchanged. Therefore this process is often called 
Pomeron-Pomeron scattering. The absence of valence quarks 
in the production process makes central production a good place
to search for glueballs. 
\end{minipage}\hfill
 \begin{minipage}[t][4cm][t]{.4\textwidth}
\vspace*{-2cm}
  \begin{center}
  \begin{tabular}{c}
    \hspace{1cm} 
    \epsfig{file=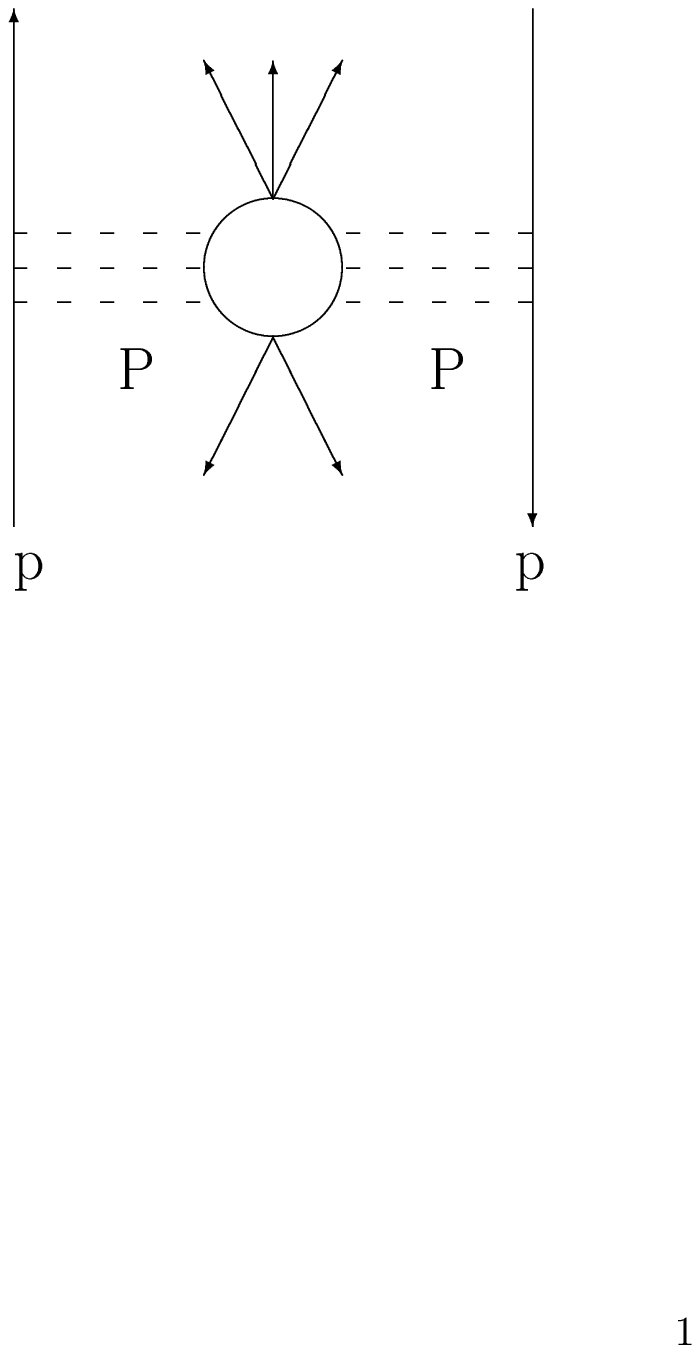,height=8cm,clip=,bbllx=110,bblly=250,bburx=340,bbury=600}\\
  \end{tabular}
  \end{center}
\end{minipage}
\vskip 1.5cm

\noindent
 \begin{minipage}[t][4cm][t]{.6\textwidth}
Finally, one can argue that in \pbp\ annihilation quark-antiquark
pairs annihilate into gluons, they will interact and may form 
glueballs. In any case, any glueball decays into hadrons and hence
hadroproduction of glueballs must be possible. Hadroproduction
experiments have the advantage - compared to J/$\Psi$ decays -
that much higher statistics can be collected. 
\end{minipage}\hfill
 \begin{minipage}[t][4cm][c]{.35\textwidth}
\vspace*{-2cm}\hspace*{2cm}
   \epsfig{file=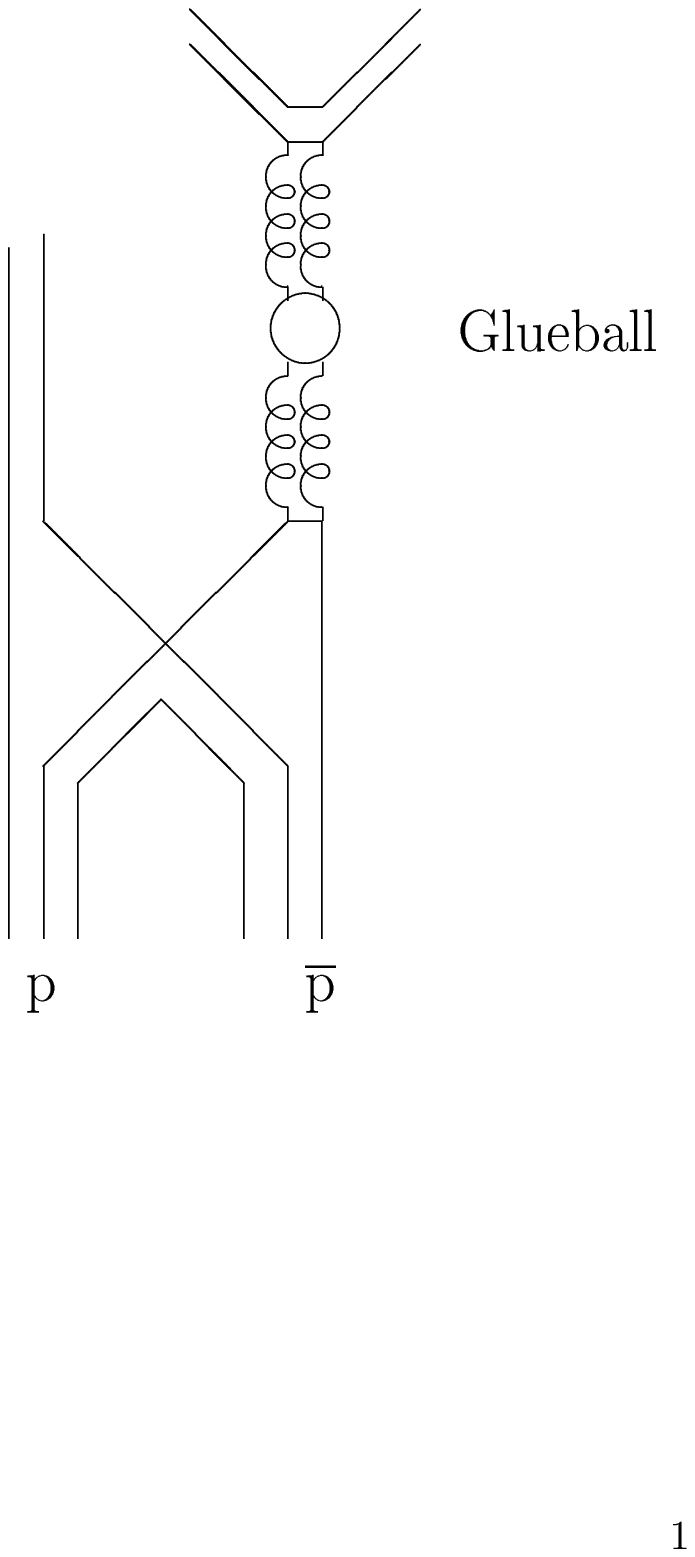,height=6cm,clip=,bbllx=110,bblly=290,bburx=340,bbury=600}\\
 \end{minipage}
%
%
\par
A further distinctive property of glueballs is their decay.
Being a flavor singlet, glueballs should decay with
flavor symmetry: thus the decay into \p\p , \etg\etg , 
\etg\etp\ and \kkb\ should scale as
3\,:\,1\,:\,0\,:\,4\,, after correcting for phase space.
As we have seen, \etg\ and \etp\ mesons may be gluish, and SU(3)
symmetry could be broken in favor of decays into these two
mesons. Also glueball-$q\bar q$ mixing will destroy this simple pattern. 
A detailed and still useful account of early suggestions how to search for
glueballs and how to identify them can be found in \cite{Robson:1977pm}.
\subsection{Scalar mesons and the Crystal Barrel experiment at LEAR} 
Our knowledge on scalar mesons was greatly improved when the data from
the Crystal Barrel experiment were analyzed. The Crystal Barrel detector
was one of the experimental installations at LEAR; its most prominent
feature are 1380 CsI crystals surrounding a H$_2$ or D$_2$
target all pointing at the target center, with a solid angle coverage 
of 98\% of 4\p . The shape of the detector -
a barrel - was chosen to house a vertex detector and an inner drift chamber
for charged particle reconstruction. The results and analysis methods
are reviewed in detail by Amsler \cite{Amsler:1998up}. 

\subsubsection{The life history of antiprotons}
Antiprotons annihilating in H$_2$ or D$_2$ undergo a series
of processes which are very relevant for the annihilation process.
Therefore we review shortly the capture process and the atomic cascade
which precedes annihilation. We use H$_2$ as example. 
\par
\begin{figure}[h!]
\epsfig{file=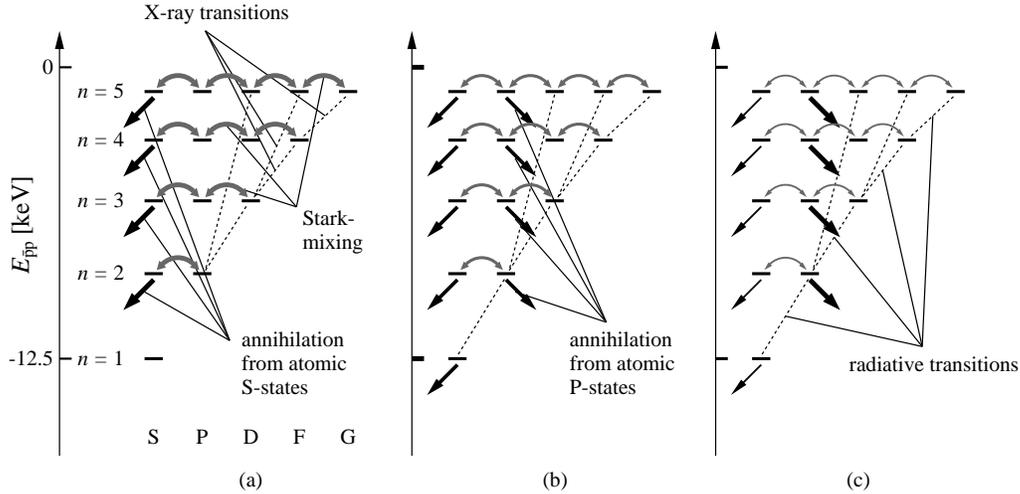,width=\textwidth}
\caption{Processes contributing to the cascade of antiprotonic
hydrogen atoms, see text. 
}
\label{cascade}
  \end{figure}

Antiprotons stopping in hydrogen are captured in the Coulomb field of 
a proton by Auger emission of an electron (or chemical dissociation
of H$_2$ and subsequent internal Auger effect)
thus forming antiprotonic hydrogen atoms. Four processes
are important: Auger ejection of electron from neighboring hydrogen
atoms, transitions between states with different angular momenta $l$
but identical principal quantum number $n$ 
due to Stark mixing in the presence of strong electric fields, 
radiative transitions to lower levels, and annihilation. Obviously
the first two processes are density dependent. At the density
of liquid H$_2$, they play a very important role. The
Stark effect mixes the angular momentum  with the effect
that $\sim$90\% of all antiprotons annihilate from high-n $S$ 
states of the \pbp\ atom. At lower density, e.g. in H$_2$ gas,
Stark mixing rates are smaller and the chance increases that \pbp\
atoms in $P$ states live long enough to annihilate before the
next collision occurs: at atmospheric pressure the chances for 
$S$ and $P$ capture are about equal. 
In low-pressure gas, Stark mixing and Auger
rates are sufficiently slow to allow radiative transitions
to the 2$P$ level from which then $P$-state capture predominates
(99\%). The strong interaction is weak in $D$ levels
and no annihilation occurs from those levels.
The processes are depicted in Fig.~\ref{cascade}. 
A review of the field of antiprotonic hydrogen atoms can be
found in \cite{Batty:1989gg}.
 
\subsubsection{Quantum numbers of the \pbp\ system}
Proton and antiproton both carry isospin
$
|\textrm{I},\textrm{I}_3> = |\frac{1}{2},\pm\frac{1}{2}>.
$
The two isospin couple to 
$
|\textrm{I}=0,\textrm{I}_3> = 0 \qquad {\rm or} \qquad 
|\textrm{I}=1,\textrm{I}_3> = 0  \qquad {\rm with} \qquad 
I_3 = 0
$
In the absence of initial state interactions between
proton and antiproton the relation
$$
\bar {\textrm{p}}\textrm{p} =\sqrt{\frac{1}{2}}\left(
|\textrm{I=1,\,I}_3=0> + |\textrm{I=0,\,I}_3=0>\right).
$$
holds. One could expect processes of the type
$\bar {\textrm{p}}\textrm{p} \to \bar {\textrm{n}}\textrm{n}$ 
in the initial state
but charge exchange - which would lead to unequal weight of the 
two isospin components - 
seems to be not very important \cite{crede}.
\par
The quantum numbers of the \pbp\ system are of course the
same as those for $q\bar q$ systems; both are bound states of fermion
and anti-fermion. Since isospin can be I=0 or I=1, every
atomic \pbp\ state may have G-parity +1 or -1.
Table \ref{ini} lists the quantum numbers of atomic levels 
from which annihilation may occur. The \pbn\ system has 
always I=1, every second level in the Table does not exist. 
For annihilation into specific final states, selection rules
may restrict the number of initial states. Annihilation into
any number of \piz\ and \etg\ mesons is e.g. allowed
only from positive-parity states.

\begin{table}[htb]        
\begin{center}
\renewcommand{\arraystretch}{1.4}        
\begin{tabular}[t]{|c l l|}
\hline
$^{2S+1}L_J$ & 
\multicolumn{2}{c|}{$I^{G} \left(J^{PC}\right)$} \\
\hline
$^1 S_0$ & $1^- (0^{-+})$ & $0^+ (0^{-+})$ \\
$^3 S_1$ & $1^+ (1^{--})$ & $0^- (1^{--})$ \\
\hline
$^1 P_1$ & $1^+ (1^{+-})$ & $0^- (1^{+-})$ \\
$^3 P_0$ & $1^- (0^{++})$ & $0^+ (0^{++})$ \\
$^3 P_1$ & $1^- (1^{++})$ & $0^+ (1^{++})$ \\
$^3 P_2$ & $1^- (2^{++})$ & $0^+ (2^{++})$ \\
\hline
\end{tabular}
\renewcommand{\arraystretch}{1.0}        
\caption{Quantum numbers of levels of the \pbp\ atom 
from which annihilation may occur.}
\label{ini}
\end{center}
\end{table}
\subsubsection{The Dalitz plot}
In a process in which the initial \pbp\ atom annihilates
into three particles, the full dynamics can be visualized
in a Dalitz plot. The three final-state particles have 
totally 12 components of four-vectors. Their masses are
known, the orientation in space is irrelevant for 
understanding the process (the orientation is given
by 3 Euler angles); energy and momentum conservation provides
4 constraints. Hence 2 variables are sufficient to
describe the full event. The two variables are often chosen
as two (squared) invariant masses of two convenient pairs of 
the 3 final-state particles, $M_{1,2}^2$ and $M_{1,3}^2$
where the squared invariant masses are calculated from
the momenta of the particles:
$$
M_{ij}^2 =
(p_i+p_j)^2 =
(E_i+E_j)^2 - 
(\vec{p_i} + \vec{p_j})^2
$$
The density distribution in a two-dimensional histogram
 $M_{1,2}^2$ versus $M_{1,3}^2$ is called Dalitz plot.
\par
\begin{figure}[bht]
\begin{tabular}{cc}
\hspace*{-1cm}\includegraphics[width=0.57\textwidth]{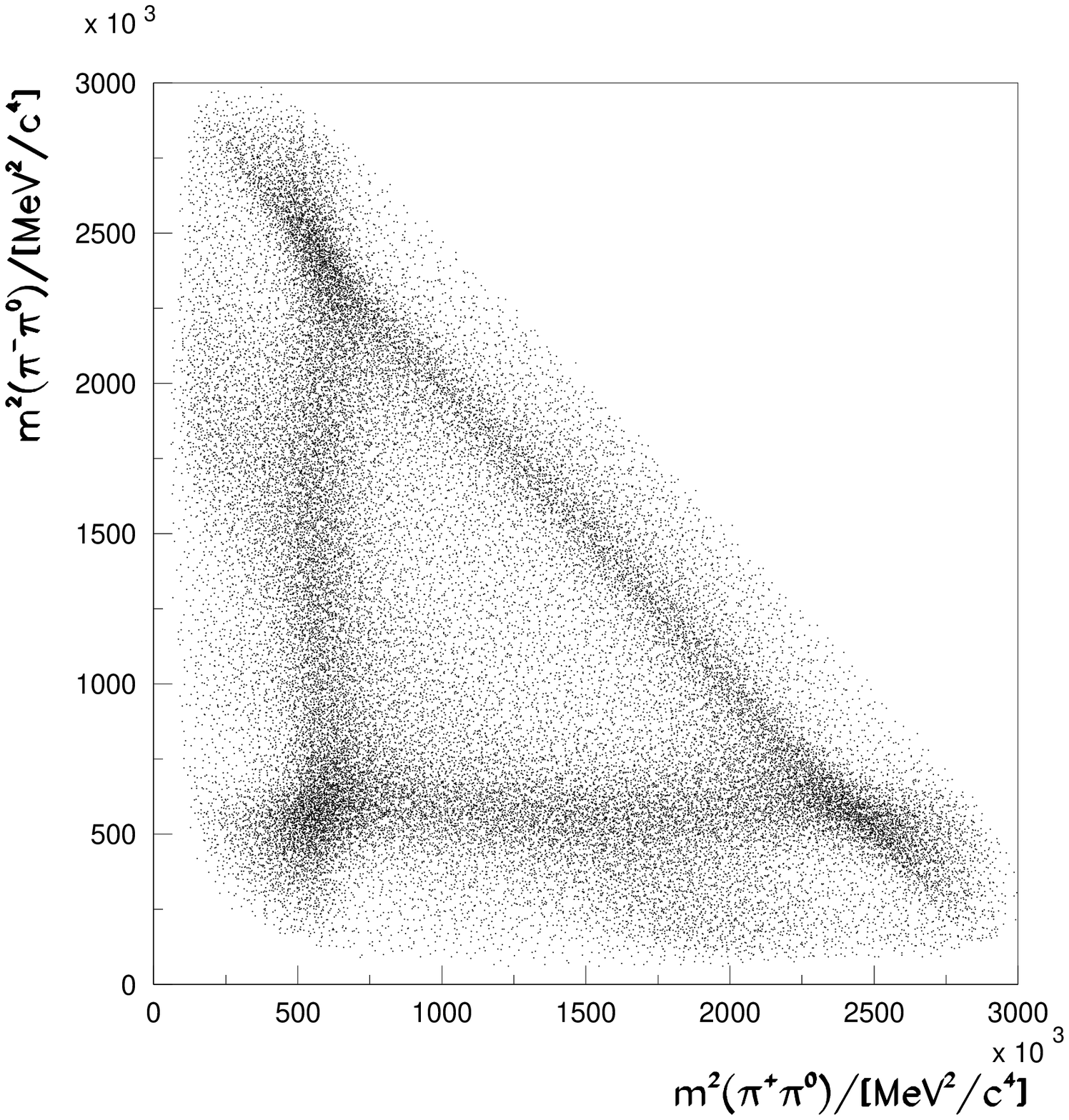}&
\hspace*{-13mm}\includegraphics[width=0.57\textwidth]{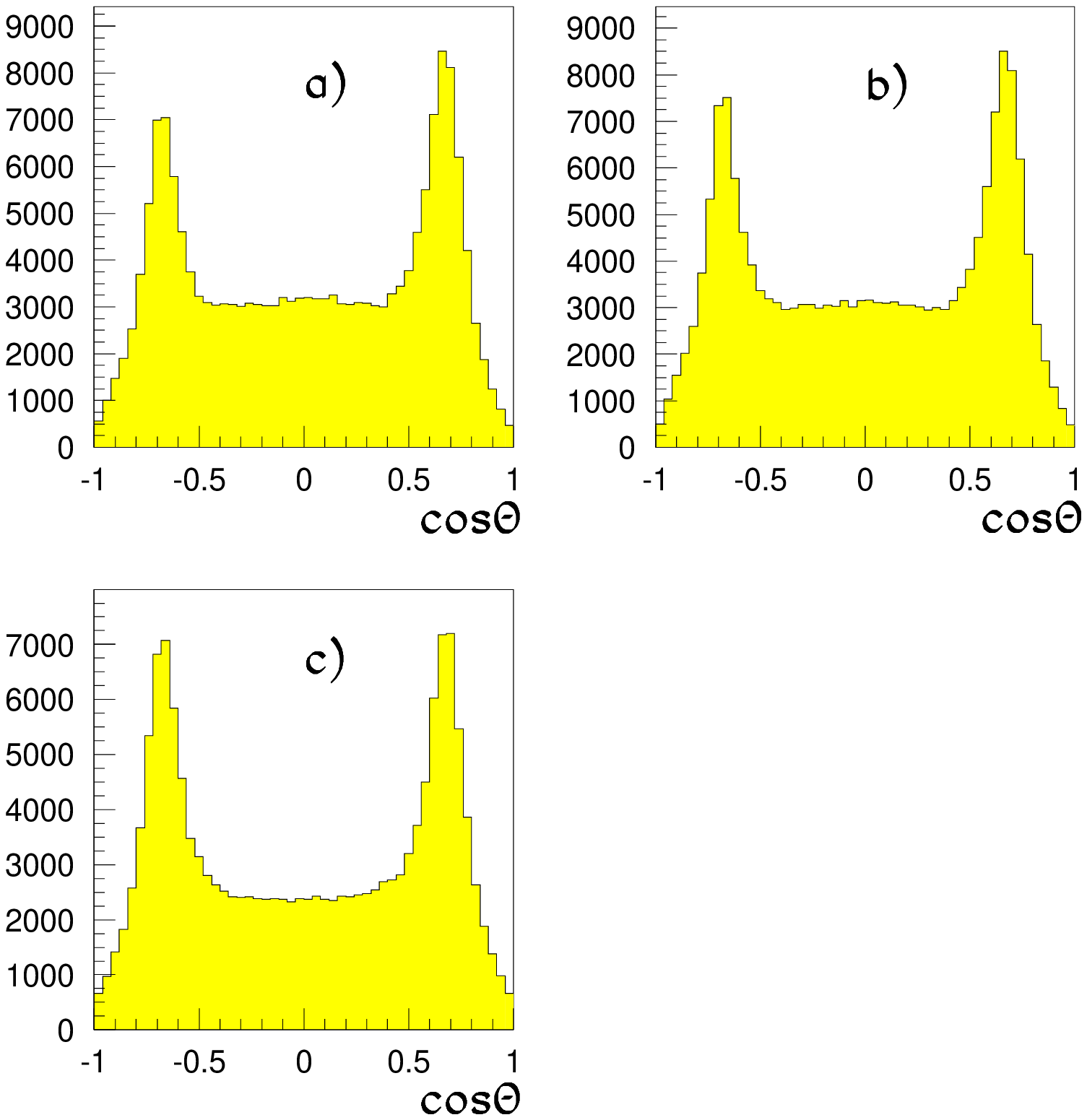}
\end{tabular}
\caption{The \pip\pim\piz\ Dalitz plot in \pbp\ annihilation
at rest, and \rhp\ (a), \rhm\ (b)
and \rhz\ (c) decay angular distributions.}
\label{dp}
\end{figure}

The following equation holds:
$$
m_{12}^2 + m_{23}^2 + m_{13}^2 = M^2 + m_1^2 + m_2^2 + m_3^2    
$$
$M$ is the mass of the \pbp\ atom, $m_i$ are the masses of the
final-state particles and $\vec{P}$ is the 3-momentum of the
initial-state (for $\pbarp$-annihilation at rest $\vec{P}=0$).
A particular simple Dalitz plot is depicted in 
Fig.~\ref{dp}. It represents the reaction \pbp\ra\pip\pim\piz .

A resonance in the subsystem of particle \pip\ and \piz\ 
(=\rhp) gives an accumulation along a vertical line, a 
\pim\piz\ resonance (=\rhm) is seen as enhancement in horizontal
direction, a \pip\pim\ resonance (=\rhz) is observed as accumulation
along the second diagonal.
\par 
The intensity distribution along the \rhp\ band gives
directly the \rhp\ra\pip\piz\ angular distribution in the
\rhp\ rest system (as a function of $\cos\theta_{\pip\piz}$).
Care has to be taken because of interferences; the 
\rhp\ and \rhm\ bands cross in the lower left corner
of the Dalitz plot. The amplitudes for the two processes
\pbp\ra\rhp\pim\ and \pbp\ra\rhm\pip\ interfere constructively
and lead to marked deviations of the observed angular distribution
from the expected one. This can be appreciated by looking at
the angular distributions in Fig.~\ref{dp}. The \rhp\ra\pip\piz\
distribution follows $\sin^2\Theta$ when
annihilation precedes from the isospin zero component in the
$^3S_1$ initial state. The crossings of the \rhp\ 
band with the \rhm\ and \rhz\ bands leads to
an increase of the intensity by a factor 4 because of 
quantum mechanical interference (the amplitudes are added\,!).
The increase of intensity in the \rhpm\ - \rhz\ crossing
is slightly smaller than for the \rhp\ - \rhm\ crossing. This
is due to a small contribution from the isovector
part of the $^1S_0$ state of the \pbp\ atom. From
this initial \pbp\ state, annihilation into \rhz\piz\ is forbidden,
the \rhpm\ decay angular distribution follows
 $\cos^2\Theta$. The  $\sin^2\Theta$ and  $\cos^2\Theta$
parts contribute a small constant distribution
which is clearly seen in the data. The
measured constant fractions in Fig. \ref{dp}a,b,c are
larger than estimated from the difference in \rhpm\ and \rhz\
intensities: this is due to annihilation from the $^1P_1$
state. A partial wave analysis identifies these 
observations, determines masses and widths of
contributing resonances and gives fractional contributions from
\pbp\ initial states given in Table \ref{ini} to
the \pip\pim\piz\ final state. 

\subsubsection{The $f_0(1500)$ in Crystal Barrel Dalitz plots}
\begin{figure}[bht]
\begin{tabular}{cc}
\hspace*{-1cm}\includegraphics[width=0.52\textwidth]{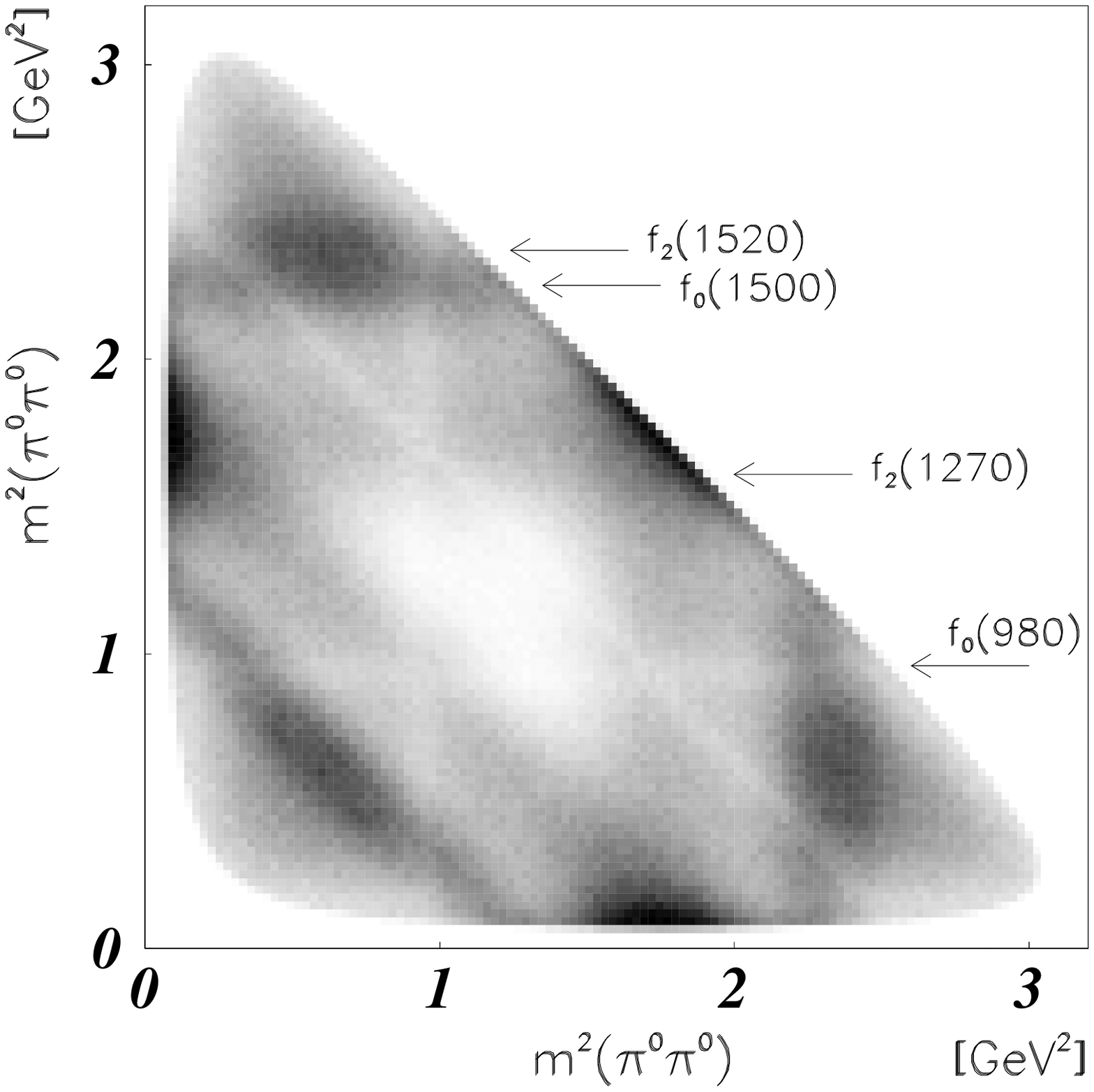}&
\hspace*{-3mm}\includegraphics[width=0.53\textwidth]{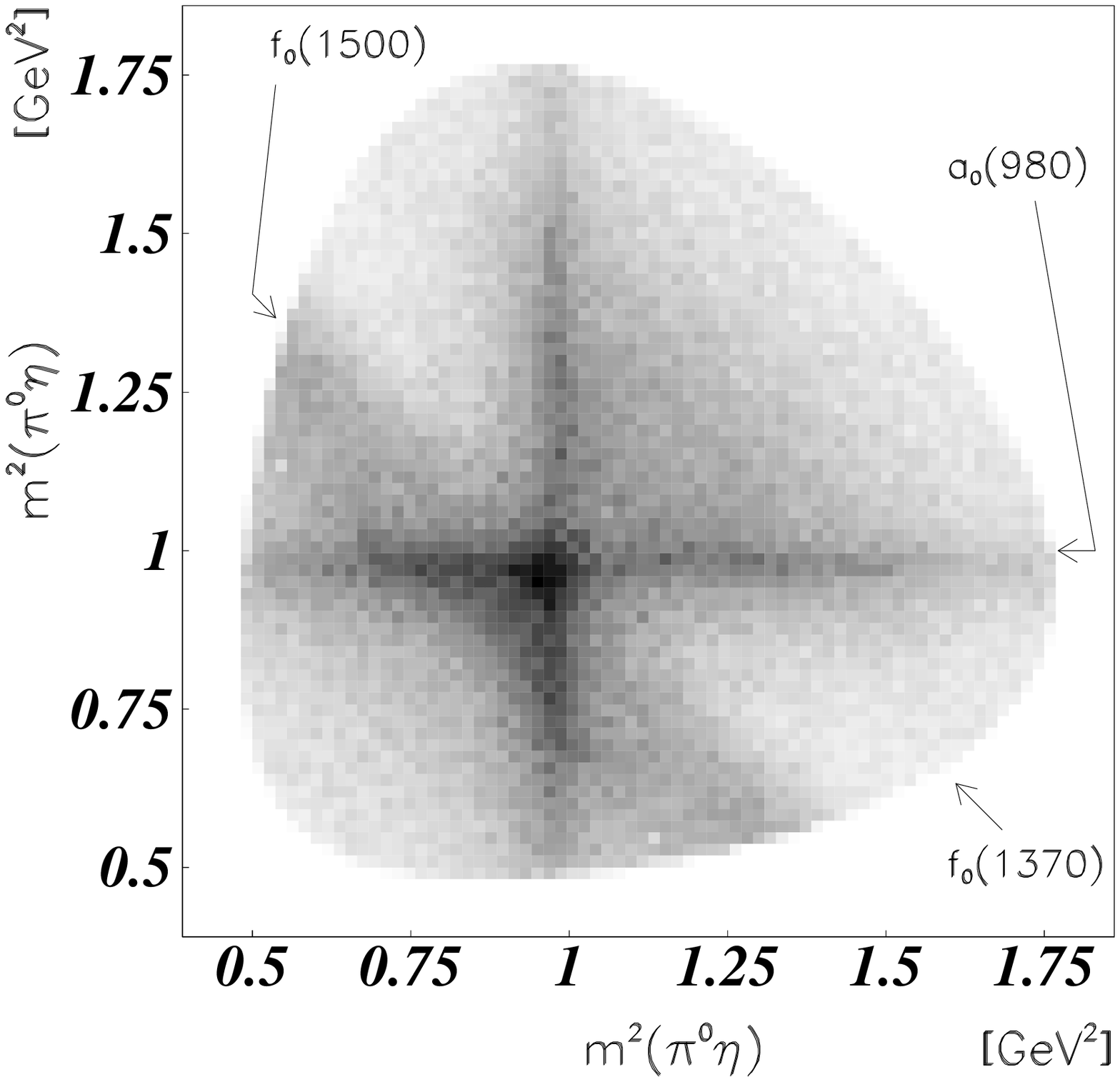}\\
\hspace*{-1cm}\includegraphics[width=0.55\textwidth]{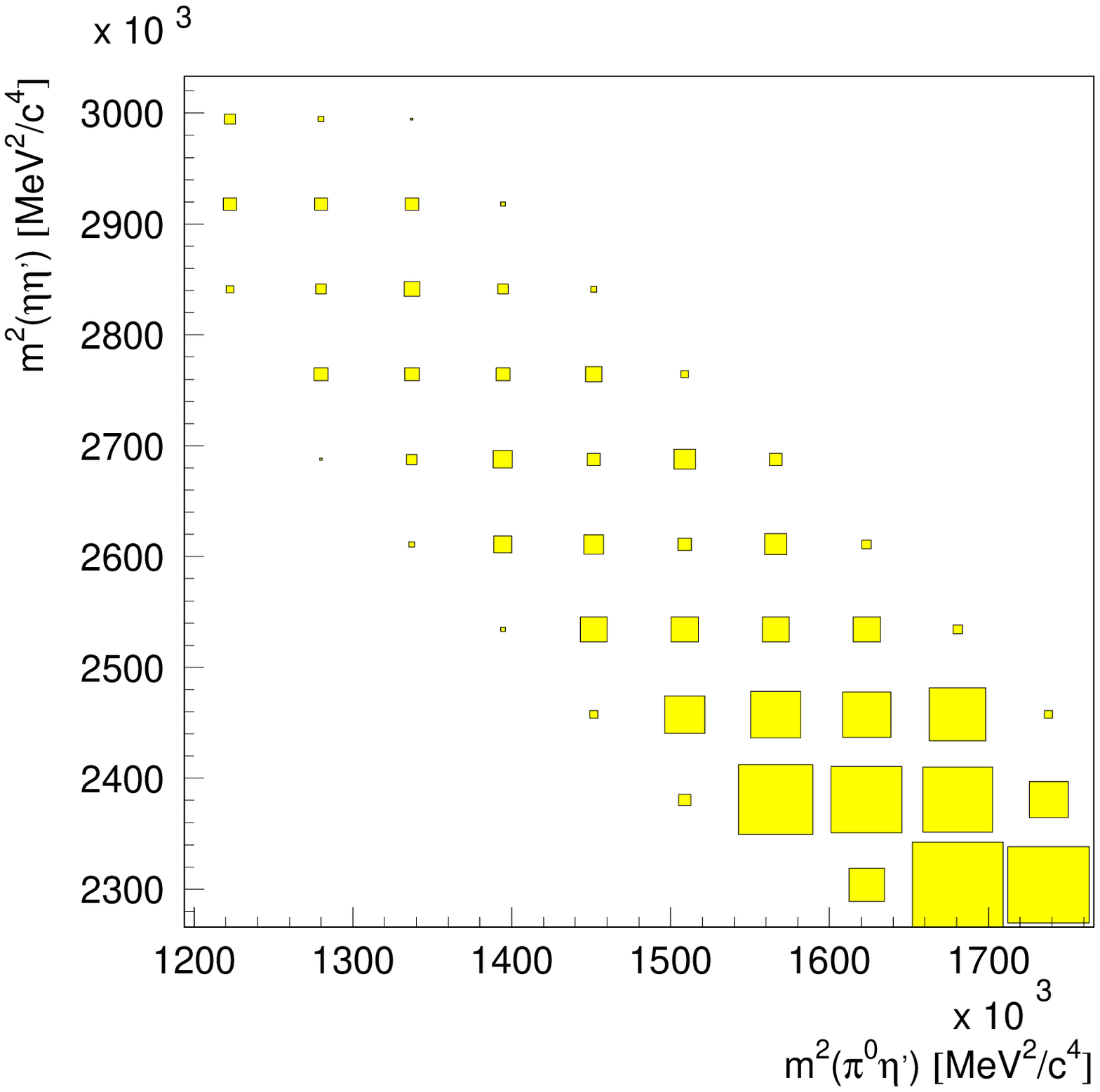}&
\hspace*{-3mm}\includegraphics[width=0.52\textwidth]{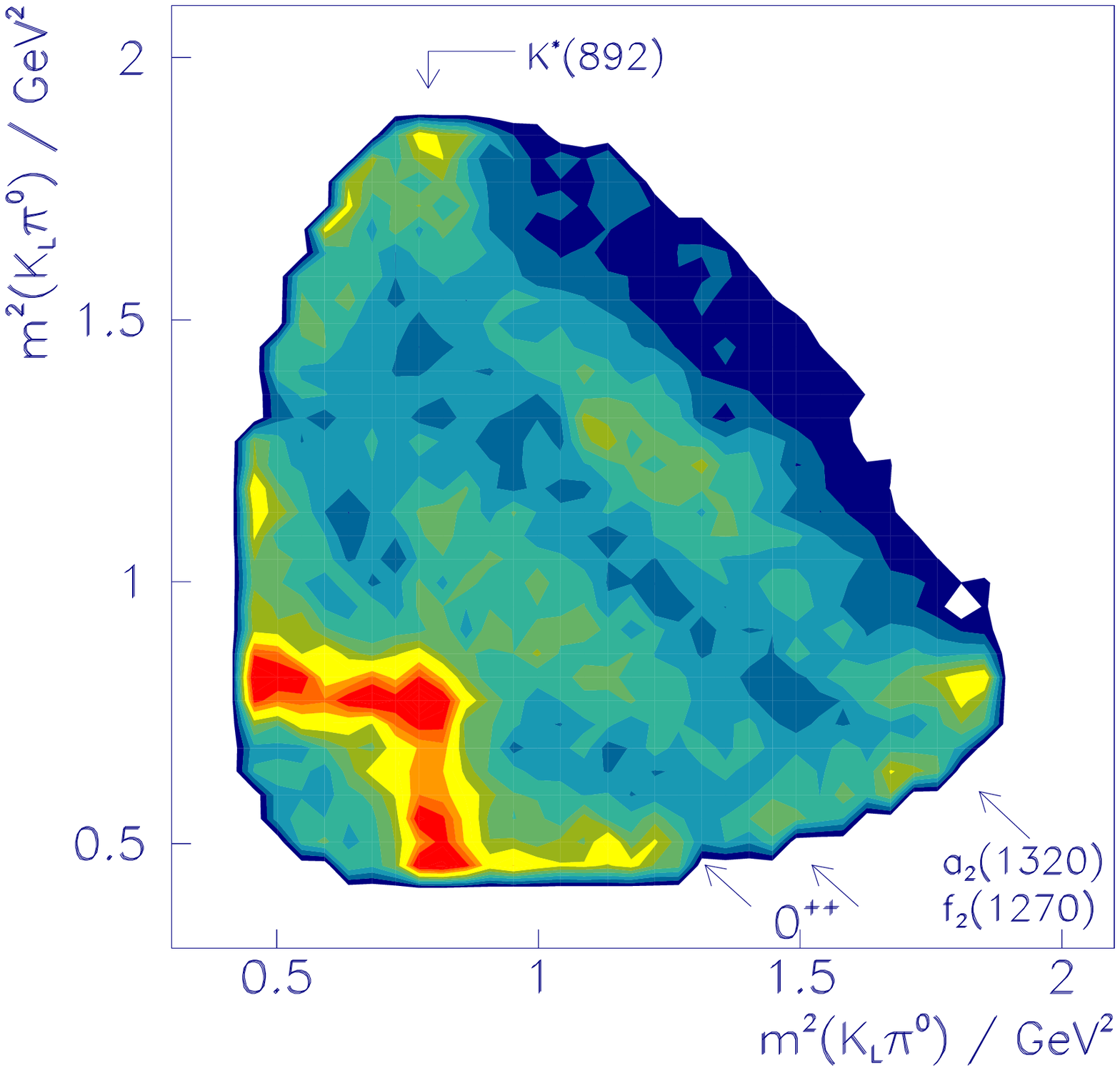}
\end{tabular}
  \caption{Dalitz plots for \pbp\ annihilation at rest
into 3\piz\ (upper left), \piz 2\etg\ (upper right),
\piz\etg\etp\ (lower left), K$_l$K$_l$\piz\ (lower right).
The $f_0(1370)$ contributes to (a,b,d), the $f_0(1500)$ 
to all 4 reactions. The  K$_l$K$_l$\piz\ is difficult to 
interprete in the black-and-white version;  the colored
Dalitz plot can be found on the web. The data
are from \cite{Amsler:1995gf}-\cite{Abele:1996nn}. }
  \label{four-dp}
\end{figure}

We now show the four Dalitz plots  
from which main properties of the $f_0(1370)$ and $f_0(1500)$ 
are derived. These two states were discovered by the Crystal Barrel
Collaboration; both play an eminent role in the present glueball
discussion. The Dalitz plots - Fig.~\ref{four-dp} -
stem from four publications 
of the Crystal Barrel Collaboration 
\cite{Amsler:1995gf}-\cite{Abele:1996nn}
\nocite{Amsler:1995gf,Amsler:1995bz,Amsler:1994ah,Abele:1996nn}.
In the 3\piz\ (upper left) and
the \piz 2\etg\ (upper right) Dalitz plots the $f_0(1500)$ is
clearly seen as band structure. In \piz\etg\etp\
a strong threshold enhancement in the \etg\etp\ invariant
mass is seen (lower left); the final state K$_l$K$_l$\piz\
has prominent K$^*$ bands; their interference with the
$f_0(1500)$ makes the intensity so large in the left corner
of the Dalitz plot (lower right).
\par
The reactions \pbp\ra \pip\pim 3\piz\ \cite{Amsler:1994rv}, 
\pbp\ra 5\piz\ \cite{Abele:1996fr}, \pbn\ra\pim 4\piz\ 
\cite{thoma1} and \pbn\ra\ 
\pim 2\pim 2\piz\pip\ \cite{thoma2}
were studied to determine decays into 4 pions.
\subsubsection{Scalar mesons} 
\begin{figure}[h]
\begin{tabular}{cc}
\hspace*{-1cm}\includegraphics[width=0.52\textwidth]{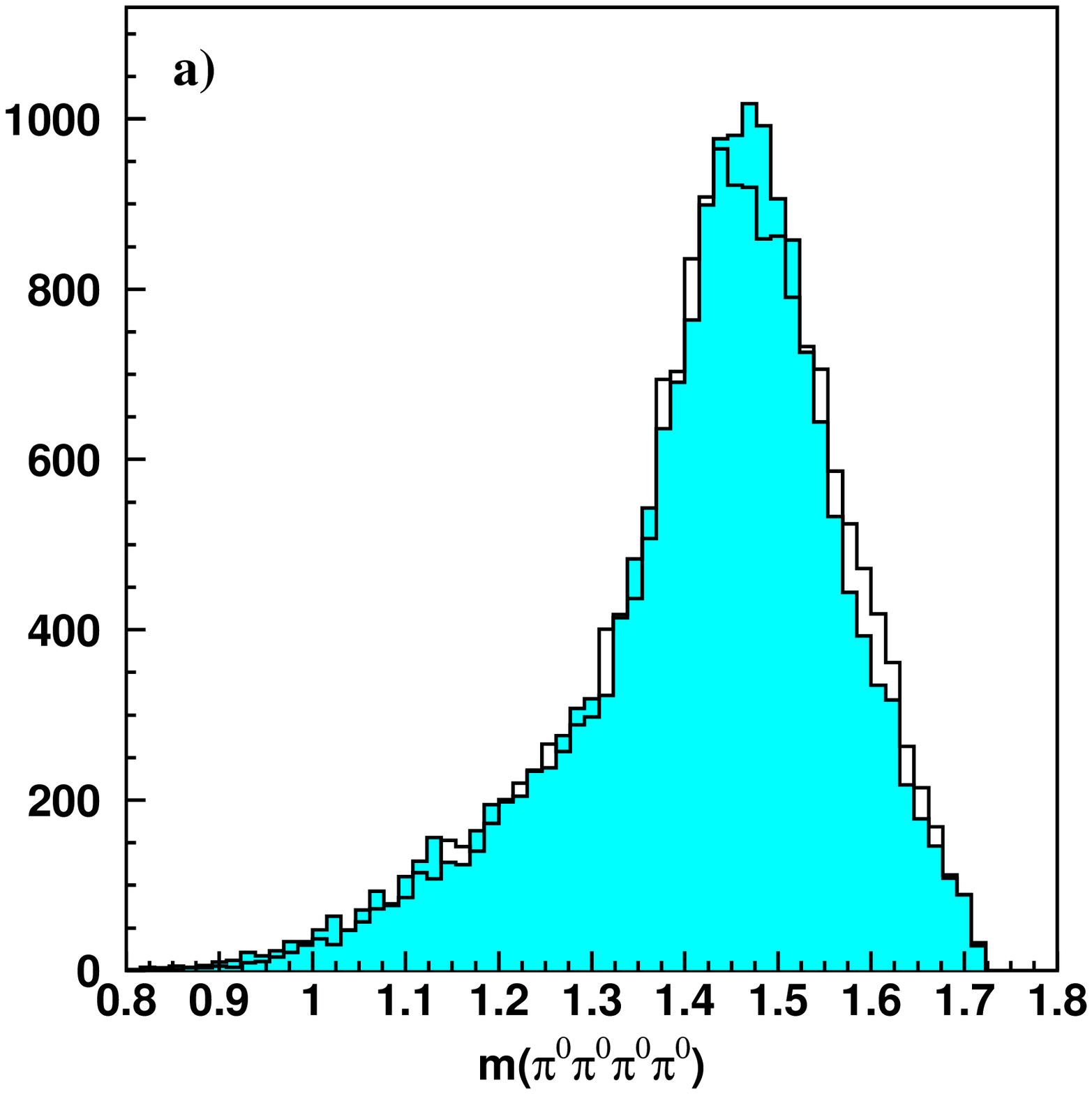}&
\hspace*{-3mm}\includegraphics[width=0.53\textwidth]{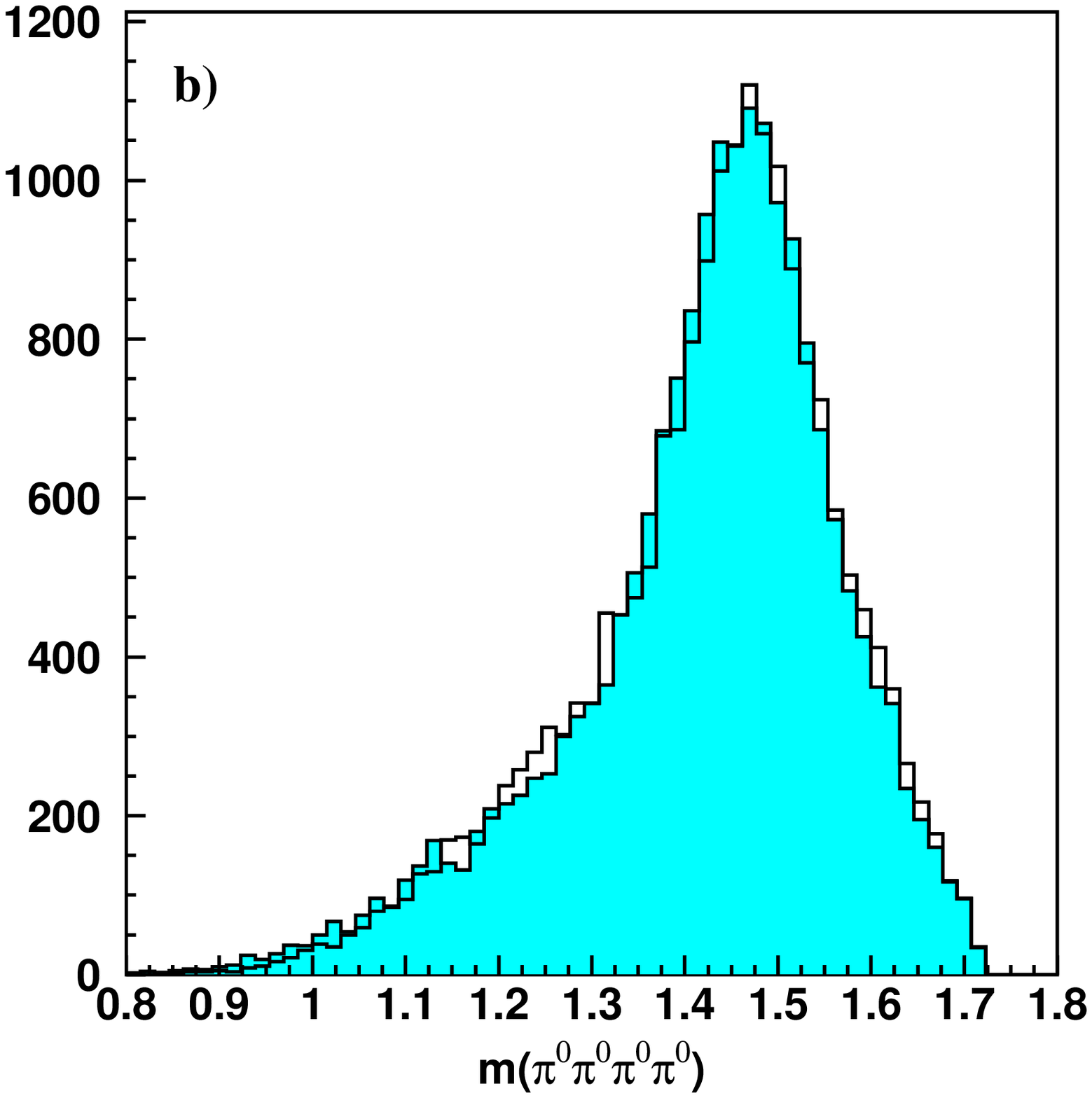}
\end{tabular}
  \caption{The 4\piz\ invariant mass in the reaction \pbn\ra\pim 4\piz
  . A fit (including other amplitudes) with one scalar state fails;
two scalar resonances at 1370 and 1500 MeV give a good fit. Note
that the full 8-dimensional phase space is fitted and not just the
mass projection shown here; (from~\cite{thoma1}).}
  \label{fivepi}
\end{figure}

We have seen that decays of mesons are constrained by SU(3)
relations.  So there is hope that the glueball nature
of a state can be un-revealed by inspecting the
coupling to various final states.
Table \ref{f0decay} lists partial widths of 
\begin{table}
\renewcommand{\arraystretch}{1.2}
\bc
\begin{tabular}{|l|cc|}
\hline           
                        & $f_0(1370)$    &  $f_0(1500)$ \\   
\hline 
$\Gamma_{tot}$   
                        &  $275\pm 55$    & $130\pm 30$   \\
$\Gamma_{\sigma\sigma}$ 
                        & $120.5\pm45.2$  & $18.6\pm12.5$ \\
$\Gamma_{\rho\rho}$     
                        &  $62.2\pm28.8$  & $8.9\pm8.2$ \\
$\Gamma_{\pi^*\pi}$     
                        & $41.6\pm22.0$     & $35.5\pm29.2$ \\ 
$\Gamma_{a_1\pi}$       
                        & $14.10\pm 7.2$   & $8.6\pm6.6$ \\
$\Gamma_{\pi\pi}$       
                        & $21.7\pm 9.9$  & $44.1\pm15.4$ \\
$\Gamma_{\eta\eta}$      
                        & $0.41\pm0.27$   & $3.4\pm1.2$   \\
$\Gamma_{\eta\eta'}$     
                        &                 & $2.9\pm1.0$   \\
$\Gamma_{\bar{K}K}$     &$(7.9\pm2.7$) to ($21.2\pm7.2$)     
                                          & $8.1\pm2.8$   \\
\hline 
\end{tabular} 
\ec
\renewcommand{\arraystretch}{1.0}
\caption{Partial decay widths of the $f_0(1370)$ and $f_0(1500)$
from Crystal Barrel data}
\label{f0decay}
\end{table}
the $f_0(1370)$ and $f_0(1500)$ as derived from the
Crystal Barrel Collaboration.
Most striking is the similarity of the partial 
decay widths for the decays into \etg\etg\ and \etg\etp\ 
and the smallness of the \kkb\ coupling of the $f_0(1500)$.
(Remember that we expect ratios for 
\p\p\,:\,\etg\etg\,:\,\etg\etp\,:\,\kkb\ of 3\,:1\,:\,0\,:\,4, 
after removal of phase space). Obviously, the $f_0(1500)$
cannot be a pure glueball, it must mix with nearby 
states\,! The $f_0(1370)$ has important couplings to
two pairs of \piz-mesons, to $\sigma\sigma$. This is evident from
Fig.~\ref{fivepi}.
\subsection{Scalar mesons and the scalar glueball}
\subsubsection{Established scalar mesons}
Below 2 GeV, 15 'established' scalar mesons are listed by 
the Particle Data Group \cite{pdg} 
which are shown in the Table below. 
\bc
\begin{tabular}{|c|c|cc|}
\hline
            &           &                      &      \\  %
\quad   I = 1/2 \quad\  & \quad  I = 1  \quad\   &  
\quad   I = 0   \quad\           &      \\  %
            &           &                      &      \\  %
\hline
            &           &                      &      \\  
            &           & $f_0(400-1200)$        &      \\  
            &           &                      &      \\  
            &           &                      &      \\  
            & {$a_0(980)$} &  
{$f_0(980)$} &      \\  
            &           &                      &      \\  
            &           &                      &      \\  
            &           &                      &      \\  
            &           &   {$f_0(1370)$}          &\\  
{$K_0^*(1430)$}   & {$a_0(1490)$} &
            {$f_0(1500)$}
            & \\ 
            &           &                      &      \\  
            &           & {$f_0(1710)$}           &      \\  
            &           &                      &      \\  
\hline
\end{tabular}
\ec
The lowest-mass entry is an $f_0(400-1200)$ representing the scalar
isoscalar \p\p\ interactions, often called $\sigma$-meson. 
For reasons discussed below,
it is likely not a $q\bar q$ meson. The two states at the \kkb\ threshold,
the $f_0(980)$ and the $a_0(980)$, have a large coupling to \kkb\ 
and a comparatively narrow width. Hence they are often 
discussed as \kkb\ molecules \cite{4}
or as four-quark states \cite{Jaffe:1977ig} 
and not as $q\bar q$ mesons.
\par
Leaving out the 5 states $f_0(400-1200)$, $f_0(980)$ and $a_{0}^{\pm ,0}(980)$,
we remain with a decuplet of states and not with nine states as
expected in the quark model. 
The three scalar isoscalar states 
$f_0(1370)$, $f_0(1500)$ and $f_0(1750)$ cannot possibly belong
to the same nonet. Since the $f_0(1370)$ is a
$u\bar u+d\bar d$ state (it decays only weakly into \kkb )
we expect ideal mixing and an $s\bar s$state at 1620 MeV,
not too far from the observed $f_0(1710)$. This
line of argument is the basis for the assignment
by the Particle Data Group \cite{pdg} of 
the $f_0(1370)$ and  $f_0(1710)$ to the 1$^3P_0$ nonet.
The $f_0(1500)$ is discussed as glueball candidate.

\subsubsection{Meson-glueball mixing}
Several authors have suggested scenarios in which a scalar glueball
mixes with two $q\bar q$ states \cite{amsclo}-\cite{gutsche}.
\nocite{lewine,li,cloki,celenza}
The mixing angles 
were (partly) determined from partial decay widths of the
scalar states. Only M.~Strohmeier-Presicek {\it et al.} 
\cite{gutsche} includes the
4$\pi$ decays into the analysis. 
You may start your mixing scheme by assuming that the
'primordial' glueball is in between a low-mass $u\bar u+d\bar d$
and the high-mass $s\bar s$state \cite{amsclo,cloki}, 
or by assuming that the
primordial glueball is above both states
\cite{lewine}. These two options
lead to different mixing schemes. Some of them
are listed in Table \ref{mix}.
\begin{table}
\renewcommand{\arraystretch}{1.4}
\bc
\begin{tabular}{|ccccccc|}
\hline
\multicolumn{3}{|c}{ 
Amsler and Close \protect\cite{amsclo}} &&&&\\
 $f_0(1370)$ &=&  $0.86\frac{1}{\sqrt{2}}(u\bar u+d\bar d)$ &+& 
  {0.13} $s\bar s$                   &-& 
  {0.50} glueball\\
 $f_0(1500)$ &=&  $0.43\frac{1}{\sqrt{2}}(u\bar u+d\bar d)$ &-& 
  {0.61} $s\bar s$                   &+&
  {\underline{0.61}} glueball\\
 $f_0(1750)$ &=&  $0.22\frac{1}{\sqrt{2}}(u\bar u+d\bar d)$ &-& 
  {0.76} $s\bar s$                   &+& 
  {\underline{0.60}} glueball\\
\hline
\multicolumn{3}{|c}{ 
Lee and Weingarten \protect\cite{lewine}} &&&&\\
 $f_0(1370)$ &=&  $0.87\frac{1}{\sqrt{2}}(u\bar u+d\bar d)$ &+& 
  {0.25} $s\bar s$                   &-& 
  {0.43} glueball\\
 $f_0(1500)$ &=& $-0.36\frac{1}{\sqrt{2}}(u\bar u+d\bar d)$ &+& 
  {0.91} $s\bar s$                   &-& 
  {0.22} glueball\\
 $f_0(1750)$ &=&  $0.34\frac{1}{\sqrt{2}}(u\bar u+d\bar d)$  &+& 
  {0.33} $s\bar s$                   &+& 
  {\underline{0.88}} glueball\\
\hline
\multicolumn{3}{|c}{ 
De-Min Li {\it et al.}\protect\cite{li}} &&&&\\
 $f_0(1370)$ &=&  $-0.30\frac{1}{\sqrt{2}}(u\bar u+d\bar d)$ &-& 
  {0.82} $s\bar s$                   &+& 
  {0.49} glueball\\
 $f_0(1500)$ &=& $+0.72\frac{1}{\sqrt{2}}(u\bar u+d\bar d)$ &-& 
  {0.53} $s\bar s$                   &-& 
  {0.45} glueball\\
 $f_0(1750)$ &=&  $+0.63\frac{1}{\sqrt{2}}(u\bar u+d\bar d)$  &+& 
  {0.22} $s\bar s$                   &+& 
  {\underline{0.75}} glueball\\
\hline
\multicolumn{3}{|c}{ 
Close and Kirk \protect\cite{cloki}} &&&&\\
 $f_0(1370)$ &=&  $-0.79\frac{1}{\sqrt{2}}(u\bar u+d\bar d)$ &-& 
  {0.13} $s\bar s$                   &+& 
  {0.60} glueball\\
 $f_0(1500)$ &=& $-0.62\frac{1}{\sqrt{2}}(u\bar u+d\bar d)$ &+& 
  {0.37} $s\bar s$                   &-& 
  {\underline{0.69}} glueball\\
 $f_0(1750)$ &=&  $0.14\frac{1}{\sqrt{2}}(u\bar u+d\bar d)$  &+& 
  {0.91} $s\bar s$                   &+& 
  {0.39} glueball\\
\hline\multicolumn{3}{|c}{ 
Celenza {\it et al.}\protect\cite{celenza}} &&&&\\
 $f_0(1370)$ &=&  $0.01\frac{1}{\sqrt{2}}(u\bar u+d\bar d)$ &-& 
  {1.00} $s\bar s$                   &-& 
  {0.00} glueball\\
 $f_0(1500)$ &=& $0.99\frac{1}{\sqrt{2}}(u\bar u+d\bar d)$ &-& 
  {0.11} $s\bar s$                   &+& 
  {0.01} glueball\\
 $f_0(1750)$ &=&  $0.03\frac{1}{\sqrt{2}}(u\bar u+d\bar d)$  &+& 
  {0.09} $s\bar s$                   &+& 
  {\underline{0.99}} glueball\\
\hline
\multicolumn{3}{|c}{ 
M.~Strohmeier-Presicek {\it et al.}\protect\cite{gutsche}} &&&&\\
 $f_0(1370)$ &=&  $0.94\frac{1}{\sqrt{2}}(u\bar u+d\bar d)$ &+& 
  {0.07} $s\bar s$                   &-& 
  {0.34} glueball\\
 $f_0(1500)$ &=&  $0.31\frac{1}{\sqrt{2}}(u\bar u+d\bar d)$ &-& 
  {0.58} $s\bar s$                   &+&
  {\underline{0.75}} glueball\\
 $f_0(1750)$ &=&  $0.15\frac{1}{\sqrt{2}}(u\bar u+d\bar d)$ &+& 
  {0.81} $s\bar s$                   &+& 
  {\underline{0.57}} glueball\\
\hline
\end{tabular}
\ec
\renewcommand{\arraystretch}{1.0}
\caption{Decomposition of the wave function of 
3 scalar isoscalar states into their quarkonium and glueball
contribution in various models.}
\label{mix}
\end{table}
Not shown in the Table is the mixing scenario suggested by 
Narison \cite{narison} who finds that all three states share 
the glueball in approximately equal portions. Anisovich et al.
\cite{anisov} believe 5 states to exist below 1.8 GeV. They
develop from the 1$^3P_0$ and 2$^3P_0$ $q\bar q$ states 
and the scalar glueball. Through mixing the glueball
strength distributes 
between the $f_0(1370)$ and $f_0(1500)$ and
a broad underlying component. SU(3) symmetry 
in the decays of $^3P_0$ $q\bar q$ states is imposed in the fits
as well as flavor-blindness of the primordial glueball.
\par
\subsubsection{The scalar glueball}
All mixing schemes 
agree in that the scalar glueball manifests itself in the scalar 
meson sector and that it has a mass, before mixing, of about 1600 MeV. 
Hence all authors agree that lattice gauge theories
are doing well in predicting a scalar glueball at this mass.  
The mixing schemes disagree how the glueball is distributed between
the three experimentally observed states. Some of the models 
assign very large $s\bar s$components to the $f_0(1370)$ or $f_0(1500)$; 
this is certainly not compatible with data.
\par

\section{Scrutinizing the  scalar glueball}
The interpretation of two resonances, of the $f_0(980)$ and
the $f_0(1370)$, plays a decisive role in the meson-glueball
mixing scenarios. Also, the nature of the $f_0(400-1200)$
is unclear. 
We discuss these 3 states in some detail. 
\subsection{Scalar mesons below 1.3 GeV}
\subsubsection{The scalar isoscalar \p\p\ interactions} 
The S-wave \p\p\ interactions 
at small energies 
are elastic. The $T$-matrix saturates unitarity, the 
inelasticity parameter $\eta_{l=0}$ vanishes.
Inelastic channels open up at the \kkb\ threshold, and
at this mass the inelasticity is large. 
At energies above 1 GeV, the inelasticity 
is small again. Results from a recent phase shift
analysis of \p\ scattering off a polarized 
target into two pions \cite{Kaminski:1997da} are shown in 
Fig.~\ref{amplphase}. 
At 980 MeV a dip is observed corresponding to
the $f_0(980)$ which has large coupling to \kkb . 
A second dip is suggested at 1500 MeV  corresponding to
the $f_0(1500)$. The amplitude reaches 
maxima at positions which correspond to the old $\sigma (600)$
which plays an important role in one-boson-exchange-potentials,
and to the  old $\epsilon (1300)$. The dips are associated 
with additional rapid phase motions. The background amplitude and its 
phase motion are correctly reproduced by 
$t$-channel $\rho$ exchange \cite{1,2}. 
\par
Adding further $t$-channel amplitudes for $\omega , 
\Phi$ and K$^*$ exchange, Speth and collaborators describe 
both the $f_0(980)$ and the $a_0(980)$ 
as generated by $t$-channel exchange dynamics
\cite{3} with the \kkb\ system forming a bound state in
isoscalar but not in isovector interactions.
Related is the interpretation
of these two resonances at the \kkb\ threshold as \kkb\ molecules
\cite{4}. These states can also be understood as
$q\bar q$ mesons with properties governed by the \kkb\ threshold
\cite{5}. And they are discussed
as $q\bar q\bar q$ resonances \cite{Jaffe:1977ig,achasov1}. 
\subsubsection{The $f_0(980)$ and $a_0(980)$ in $Z^0$ fragmentation}
At LEP the fragmentation of quark- and gluon jets has been
studied intensively \cite{bohrer}. In particular the inclusive
production of the $f_0(980)$ and $a_0(980)$ provides new insight
into their internal structure. The OPAL collaboration
searched for these and other light meson resonances in a 
data sample of 4.3 million hadronic $Z^0$ decays. For the 
$f_0(980)$ a coupled channel analysis was made by
simultaneously fitting the inclusive $\pi\pi$ and \kkb\ mass 
spectra. Some total inclusive rates are listed in Table
\ref{Z0decay}. We notice that the three mesons \etp , $f_0(980)$
and $a_0(980)$ - which have very similar masses - also have 
production rates which are nearly identical (the two charge 
modes of the $a_0(980)^{\pm}$ need to be taken into
account). Hence there is primary evidence that the 
three mesons have the same internal structure, that they are all
three $q\bar q$ states. 
\par
This conclusion can be substantiated
by further studies~\cite{opal}.
The production characteristics of the $f_0(980)$ 
are compared to those of f$_2(1270)$ and $\Phi(1020)$ mesons,
and with the Lund string model of hadronization 
within which the $f_0(980)$  is treated as a conventional meson.
No difference is observed in any of these comparisons between
the $f_0(980)$ and the f$_2(1270)$ and $\Phi(1020)$. 
We emphasize that it would be extremely useful if 
the studies could be extended to include other scalar
particles.  
Of course, background problems become more important 
for higher-mass particles and their production is reduced.
\begin{table}
\renewcommand{\arraystretch}{1.3}
\vspace*{-3mm}
\bc 
\begin{tabular}{|lc|}
\hline
\piz          & $ 9.55\pm 0.06\pm 0.75$ \\
\etg          & $ 0.97\pm 0.03\pm 0.11$ \\
\etp          & $ 0.14\pm 0.01\pm 0.02$ \\
$a_{0}^{\pm}(980)$& $ 0.27\pm 0.04\pm 0.10$\\
$f_0(980)$    & $ 0.141 \pm 0.007 \pm 0.011$\\
$\Phi (1020)$ & $ 0.091\pm0.002\pm0.003$\\
$f_2(1270)$   & $ 0.155\pm0.011\pm0.018$ \\
\hline
\end{tabular}
\ec
\vspace*{-3mm}
\caption{Yield of light mesons per hadronic $Z^0$ decay;
(from~\cite{bohrer},\cite{opal}).}
\renewcommand{\arraystretch}{1.0}
\label{Z0decay}
\end{table}   

  \begin{figure}[h]
\begin{tabular}{cc}
\hspace*{-1cm}\includegraphics[width=0.60\textwidth]{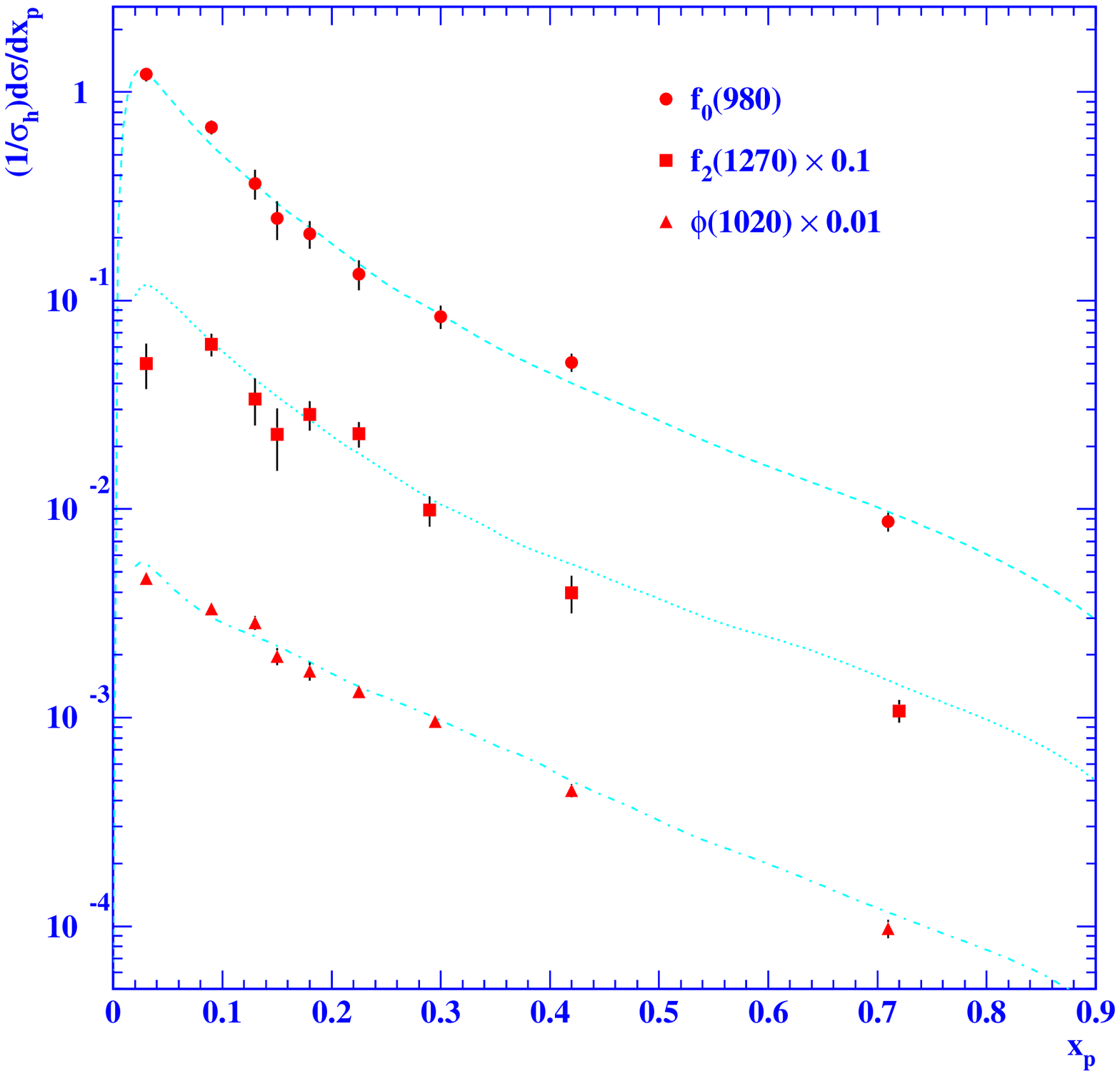}&
\hspace*{-1.3cm}\includegraphics[width=0.60\textwidth]{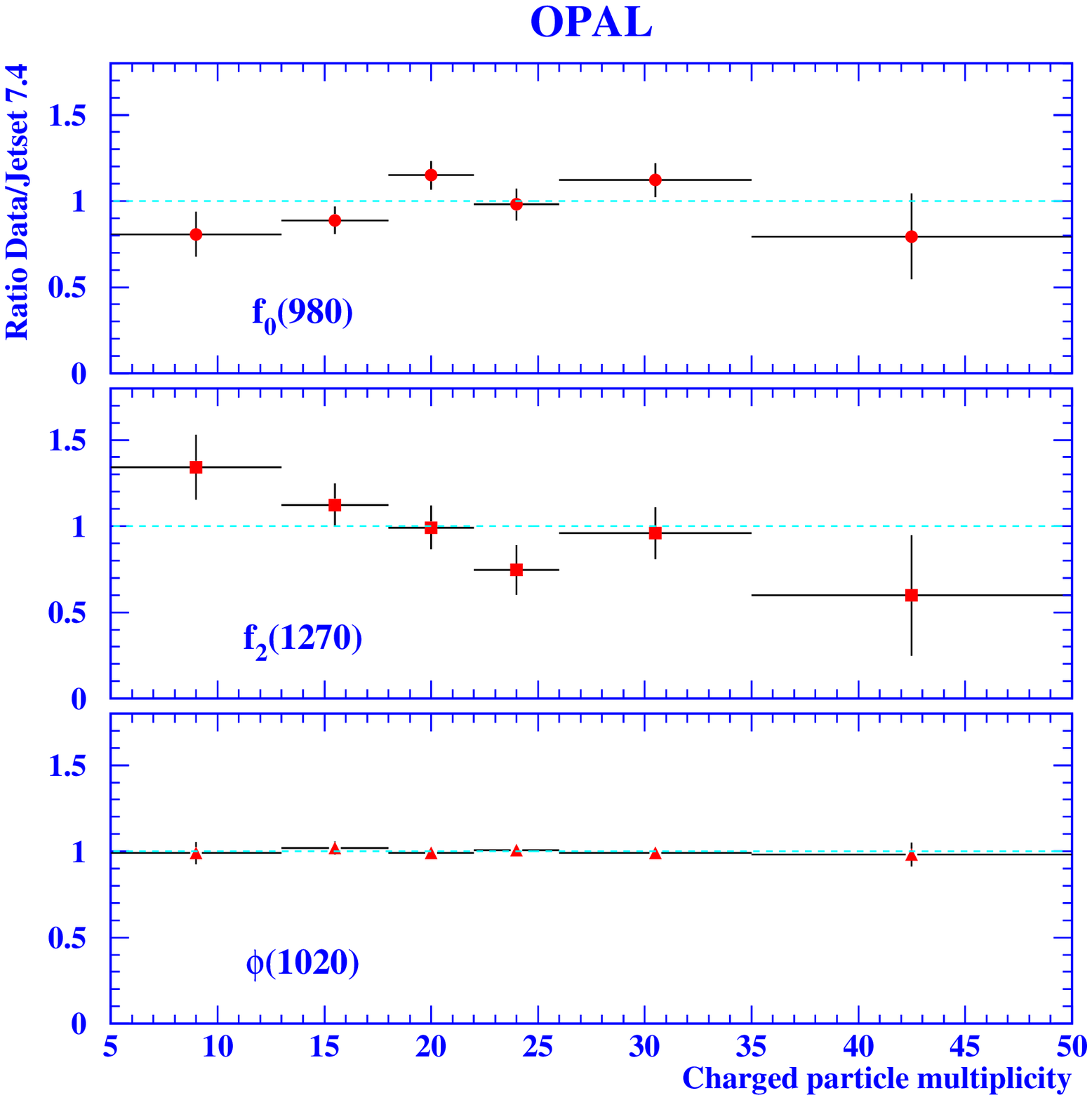}
\vspace*{-1cm}\\
\hspace*{-1cm}\includegraphics[width=0.60\textwidth]{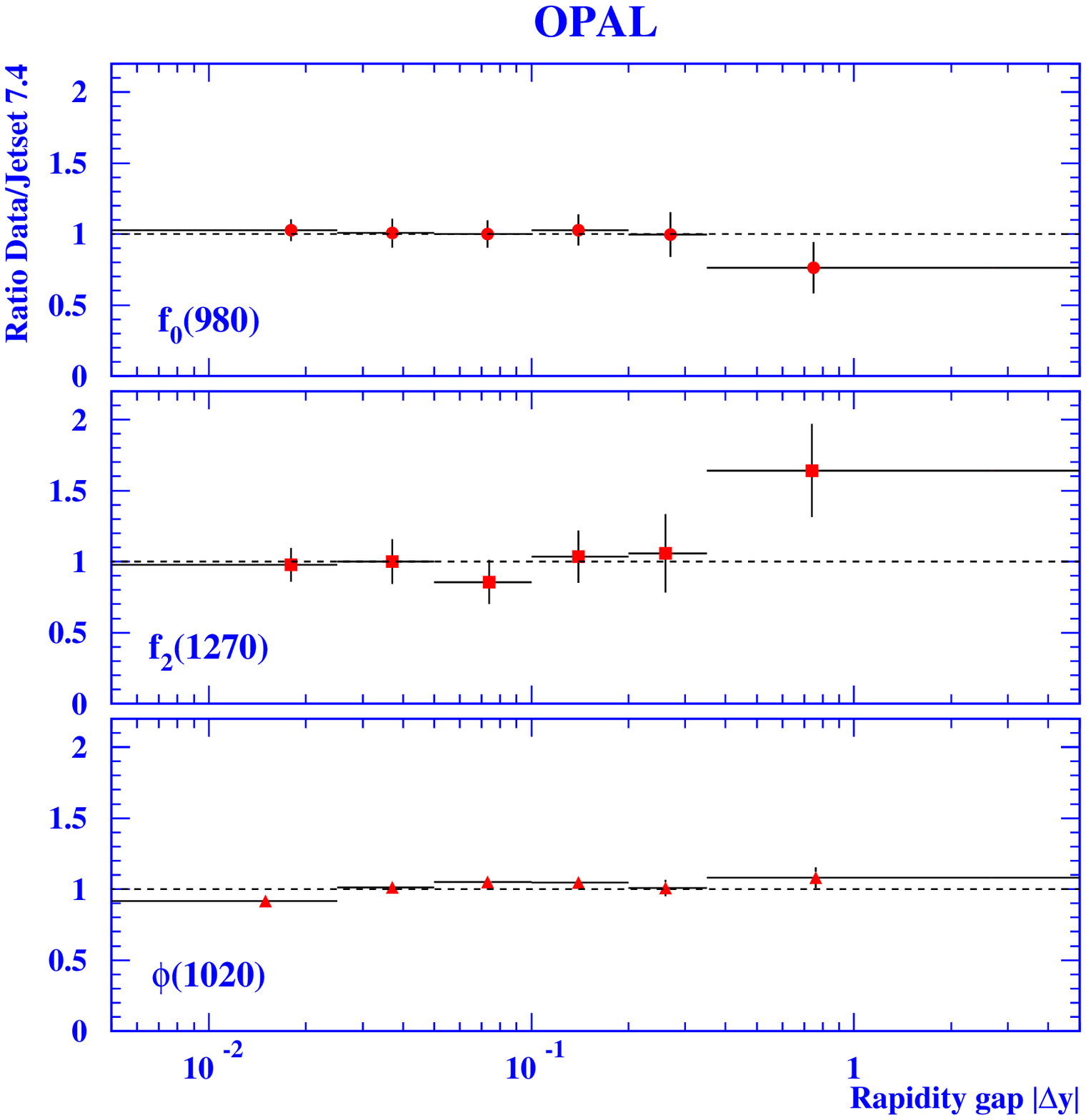}&
\hspace*{-1.3cm}\includegraphics[width=0.60\textwidth]{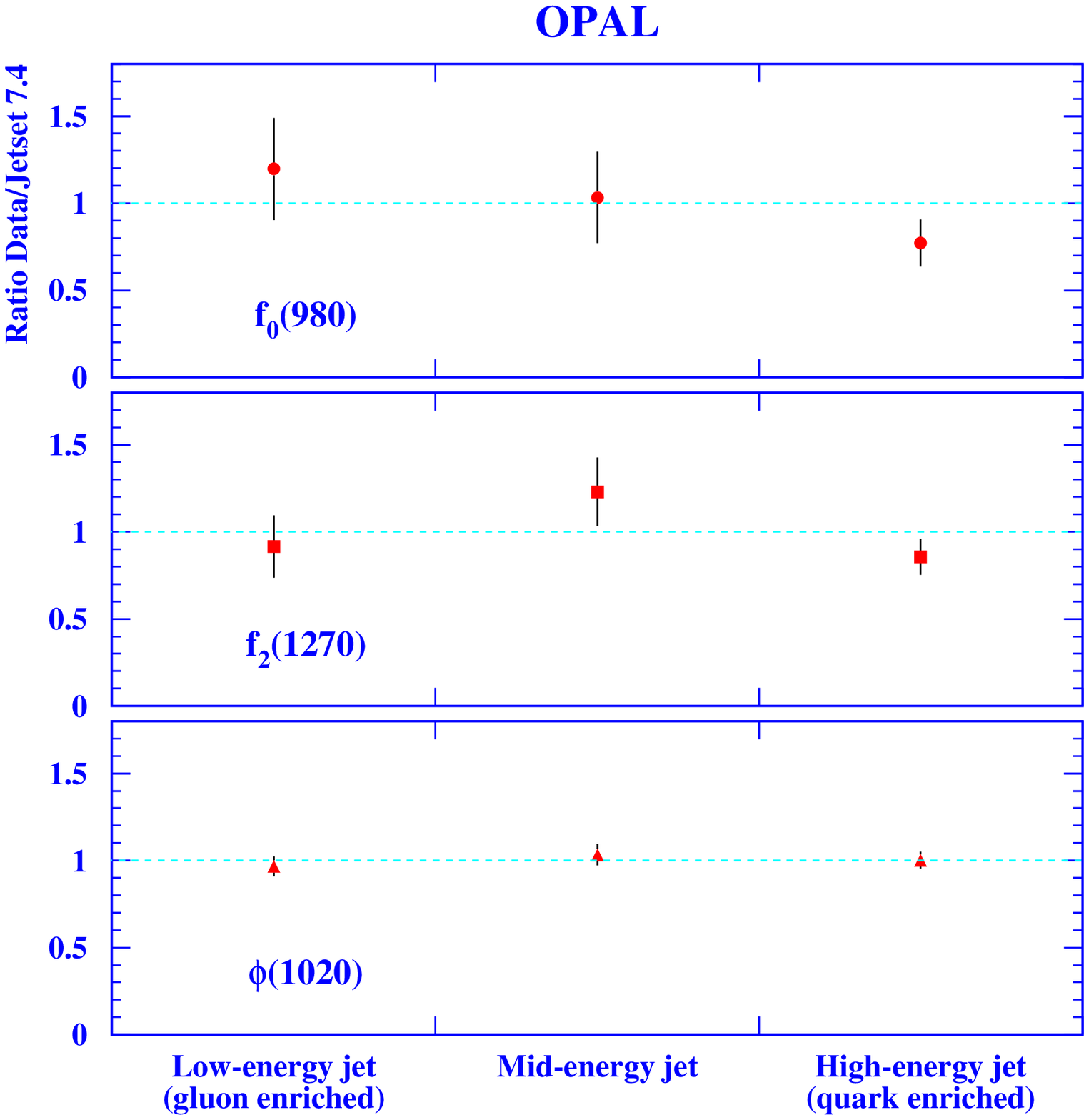}
\vspace*{-2cm} \\
\end{tabular}
  \caption{Fragmentation of $Z^0$'s into jets containing
 $f_0(980)$, f$_2(1270)$ and $\Phi(1020)$ mesons as functions 
of $x_p$ (a), of the charged-particle multiplicity (b),
of the rapidity gap between the meson and the nearest 
charged particle (c) and of the jet energy (d).
The lines correspond to simulations ~\protect\cite{opal} 
based on the Lund string model of hadronization.}
  \label{opal}
\end{figure}

\subsubsection{The two-photon widths of the $f_0(980)$ and $a_0(980)$}
In a recent report, Boglione and Pennington \cite{Boglione:1999rw}
reexamined 
data on two-photon production of scalar mesons. The experimental
information from different experiments is sometimes inconsistent
resulting in large errors. On the other hand,    
two-photon decays provide deep insight into the 
internal structure \cite{minkochs}. The ratio \cite{pdg}  
\begin{equation}
R_{\gamma\gamma} = \frac{\Gamma_{f_0(980)\ra\gamma\gamma}}
{\Gamma_{a_0(980)\ra\gamma\gamma}} = 1.38\pm 0.63
\end{equation}
is related to the scalar mixing angle (with $f_0(980)$
being the singlet for $\Theta = 0$) by
\begin{equation}
R_{\gamma\gamma} = \frac{1}{3}\cdot\left(\sin\Theta
+ 3\sqrt 2\cdot\cos\Theta \right)^2
\end{equation}
There are two solutions: 
$\Theta = (66.8 \pm 15)^{\circ}$ and
 $\Theta = (-27.4 \pm 15)^{\circ}$. 
Only the latter value is
compatible with the result obtained in  \cite{minkochs}.
The angle $\Theta = 60^{\circ}$ 
corresponds to a wave function
\begin{equation}
f_0(980) \sim \sqrt{\frac{1}{9}}\frac{1}{\sqrt 2}(u\bar u+ d\bar d ) + 
\sqrt{\frac{8}{9}}s\bar s
\label{f980}
\end{equation}
which has a larger $s\bar s$content than a isosinglet state
would have, likely due to a strong \kkb\ component in the
wave function.

\subsubsection{The $\Phi\ra\gamma f_0(980)$ decay} 

The $\Phi$ radiative decay rate into the  $f_0(980)$ is
surprisingly large \cite{achasov2},  
\be
\frac{\Gamma_{\Phi\ra\gamma f_0(980)}}{\Gamma_{\Phi{\rm tot}}} 
= (3.5\pm 0.3^{+0.8}_{-0.3})\cdot 10^{-4}
\ee
Early predictions \cite{closeisgur}
assuming different structures for the $f_0(980)$, $q\bar q$, \kkb\ or
four-quark, were all well below the recent
experimental value \cite{achasov2}. Recently,
the reaction was studied by Markushin 
\cite{Markushin:2000fa} and by Marco {\it et al.}
\cite{Marco:1999df}. They found that kaonic loops play a 
decisive role and that, including these, rate and 
\p\p\ invariant mass distributions are well
reproduced. The $f_0(980)$ resonance corresponds
to a $T$-matrix pole close to the \kkb\ threshold;
a good description of the data is achieved assuming
that the pole is of dynamical origin
and represents a molecular-like \kkb\ state. 
An underlying $q\bar q$ component is possible but not required. 
\par
\vspace*{-0.2cm}
\begin{figure}[h]
 \begin{minipage}[b][4cm][t]{0.55\textwidth}
\epsfig{file=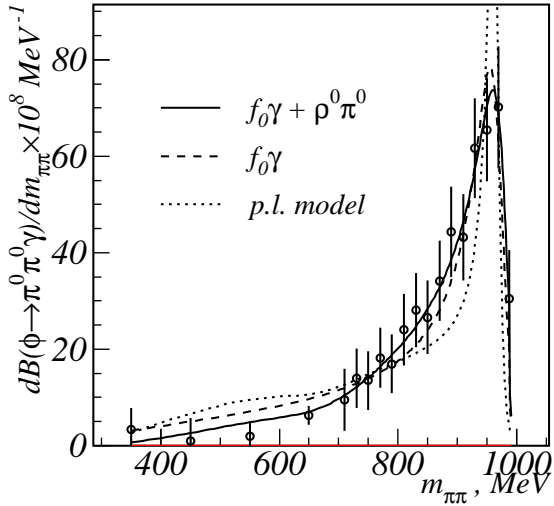,width=\textwidth}
\end{minipage}\hfill 
 \begin{minipage}[t][4cm][t]{.4\textwidth}
\vspace*{-1cm}
\caption{The \p\p\ invariant mass distribution
from $\Phi\ra\pi\pi\gamma$. The fit assumes 
that the $f_0(980)$ is a four-quark state; 
(from~\protect\cite{achasov2}). 
Other models give similar agreement between data
and fit. }
\end{minipage}
\label{f0pipigam}
\end{figure}
\vspace*{-1cm}
\par
The rate for 
$\Phi\ra\gamma a_0(980)$ is smaller by a factor
$\sim 4$ \cite{Achasov:2000ku}
than the rate for $\Phi\ra\gamma f_0(980)$
which seems difficult to reproduce if both
mesons are \kkb\ mole\-cules. In this case their
rates should be equal \cite{closeisgur}. 
If the $f_0(980)$ has a structure
as given in (\ref{f980}) and the $a_0(980)$ is
$\frac{1}{\sqrt 2}(u\bar u -d\bar d )$,  
the decay chain 
$\Phi\ra\gamma f_0(980);  f_0(980)\ra\pi\pi$ should be
much larger than $\Phi \ra \gamma a_0(980)$;  
$a_0(980)\ra\eta\pi$. The data are in-between
these two extreme values. This may suggest that
both pictures are oversimplified. Isospin-breaking 
mixing between $f_0(980)$ and $a_0(980)$ due to the
mass splitting between the \kp\km\ and \kn\knb\ 
thresholds \cite{spetmix},\cite{Close:2000ah}
are too weak to be responsible for
the large $\Phi\ra\gamma a_0(980)$ rate. 
\par
\subsubsection{$D_s$ decays into three pions} 
\begin{figure}[h]
 \begin{minipage}[t][4cm][t]{.55\textwidth}
  \begin{center}
  \begin{tabular}{c}
\epsfig{file=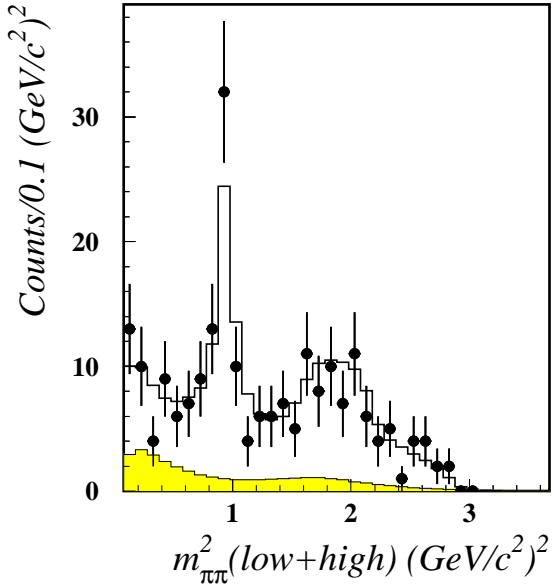,width=0.97\textwidth}
  \end{tabular}
  \end{center}
\end{minipage}\hfill
 \begin{minipage}[t][4cm][t]{.4\textwidth}
\vspace*{0.5cm}
  \caption{Dalitz plot for OZI violating 
   \protect{$D_s$} decays into three pions.
The low-mass peak is due to the $f_0(980)$
decaying into \pip\pim , the broad
enhancement is assigned to the  $f_0(1500)$
(in a partial wave analysis). }
  \label{ds}
\end{minipage}
\end{figure}

$D_s$ decays
into three pions provide further insight into the spectrum  
of isoscalar scalar resonances. The comparatively large rate for
three-pion production is surprising: consider the reaction 
$D_s^+$\ra 2\pip\pim . The quark content of the $D_s^+$
is $c\bar s$. In the decay, 
the $c$ undergoes a transition to an $s$, the $W^+$ converts 
into a \pip . Hence an $s\bar s$ state is produced which decays
into \pip\pim . This is OZI rule violating, and the OZI violation is
strong:
\begin{equation}
\frac{\Gamma_{\pi^+\pi^-\pi^0}}{\Gamma_{K^+K^-\pi^0}} = 0.23 \pm 0.04
\end{equation}
The three-pion Dalitz plot has moderate statistics only, but the 
$f_0(980)$ is clearly seen and the partial wave analysis finds a 
second scalar state at $f_0(1470)$ which we
identify with the $f_0(1500)$. The two states $f_0(980)$ and 
$f_0(1500)$ then both decay into \pip\pim . The data are
shown in Fig.~\ref{ds}.
\par

\vspace*{4cm}
\par
We note two aspects: first, the two states $f_0(980), f_0(1500)$ 
are produced in a similar way and - taking phase space into
account - with similar couplings. Second, both mesons do not
respect the OZI rule. This is similar to the \etg\ and 
\etp . The wave functions of both, $f_0(980)$ and $f_0(1500)$,
must contain $u\bar u$  + $d\bar d$ and $s\bar{s}$ components.
\subsubsection{The $f_0(980)$ and $a_0(980)$: $q\bar q$ or \kkb\,? }
The situation is confused: when low-energy phenomena are
discussed, the \kkb\ molecule interpretation provides for
a very good description of the data with no $q\bar q$ component
being required. When you start from the $q\bar q$ picture
and place a $q\bar q$ state in the vicinity of the \kkb\ 
threshold, the state is attracted by the threshold when
the coupling to the \kkb\ channel is taken into account
\cite{Boglione:1997aw} and a large \kkb\ component
develops. Reactions like Z$^0$ fragmentation
and D$_s$ decays point at a $q\bar q$ nature. 
We conclude that there are good reasons to believe that
the  $f_0(980)$ and  $a_0(980)$ should be counted as $q\bar q$ 
1$^3P_0$ states. Of course, the vicinity of the \kkb\ 
threshold plays a significant role and a large \kkb\ component
is to be expected as part of their wave functions
\cite{5}. 

\subsection{Scalar mesons above 1.3 GeV} 
\subsubsection{$D_s$ decays} 

$D_s$ decays into three pions show no evidence
for the $f_0(1370)$. As discussed, only the two states $f_0(980)$ and $f_0(1500)$
are produced. 
\subsubsection{The two-photon widths}
The Aleph Collaboration searched for two-photon production of
the $f_0(1500)$ and the $f_0(1710)$ \cite{Barate:2000ze}. 
No signal was seen. From the absence the authors concluded that
\begin{eqnarray}
\Gamma_{\gamma\gamma\ra f_0(1500)}\cdot BR(f_0(1500)\ra\pip\pim ) < 
0.31 {\rm keV} 
\label{aleph} \\
\Gamma_{\gamma\gamma\ra f_0(1710)}\cdot BR(f_0(1710)\ra\pip\pim ) < 
0.55 {\rm keV} \nonumber 
\end{eqnarray}
at 95\% confidence level. 
\par
At HADRON97, Barnes estimated that the two-photon width should
be of the order of 8\,keV. Taken at face value the upper limit
in (\ref{aleph}) claims that the $q\bar q$ content of the  $f_0(1500)$
should be at a few \% level!

\subsubsection{Radiative J/$\psi$ decays}
\begin{figure}[t]
\centerline{
\epsfig{file=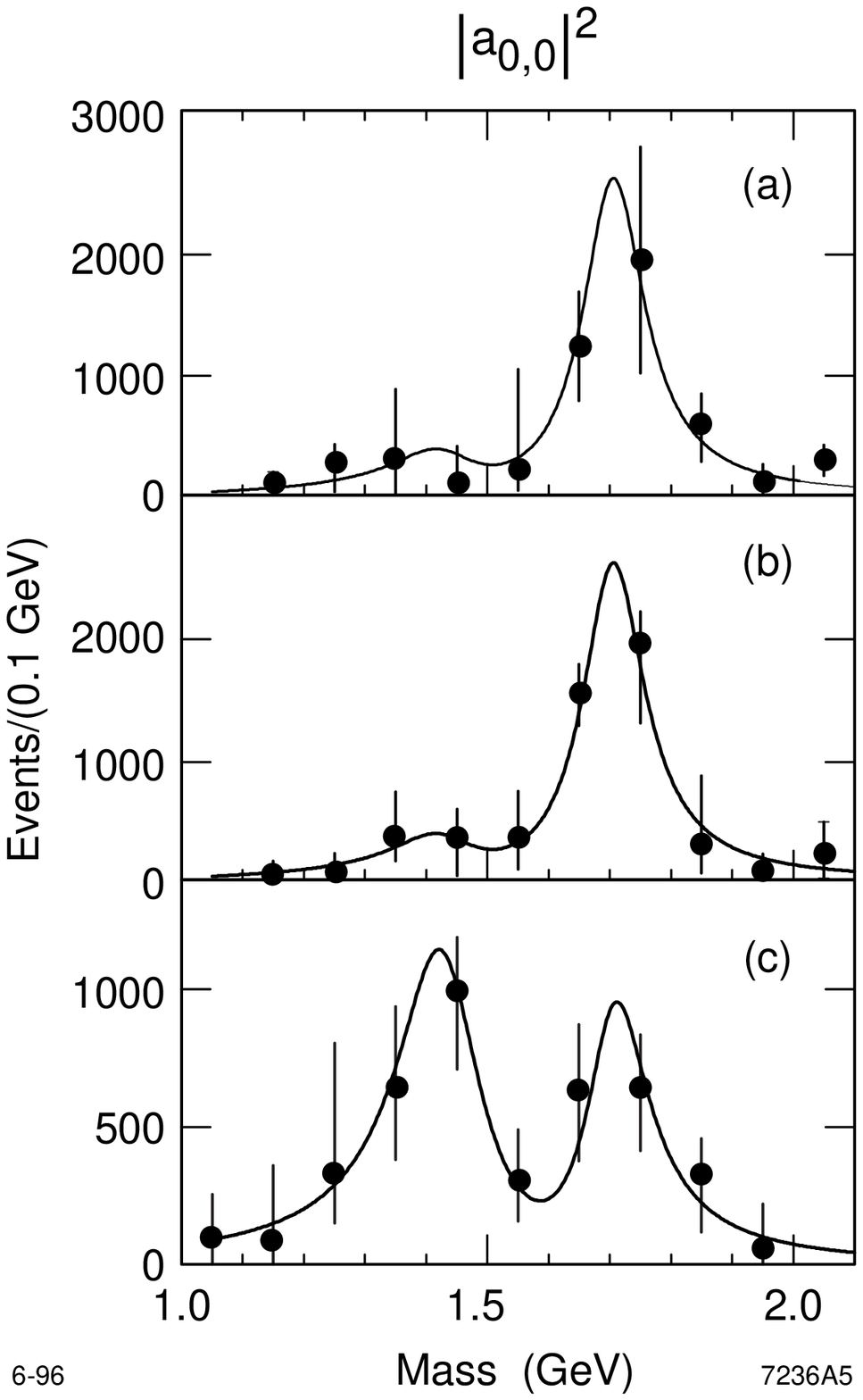,height=2.5in,width=2.8in}\hfill
\epsfig{file=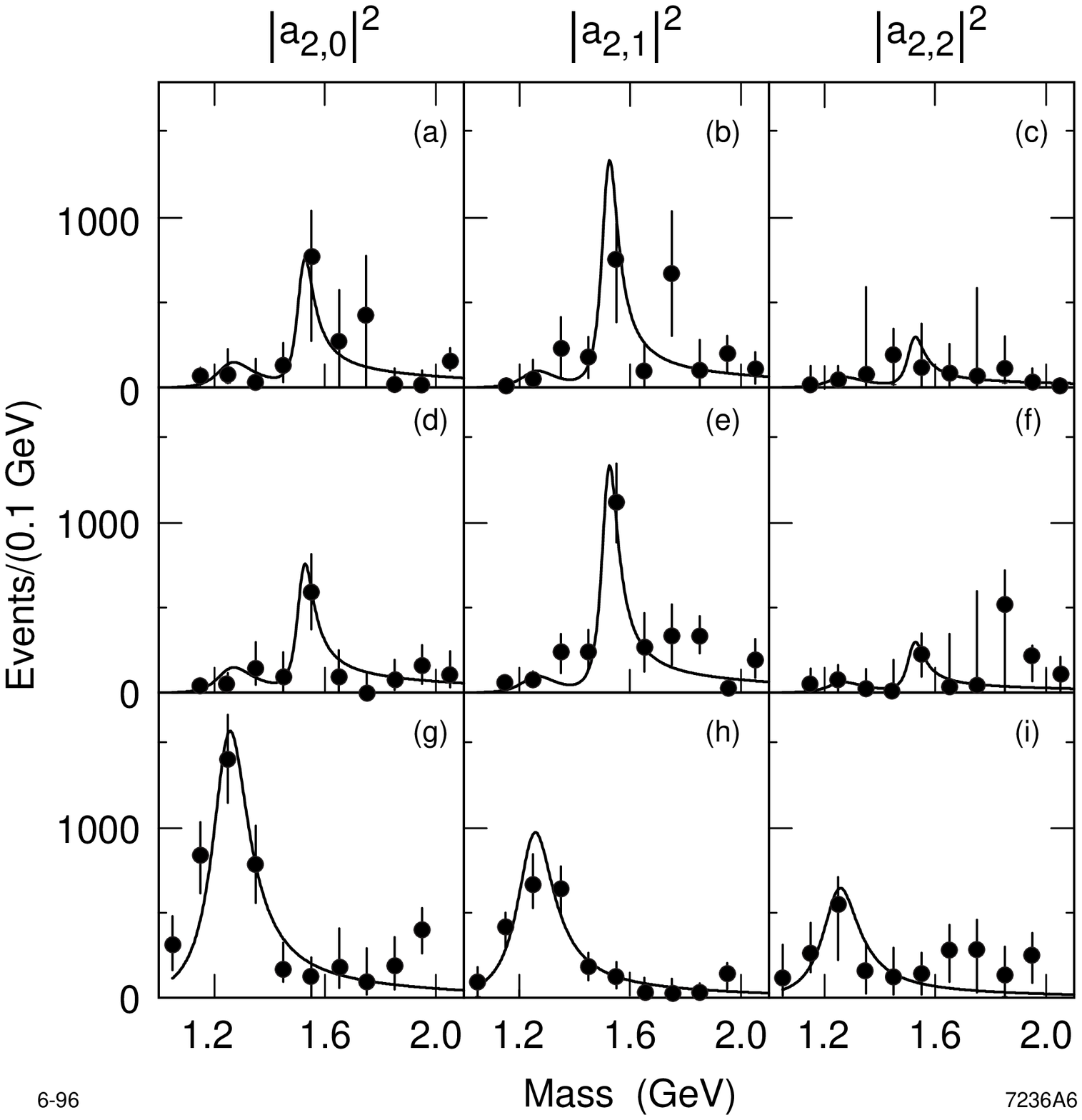,height=2.5in,width=2.8in}}
\vspace*{-3mm}
\caption{Partial wave analysis of $J/\psi$ radiative decay to two
pseudoscalar mesons, from the Mark~III collaboration.  
$S$-wave (left) and $D$-wave (right).  The top row is the analysis
of $J/\psi\to\gamma K_SK_S$, middle row for $J/\psi\to\gamma K^+K^-$,
and bottom row for $J/\psi\to\gamma \pi\pi$.  The structure near
$1700$~MeV/$c^2$ is clearly dominated by an $S$-wave state.}
\label{dunnw}
\vskip 5mm
\epsfig{file=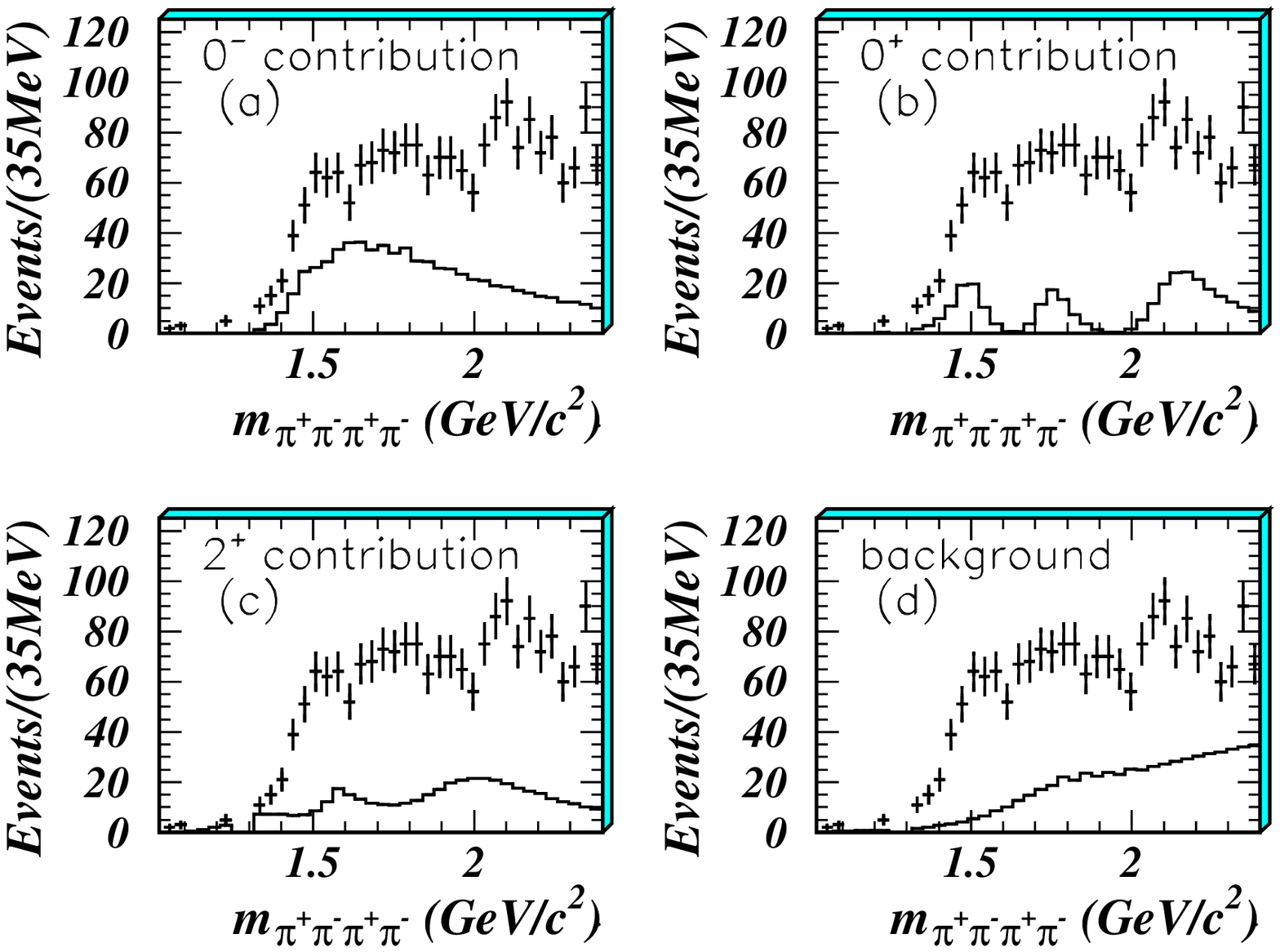,width=0.90\textwidth}
\vspace*{-3mm}
\caption{Partial wave decomposition of radiative J/$\psi$ 
decays into 2\pip 2\pim\ (from \protect\cite{bes4p}).}
\label{jto4pi} 
  \end{figure}

Glueballs have to show up in radiative J/$\psi$ decays. In these
decays, one photon and two gluons are emitted by the annihilating 
$c\bar c$ system, the two gluons interact and must form glueballs
 - if a glueball exists in the accessible mass range. A most prominent
- possibly scalar - signal in radiative J/$\psi$ decays into
2\etg\ is the old $\Theta (1690)$ with spin 0 or spin 2, 
which might have a 
large fraction of glue in its wave function. In a recent
reanalysis, the state was shown to be of scalar nature
\cite{dunnwoodie}. The scalar part in Fig.~\ref{dunnw}
shows two resonances,
at 1430 MeV and at 1710 MeV. The latter has strong coupling to 
\kkb . The mass shift from 1500 to 1430 MeV will be discussed below.
Three scalar resonances 
are observed at BES in radiative J/$\psi$ decays into 2\pip 2\pim\
\cite{bes4p}. The results of a partial wave analysis (see 
Fig.~\ref{jto4pi})
show a slowly rising instrumental background and 
3 important contributions with scalar, pseudoscalar and 
tensor quantum numbers. 

Of particular importance here is 
the scalar part. It is seen to contain 3 resonances, at 1500, 1740
and 2100 MeV. This pattern of states was already suggested in 
a reanalysis of MARKIII data \cite{tokibugg}. 
The $f_0(1500), f_0(1740)$ and the $f_0(2100)$ have a similar 
production and decay pattern. Neither a  $f_0(1370)$ nor a
'background' intensity is assigned to the scalar isoscalar 
partial wave. 
\subsubsection{\pbp\ annihilation in flight}
The \etg\etg\ invariant mass spectrum produced 
in \pbp\ annihilation in flight into \piz\etg\etg\ 
\cite{E760} exhibits three peaks at 
$f_0(1500), f_0(1750)$ and $f_0(2100)$ MeV, fully compatible 
with the findings in radiative J/$\psi$ decays into four pions.  
The data were not decomposed into partial waves
in a partial wave analysis, so the peaks could have 
J$^{\rm P\rm C}=0^{++}$ or $2^{++}$. If the states would have
J$^{\rm P\rm C}=2^{++}$, their decay into \etg\etg\ would 
be suppressed by the angular momentum barrier. 
The fact that the peaks are seen so clearly suggests 
$0^{++}$ quantum numbers, and this is the result of
the partial wave analysis of the J$/\psi$ data. 
Hence we believe that the 3 peaks are scalar
isoscalar resonances.
\subsubsection{Central production}
\begin{figure}[h]
\includegraphics[width=\textwidth]{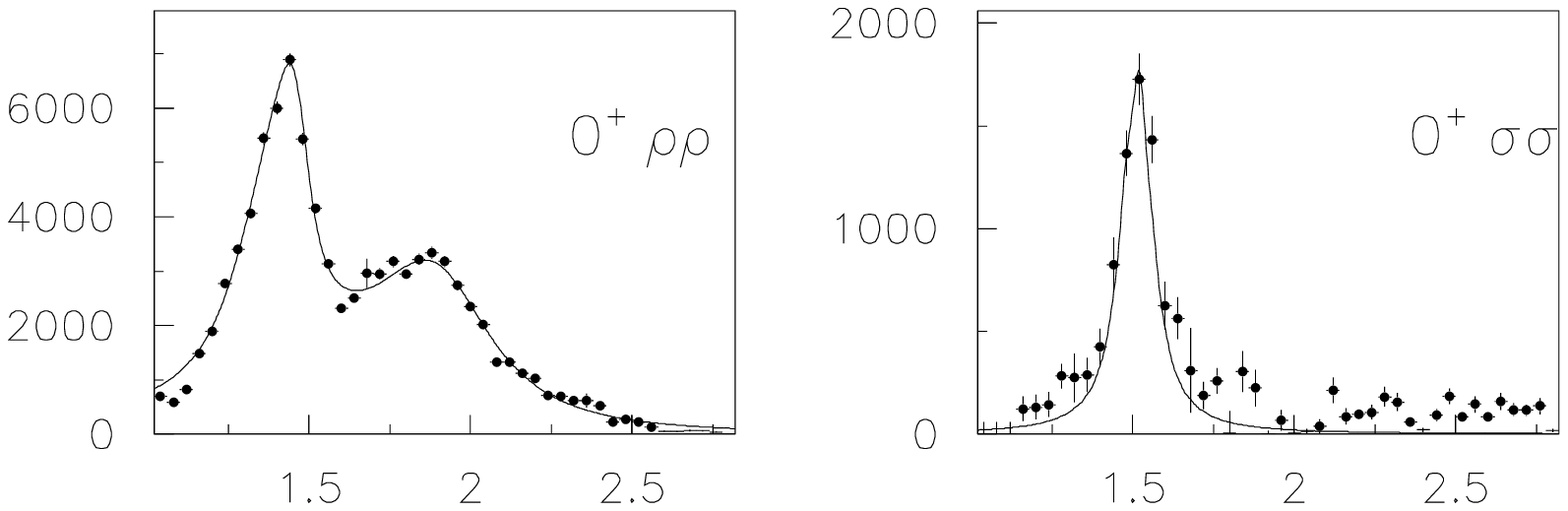}\\
\vspace*{-2cm}

\includegraphics[width=\textwidth]{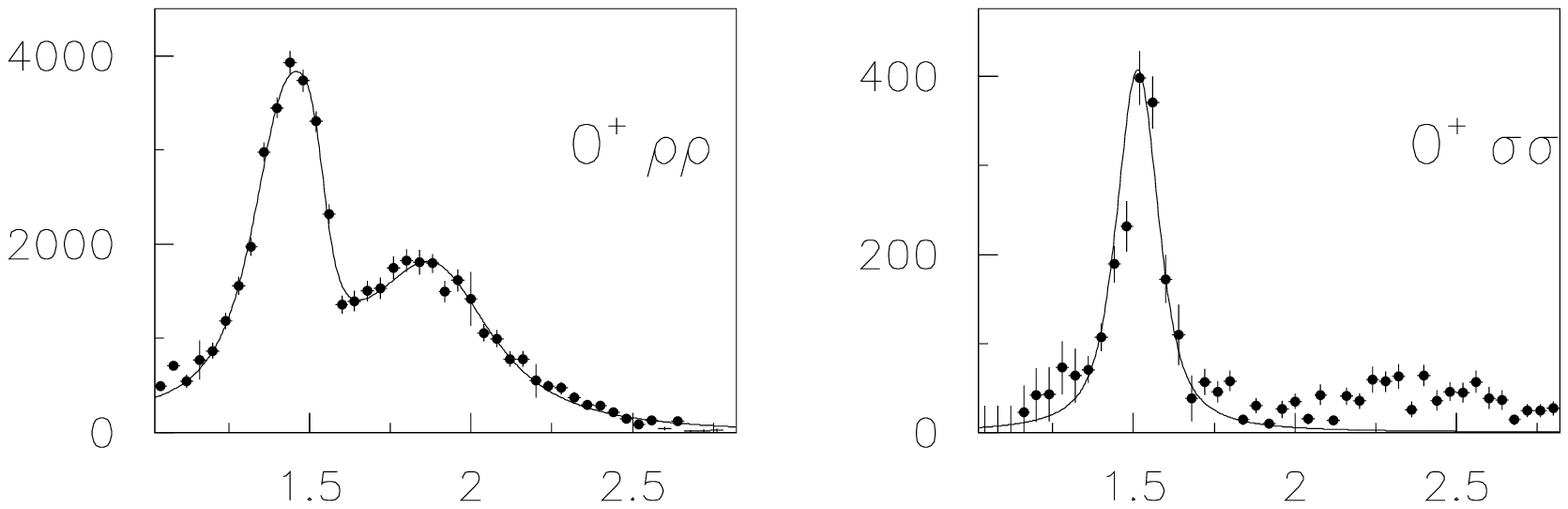}\\
\vspace*{-2cm}

\begin{minipage}[t][4cm][t]{0.65\textwidth}
\hspace*{-1.5cm}
\includegraphics[width=\textwidth]{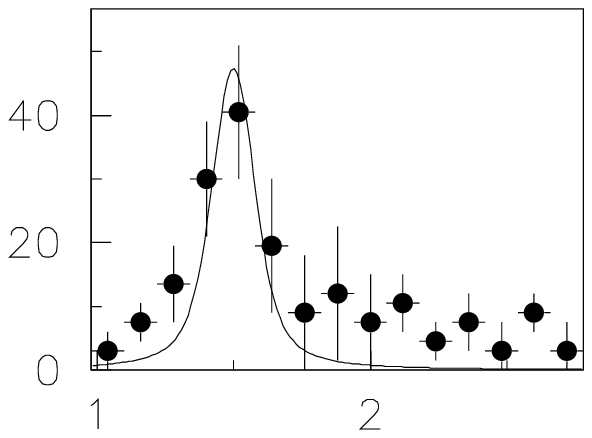}
\end{minipage}

\vspace*{-6cm}

\hspace*{-16cm}\hfill
\begin{minipage}[t][4cm][t]{.4\textwidth}
\caption{4\p\ invariant mass (in GeV) spectra from central
production. First row: 2\pip 2\pim ; second row \pip\pim 2\piz ;
left: \rh\rh\ S-wave; right: $\sigma\sigma$ S-wave. 
Third row: $\sigma\sigma$ S-wave in 4\piz ; (from 
\protect\cite{Barberis:2000em}).}
\label{wa4pi}
\end{minipage}
  \end{figure}

Central production is believed to be a good 
place for a glueball search. Fig.~\ref{wa4pi}
shows 4$\pi$ invariant mass  
spectra from the WA102 experiment 
\cite{Barberis:2000em}. 
A large peak at 1370 MeV is seen, followed by a dip in the
1500 MeV region and a further (asymmetric) bump. 

The partial
wave analysis decomposes this structure into several scalar
resonances, the $f_0(1370), f_0(1500)$ and $f_0(1750)$ and
a new $f_0(1900)$. We note that the partial wave analysis 
finds $f_0(1370)$ decays into $\rho\rho$ but not
into $\sigma\sigma$ while the $f_0(1500)$ shows both decay 
modes. In the Crystal Barrel experiment the $f_0(1370)$
decays into $\rho\rho$ and into $\sigma\sigma$ with similar
strength, see Table \ref{decayvergl}. 
\par
The upper limit for $f_0(1370)\rightarrow\sigma\sigma$ 
of the WA102 experiment is not very
restrictive. 
\begin{table}[h!]
\renewcommand{\arraystretch}{1.3}
\bc 
\begin{tabular}{|lclc|}
\hline 
$f_0(1370)\rightarrow\sigma\sigma /f_0(1370)\rightarrow 4\pi  $&=&
$\leq 0.23 $ & WA102 \\
$f_0(1500)\rightarrow\sigma\sigma /f_0(1500)\rightarrow 4\pi  $&=&
$0.23 - 0.50$ & WA102\\
\hline
$f_0(1370)\rightarrow\sigma\sigma /f_0(1370)\rightarrow 4\pi  $&=&
$0.51\pm 0.09 $ &CBAR\\
$f_0(1500)\rightarrow\sigma\sigma /f_0(1500)\rightarrow 4\pi  $&=&
$0.26\pm 0.07$ & CBAR \\
\hline
\end{tabular}
\ec
\caption{Decay fractions into $\sigma\sigma$ of scalar mesons 
from the WA102 and CBAR experiments.}
\renewcommand{\arraystretch}{1.0}
\label{decayvergl}
\end{table}
In the partial wave analysis 
representing the preferred solution, the upper limit
for $f_0(1370)\rightarrow\sigma\sigma /f_0(1370)\rightarrow 4\pi  $
is certainly smaller. On the other hand, in \pbp\ annihilation
the $\sigma\sigma$ decay mode is certainly present and strong. 
This is an important observation and
provides a clue for the interpretation of the spectrum of
scalar mesons. In any case, there are large differences
between the \rh\rh\ and $\sigma\sigma$ mass distributions
which need to be explained.
\par
\subsection{The {\em red dragon} or $f_0(1000)$}
Before we continue the discussion
we have to introduce a further concept:
\subsubsection{$\bf s$-channel resonances and $\bf t$-channel
exchanges}
\begin{figure}[h!]
\epsfig{file=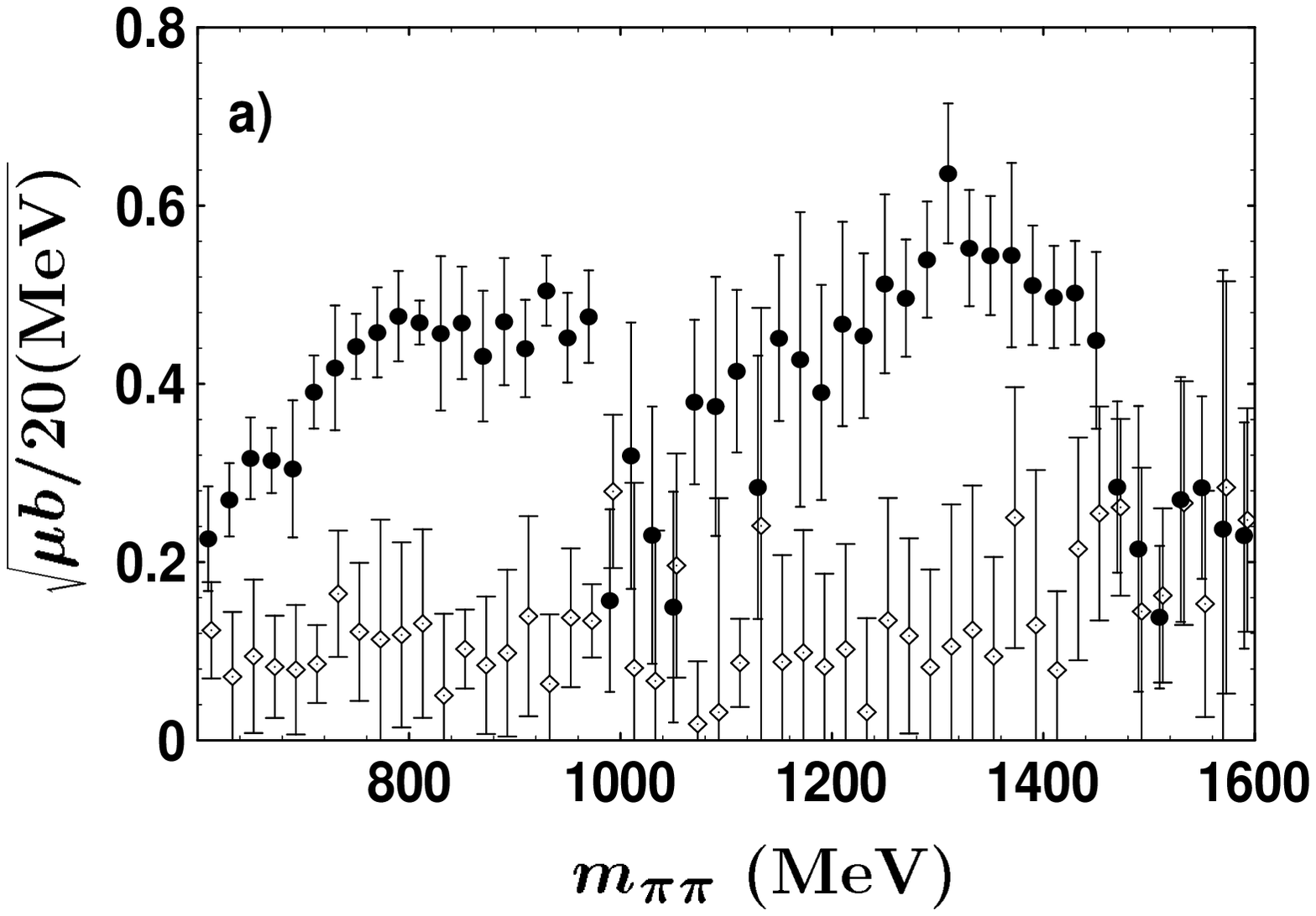,width=0.92\textwidth}
\vspace*{-10mm}
\hspace{-2mm}\epsfig{file=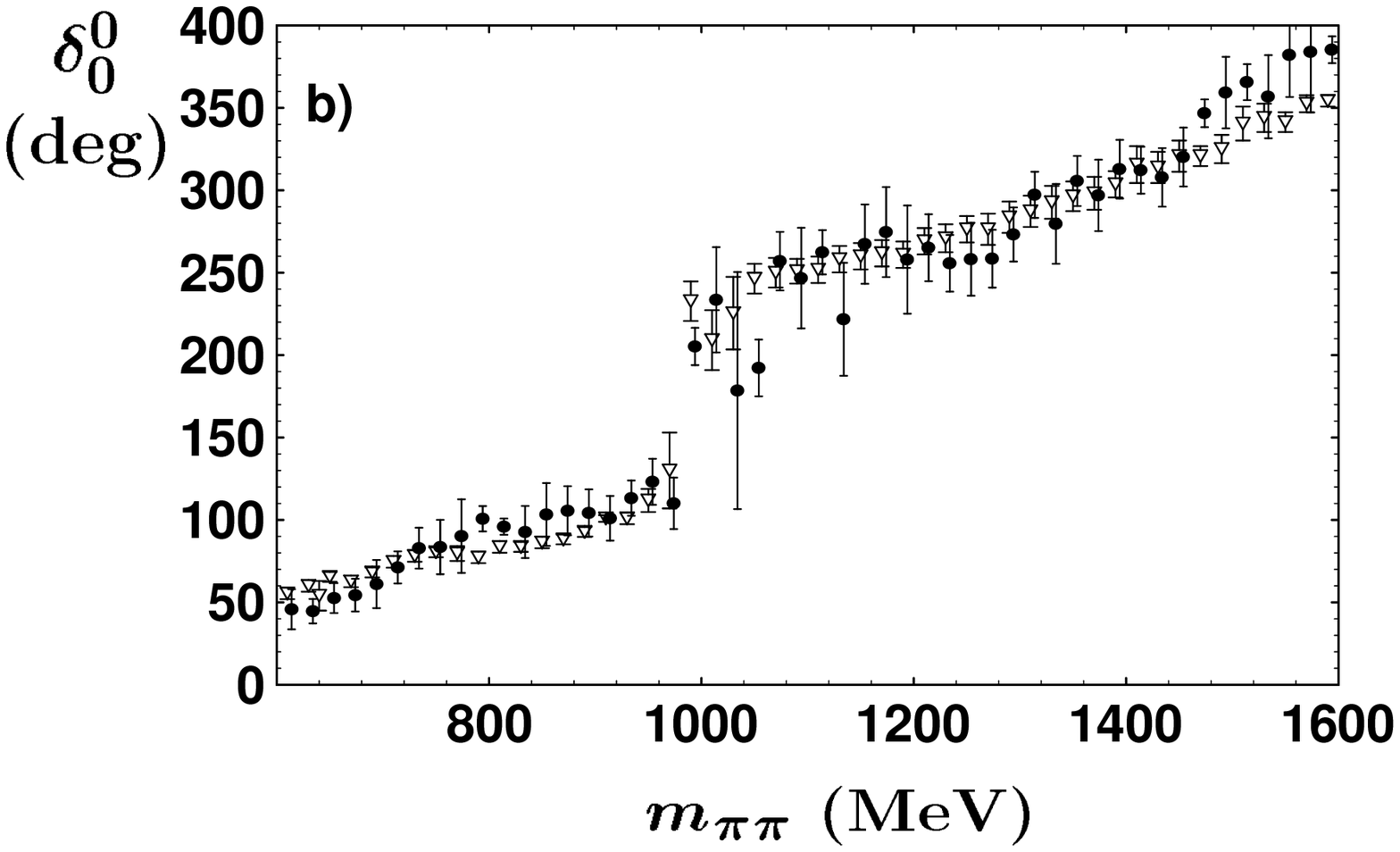,width=0.945\textwidth}
\caption{Amplitude and phase from \p\p\ scattering (black dots);
open circles: \p$a_1(1260)$ scattering. Shown is the so-called
down-flat solution; (from \protect\cite{Kaminski:1997da}).}
\label{amplphase}
 \end{figure}

Fig.~\ref{amplphase} shows the $\pi\pi$ scattering 
amplitude. The phase rises slowly, then there is
a sudden phase increase at 980 MeV indicating the
presence of the $f_0(980)$. The modulus of the amplitude 
shows a dip at the mass of the  $f_0(980)$: intensity
is taken from \p\p\ scattering to the \kkb\ inelastic
channel. The peak in the scattering amplitude at 
low energy is often called $\sigma$-meson; the second
bump at 1300 MeV was called $\epsilon (1300)$. 
There is an on-going discussion on whether the 
$\sigma$ should be considered as genuine meson or
not; the interested reader may 
consult~\cite{Pennington:1999fa}.
\par
\begin{figure}[h!]
\epsfig{file=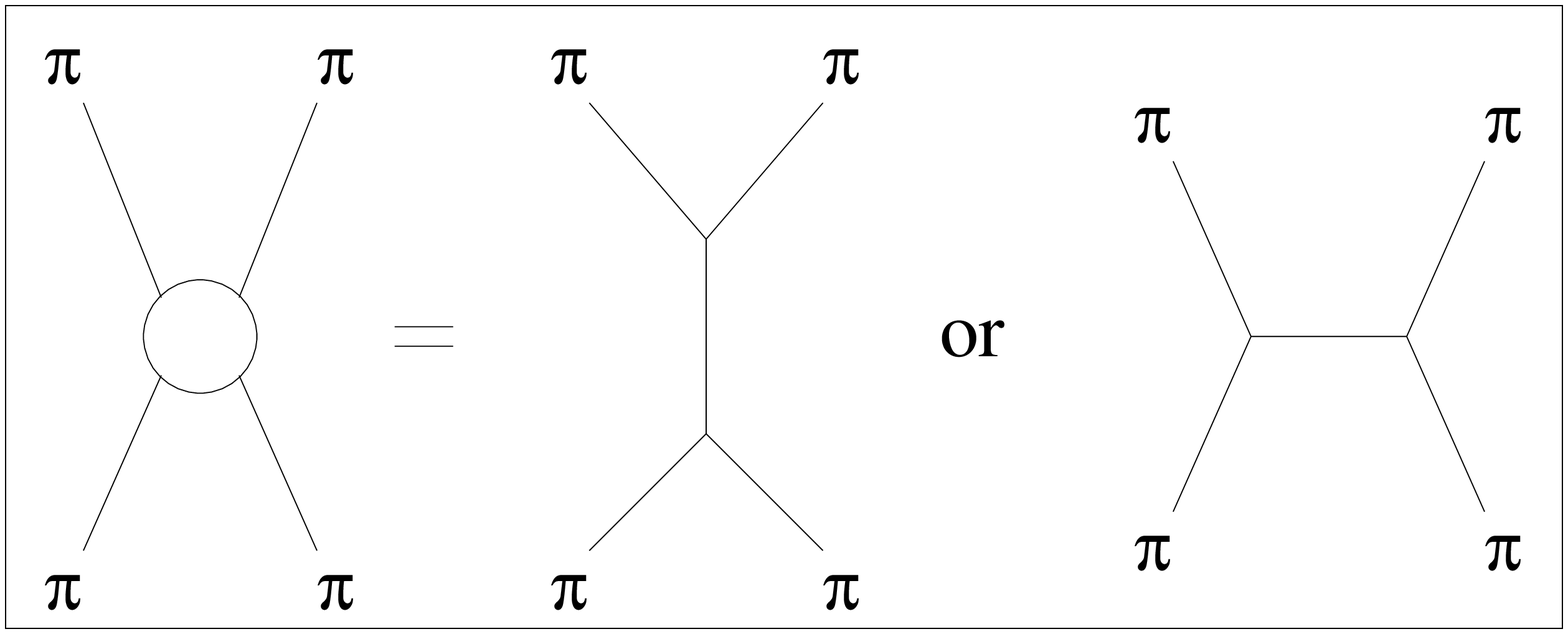,width=\textwidth}
\caption{Scattering of two pions (left) via $s$-channel 
resonances (center) and $t$-channel exchange.
}
\label{pipiscatt}
  \end{figure}

There are two processes which may contribute to the \p\p\
scattering amplitude: formation of $s$-channel resonances and
scattering via $t$-channel exchanges. They are schematically
drawn in Fig.~\ref{pipiscatt}. Scattering processes or more
precisely the scattering matrix can be expressed
by a sum of $s$-channel resonances or $t$-channel
exchanges; in Regge theory this is called duality
and is the basis for the Veneziano model. 
So you may analyze a data set and describe the data by a sum
over $s$-channel resonances and get a very good description
with a finite number of complex poles in the
\p\p\ S-wave scattering amplitude. 
You could also analyze the data by a summation over $t$-channel
exchange amplitudes and also get a good fit. If you add 
amplitudes for both processes, you run the risk of double 
counting. So in fits you should avoid to mix the two schemes.
\par
\par
There is a common believe that the interpretation of 
a pole in the complex scattering energy plane as
originating from $s$-, $t$- or $u$-channel phenomenon
is a matter of convenience. I do not share this view:
only $s$-channel resonances have defined couplings to 
different final states,  $t$- or $u$-channel
exchanges have not. The $q^2$ dependence of
$t$- or $u$-channel exchange processes reflects 
interaction ranges, the $q^2$ dependence of
$s$-channel resonances reflects the spread
of the wave function. When 'counting'
the number of $q\bar q$ states (to argue that we have a decuplet
of scalar states instead of a nonet), we have to
count the number of 'true' 
$s$-channel resonances and have to
suppress poles in the scattering plane which originate
from $t$-channel exchange processes. 
The $f_0(400-1200)$, 
e.g., is certainly present in scattering data with a 
pole in the complex energy plane. And you may choose
to describe this pole as $s$-channel resonance even
if its true origin might be $t$-channel exchange.
\par
We are searching for resonances in the $s$-channel, for
poles originating from  $s$-channel resonances. But there may also be
poles due to $t$-channel exchanges, poles originating from meson-meson 
interactions. They should rather be interpreted as mesic molecules 
and not as $q\bar q$ mesons. So, how does one decide if a
particular pole in the scattering plane is due to a $s$-channel 
resonance or to $t$-channel exchanges\,? 
\par
$s$-channel resonances have always the same ratio of 
couplings to different final states. The partial widths of the
$f_0(1500)$ must not depend on the way how it was produced. This
is different for $t$-channel exchanges: assume the $f_0(980)$
would be produced by  $t$-channel exchanges only, then it
could show up differently in \p\p\ and in \kkb\ scattering. 
Properties of a 'resonance' which depend strongly 
on the production process suggest that the resonance may
originate from $t$-channel exchanges. 
\subsubsection{The {\em red dragon} in $\bf\pi\pi$}
The \p\p\ scattering amplitude exhibits a continuously and slowly
rising phase and a sudden phase increase at 980 MeV. The rapid phase
motion is easily identified with the $f_0(980)$, the slowly rising
phase can be associated with an $s$-channel resonance which was
called $f_0(1000)$ by Morgan and Pennington \cite{MP}. It extends
at least up to 1400 MeV. Minkowski and Ochs \cite{minkochs} suggested
that this broad enhancement which they call {\it the red dragon}
is the scalar glueball. However, we have seen 
that this broad background amplitude - including
the monotonously rising phase - can well be reproduced
by a \rh\ exchange amplitude in the $t$-channel. From
a fit to the \p\p\ S-wave scattering data even mass and width
of the \rh\ exchanged in the $t$-channel
can be determined. So this background amplitude
is likely not a  $f_0(1000)$ $q\bar q$ state, it is 
likely not two mesons,
the old $\sigma (550)$ and $\epsilon (1300)$, it is caused
by \rh\ (and possibly other less important) exchanges
in the $t$-channel. 
\par
\begin{figure}[h]
\includegraphics[width=\textwidth]{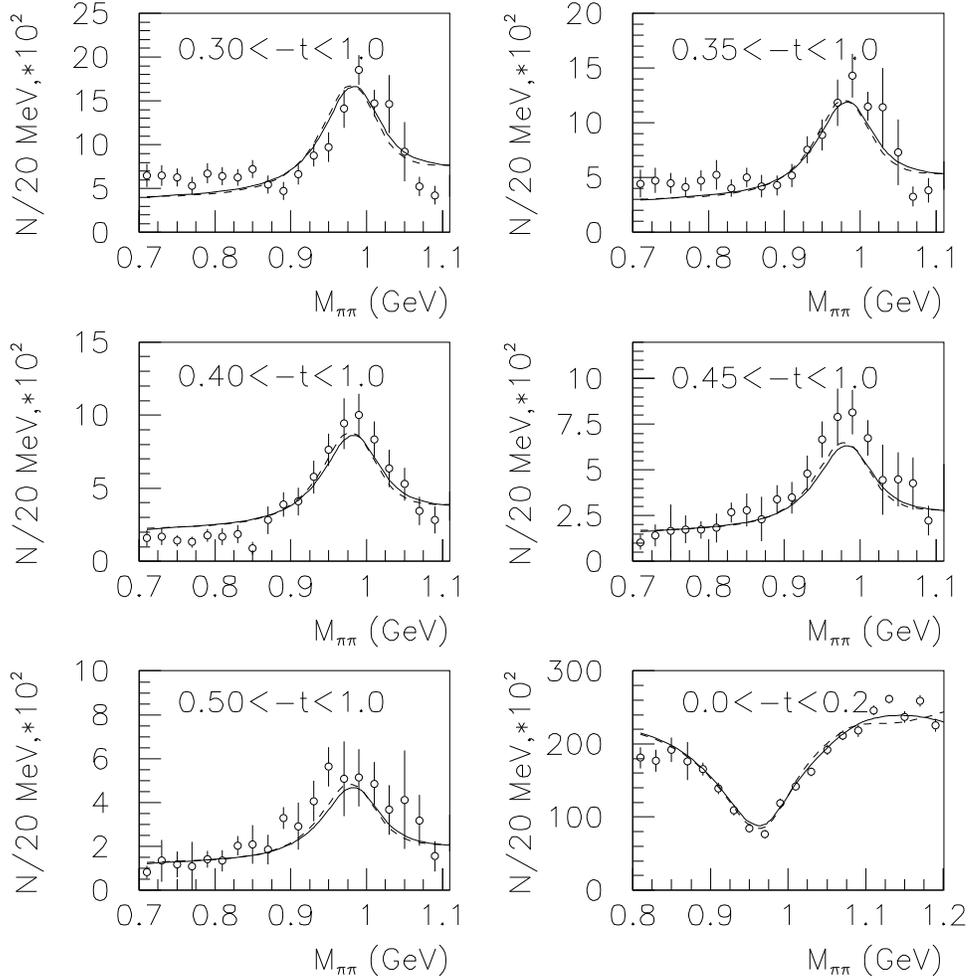}
\caption{The \p\p\ invariant mass distribution
for different cuts in the momentum transfer t
$=-q^2$ in the reaction \pim p\ra\piz\piz n. 
At small t the $f_0(980)$ is observed as dip,
for large momentum transfer as peak. The data are
from \cite{Alde:1995jj}, the fit from \cite{Anisovich:1996tk}.}
\label{GAMS}
\end{figure}

Fig.~\ref{GAMS}) shows data of the GAMS collaboration 
on \p\p\ scattering from the charge exchange reaction
\pim\ p\ra 2\piz n at 40 GeV/c. 
At small momentum transfer, the  $f_0(980)$ is produced
as a dip. But at large momentum transfer, the dip disappears
and the $f_0(980)$ shows up as a peak. I interprete
this behavior as evidence that 'soft' processes like $t$-channel
exchange dominate the scattering amplitude at small
momentum transfer. At large momentum transfers the $q\bar q$ nature
of the  $f_0(980)$ core become the main feature
of the data. The 'background' amplitude disappears
at large $t$: the background is due to soft processes 
expanded over a larger volume. It is certainly not a compact and well
localized glueball with properties as predicted in lattice 
gauge calculations. 
\subsubsection{The {\em red dragon} in $\bf 4\pi$}
A similar question arises in the 4\p\ final state. Is the 
enhancement seen in the left parts of Fig.~\ref{wa4pi} 
a $q\bar q$ resonance\,? Or can it be traced to 
\rh\ and other exchanges in the $t$-channel\,? We now
argue that the latter is indeed the case.
\par
\begin{figure}[h]
\includegraphics[width=\textwidth]{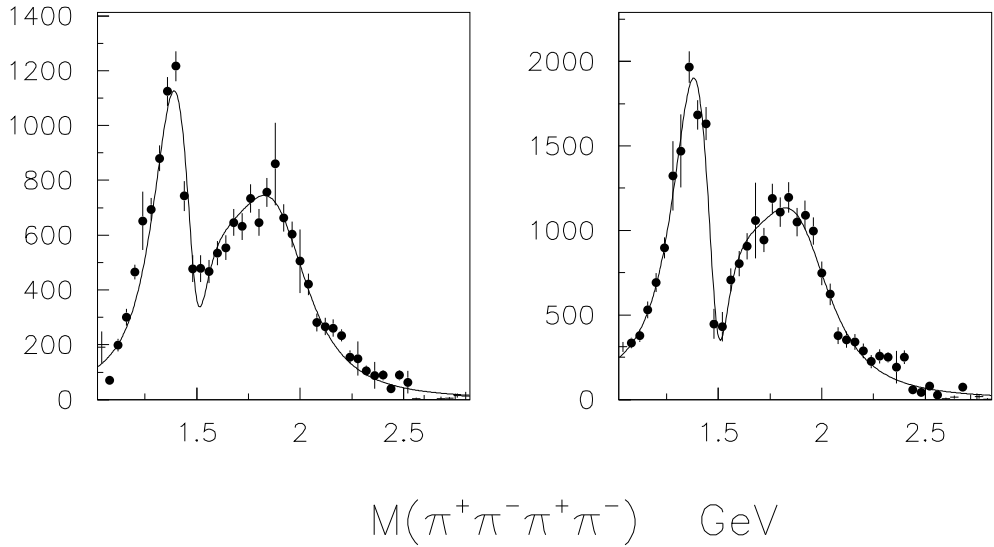}

\vspace*{-0.5cm}

\caption{4\p\ invariant mass spectrum produced by two protons in
central collisions. A cut is made on
the angle in transverse direction between the two 
outgoing protons. Left: 90-135$^{\circ}$, Right: 
135-180$^{\circ}$. 
The latter setting corresponds to the so-called
glueball filter. The $f_0(1500)$ shows up as a dip
just like the  $f_0(980)$ in \p\p\ scattering. 
From \cite{Barberis:2000em}}
\label{wa_dip}
\end{figure}

First, consider Fig.~\ref{wa_dip}. On the right side,
a selection is made for small momentum transfers to the
4\p\ system. At small momentum transfer, the $f_0(1500)$ is
seen as a dip. This resembles very much the data of the 
GAMS collaboration on \p\p\ scattering (Fig.~\ref{GAMS}).  
But it also reminds us of the
J/$\psi$\ra\gam\pip\pim\ data (see Fig.~\ref{dunnw})
where a dip is seen at 1500 MeV instead of a
peak.
\par
Consequently, we now make the attempt to 
interprete the 4\p\ mass spectra of the WA102 experiment
as $t$-channel exchange phenomenon in order to investigate if
we gain further insight. We assume 
that Pomeron-Pomeron scattering can also proceed via $\rho$ 
exchange in the $t$-channel.  
This $t$-channel amplitude then interferes with the production of the
$q\bar q$ state $f_0(1500)$ producing a dip, very much alike the dip
seen at 980 MeV in $\pi\pi$ scattering. Isospin conservation
does not allow $\sigma\sigma$ production from 
$\rho$ exchange in the $t$-channel for Pomeron-Pomeron scattering. 
We predict that at large momentum transfer the $f_0(1500)$
will also show up as peak. 
\par
Here we have made a first important step: we now understand why the
left-hand spectra of Fig.~\ref{wa4pi} differ so much from
the right-hand spectra. The $\sigma\sigma$ final state can 
be reached only via $s$-channel resonances and there is only 
one: the $f_0(1500)$. The \rh\rh\ final state is produced
by $t$-channel exchanges; they generate the broad enhancement
extending over the full accessible mass range. It rises
at threshold for 4\p\ production and falls off because
of the kinematics of central production: high mass
systems are suppressed with 1/M$^2$. 
\par
In contrast to Pomeron-Pomeron scattering,  
\pbp\ annihilation may also start from \pbp\ra$\rho\rho\pi$
which then converts via $\rho$ exchange in the $t$-channel into
$\sigma\sigma$. Hence we may expect \rh\rh\ and $\sigma\sigma$
to contribute to the scalar isoscalar 4\p\ mass spectrum.
\par
We notice the similarity of the  $f_0(1370)$ and the old
$\epsilon (1300)$. The relation between these two phenomena
is not well understood. The reason that the old 
$\epsilon (1300)$ was not identified with the $f_0(1370)$ 
lies just here: the $\epsilon (1300)$ was seen in \p\p\
scattering with a small inelasticity, i.e. small coupling
to 4\p\ while the $f_0(1370)$ has small coupling to \p\p\
and a large one to 4\p . This is naturally explained
when the 1300 MeV region interacts via $t$-channel exchange.
Then \p\p\ goes to \p\p , \kkb\ to \kkb ,
Pomeron-Pomeron to \p\p\ by pion
exchange, to \rh\rh\ via \rh\ exchange, etc. 
\par
There is one counter argument: the 
$\epsilon (1300)/f_0(1370)$ is not only
a peak but it has also an intrinsic phase which varies 
by 180$^{\circ}$. However, it is in the middle
of two $q\bar q$ resonances. Two successive resonances in one
partial wave have opposite phases. So there must be a phase advance
of about 180$^{\circ}$. This is not only the case
for \p\p\ scattering but must also be true for
\rh\rh\ scattering. $f_0(980)$ as scalar state must
have a $\rho\rho$ coupling even though it cannot 
decay into \rh\rh\ because of phase space limits.
Hence the phase of the $\rho\rho$ scattering 
amplitude should raise
from 980 to 1500 MeV by 180$^{\circ}$. Due to the  
$\rho\rho$ threshold and the destructive interference with the
$f_0(1500)$ the $\rho\rho$ scattering amplitude has a peak
between 1000 and 1500 MeV: the most natural and economic
description is by use of a Breit-Wigner resonance. But its
true nature is of molecular character. 
\par
The interpretation suggested
here can be tested: if the $f_0(1370)$ is a $t$-channel phenomenon
it is produced with a phase of $\pm \pi /2$ with respect to the
$f_0(980)$ and $f_0(1500)$. 
\par
We conclude that the $f_0(1370)$ is not produced
in hard processes like J/$\psi$ radiative decays, $D_s$ decays or
\pbp\ annihilation in flight, it is seen only in peripheral 
processes. The production and decay pattern in central production
suggests that it is a $t$-channel phenomenon originating from
meson-meson interactions. 
\par
\subsubsection{Is the \p\p\ low-mass enhancement the glueball\,?} 
In Pomeron-Pomeron scattering there is ample production of
two-pions in S-wave. Data on \pip\pim\ and \piz\piz\
production for S-wave di-pions show a huge enhancement above
the two-pion threshold. The \p\p\ production in P-wave is 
strongly suppressed compared to \p\p\ production in S-wave.
This suppression supports the interpretation that central
production is dominated by Pomeron-Pomeron scattering,
that it is a gluon-rich process. Could this low-mass 
enhancement be the scalar glueball\,? 
\par
First we note that the question if the low-mass \p\p\
interactions should be interpreted as $s$- or $t$-channel
phenomenon cannot be decided by comparing different
production and decay rates; below the opening of an
inelastic channel the dependence of couplings to different 
final states cannot be investigated. On the other
hand we know that the probability of 
Pomeron-Pomeron scattering scales with 1/$M^2$ in
the invariant mass. Taking this scaling into account,
the data are fully compatible with the data from \p\p\
scattering \cite{kirkpriv}. 
These data do not contain more glue
than 'normal' \p\p\ scattering data which can be understood 
quantitatively by \rh\ exchange amplitudes.
The strong \p\p\ production above threshold in
central production does not evidence its glueball
nature. 

\section{Interpretation}
\subsection{The spectrum of scalar mesons}

After having argued that the $f_0(400-1200)$ and $f_0(1370)$,
the red dragon of scalar isoscalar interactions, are
likely generated by $t$-channel exchange amplitudes, we
are left with 18 scalar states which can be grouped into 2 nonets.
This is done in Table \ref{scalar12}. 
\par
\begin{table}
\bc
\begin{tabular}{|c|c|cc|}
\hline
                 &               &                      &      \\  
                 &               & $f_0(400-1200)^{1}$  &      \\  
                 &               &                      &      \\  
                 &               &                      &      \\  
                 &$a_0(980)^{2}$ &  $f_0(980)^{2}$      &      \\  
                 &               &                      &      \\  
                 &               &                      &      \\  
                 &               &  $f_0(1370)^{1}$     &      \\  
                 &               &                      &      \\  
$K_0^*(1430)^{2}$&$a_0(1490)^{3}$&  $f_0(1500)^{2}$     &      \\  
                 &               &                      &      \\  
                 &               &  $f_0(1750)^{3}$     &      \\  
                 &               &                      &      \\  
$K_0^*(1950)^{3}$&               &                      &      \\  
                 &               &  $f_0(2100)^{3}$     &      \\  
\hline
\end{tabular}
\ec
\caption{The scalar mesons and their interpretation:
$^1$: generated by $t$-channel exchanges. $^2$: The $1^3P_0$ ground state
scalar meson nonet. $^3$: The $2^3P_0$ first radially excited
scalar meson nonet.}
\label{scalar12}
\end{table}
\par
It must be emphasized again that the $f_0(980)$ and $f_0(1500)$ 
show a similar production and - partly also - decay pattern
in \p\p\ scattering and in D$_s$ decays. In D$_s$ 
decays it is appearant that the $f_0(980)$ and $f_0(1500)$ 
have no simple $u\bar u +d\bar d$ or $s\bar s$ structure: both 
are strongly produced in an $s\bar s$initial state and 
decay strongly to \p\p . This strong OZI rule violation
is also observed in J/$\psi$ decays into $\Phi$\p\p\ and 
$\Phi$\kkb .
\par
The three states $f_0(1500)$, $f_0(1740)$ and $f_0(2100)$ have 
striking similarities in radiative J/$\psi$ decays into 
4 pions and \pbp\ annihilation in flight into \etg\etg . 
These two reactions also show that the scalar isoscalar states 
are isolated non-overlapping resonances. The lowest state, the 
$f_0(980)$, is obviously strongly influenced by the \kkb\
threshold and its wave function must contain a large
\kkb\ contribtion of molecular character. However, at large
momentum transfer reactions, the $q\bar q$ nature of its
core wave function becomes visible. The $f_0(980)$ should be
produced in radiative J/$\psi$ decays; it has however not
been observed. This is certainly a hint which supports the
interpretation of the $f_0(980)$ as \kkb\ molecule.
On the other hand, also the $f_0(1500)$ 
is - at best - difficult to see in J/$\psi$ decays. 
We have to wait for better data. 
\subsection{Instanton interactions}
As we have discussed, the \etg\ and \etp\ mesons are often considered as 
{\it gluish}. Let us discuss why this conviction is widely spread. 
\par
Most nonets have mixing angles which are close to the ideal one: there is
a mainly $n\bar n  = \sqrt{\frac{1}{2}}(u\bar u+d\bar d)$ 
state and a $s\bar s$state.
This is not true in the case of the pseudoscalar nonet: the \etg\ and
\etp\ mesons are not $n\bar n$ and $s\bar s$states but rather SU(3) octet and
singlet states. You may try to understand this perturbatively. In 
vector mesons the $s\bar s$component needs at least 3 gluons to 
convert into light quarks, in pseudoscalar mesons two gluons are sufficient.
Hence the coupling between $n\bar n$ and $s\bar s$is much stronger for
pseudoscalar (or scalar) mesons while it is weak for the other nonets.
In other nonets like the tensor mesons
you may invoke angular momentum barrier
arguments to justify ideal mixing. 
\par
This picture has led to the conjecture that the \etg\ and
\etp\ wave functions should contain a glueball fraction.
This conjecture is not supported experimentally. Also, the argument above
does not consider the Goldstone nature of pseudoscalar mesons. In the limit
of massless quarks, the pseudoscalar octet
mesons should have vanishing masses, and the pseudoscalar
mesons have indeed masses which are small compared to the
nucleon mass. Only
the \etp\ has obviously not a
small mass. Its mass originates from the coupling of the SU(3)
singlet state to the gluon field. But this does not entail
that there is a pure-glue component in the wave function in any
point in space and time. The coupling between the \etp\ and the
gluon field can be described by introducing
a new type of interactions based on instanton effects. 
\par
Quantization of the color fields leads to strong local fluctuations 
of the expectation values for $F^{\mu\nu}F_{\mu\nu}$ which 
can be localized as instanton solutions. 
The color fields generate a series of degenerate minima of the 
vacuum Hamiltonian (carrying different winding numbers) 
determining the propagation of massless quarks. Instantons and 
anti-instantons induce tunneling between the vacuum states.
The tunneling to a state with a different winding number
requires a flip of the helicities:
$$
(S,M_s)\ra (S,-M_s)
$$
and are hence (in first order) restricted to interactions in
pseudoscalar and scalar mesons since only here the total
angular momentum $J$ can remain unchanged when the spin is 
flipped. The strength of instantons changes sign for scalar mesons
compared to pseudoscalar mesons but keeps its absolute value. 
\par
Instantons can be identified
in lattice QCD and are found to have typical sizes of 
0.36 fm. They are frequent: they are observed with a density
(in space and time) of 1.6 fm$^{-3}\cdot c$fm$^{-1}$. Instantons
induce spontaneous breaking of chiral symmetry leading to 
effective quark masses. They contribute to the vacuum condensate and
provide an explanation for the axial anomaly.
\par
Obviously, instantons mediate a very strong coupling between 
gluon fields and Goldstone bosons, in particular pions.

\subsubsection{pseudoscalar mesons and Goldstone bosons}
\begin{figure}[h]
\epsfig{file=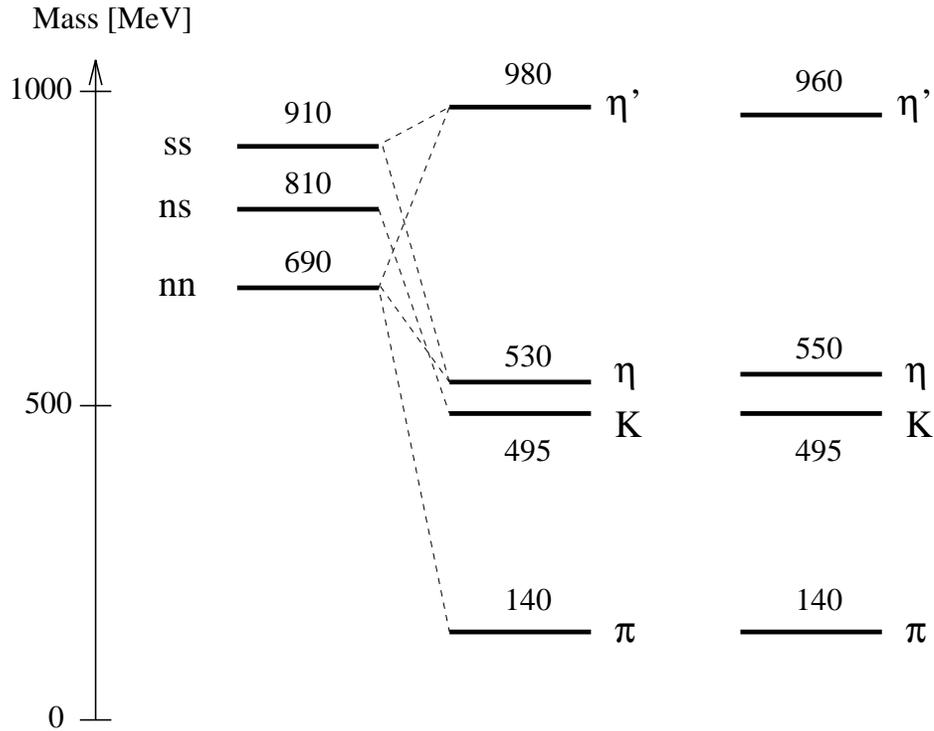,width=\textwidth}
\caption{The mass spectrum of pseudoscalar mesons in a relativistic
quark model using a linear confinement potential fitted to
reproduce the Regge behavior and instanton interactions.
Left: without, right including instanton interactions.}
\label{psinst}
  \end{figure}

Let us now discuss the influence of instanton induced interactions
on the pseudoscalar mass spectrum. Fig.~\ref{psinst} shows that
instanton interactions bring the pion mass down and the \etp\
mass up: the experimental masses of pseudoscalar mesons
are very well reproduced. 
The calculated masses come from quark model calculations
solving the Bethe Salpeter equation for a linear confinement 
potential the parameters of which are fitted to reproduce 
Regge trajectories \cite{Klempt:1995ku}. 
There is no gluon exchange in the model; instead
instanton interactions are used with a strength which is fitted 
to reproduce pseudoscalar meson mass spectrum. The confinement potential
has a Lorentz structure which is chosen as 
$\frac{1}{2}(1\otimes 1 - \gamma_0\otimes\gamma_0)$. 
Alternatively, also a Lorentz structure
given by  $\frac{1}{2}(1\otimes 1 - \gamma_{\mu}\otimes\gamma^{\mu}
- \gamma_{5}\otimes\gamma_{5})$  was used. 

\subsubsection{Scalar mesons and instanton interactions}
\begin{figure}[h]
\epsfig{file=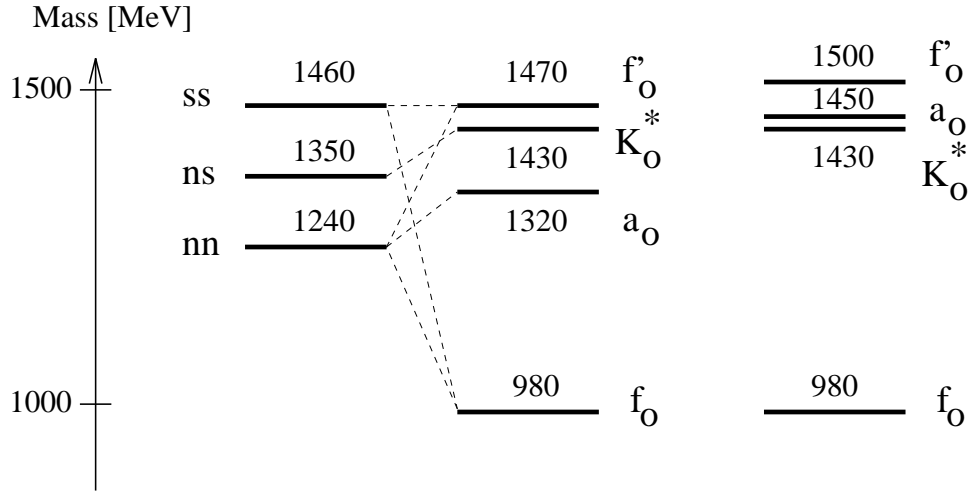,width=\textwidth}
\caption{The mass spectrum of scalar mesons in a relativistic
quark model using a linear confinement potential fitted to
reproduce the Regge behavior and instanton interactions.
Left: without, right including instanton interactions.}
\label{scinst}
  \end{figure}

Fig.~\ref{scinst} shows the spectrum of scalar mesons 
as it is calculated now without
using any new parameter \cite{Klempt:1995ku}. 
The two predominantly isoscalar mesons are
calculated to have masses of 980 and 1470 MeV, respectively.
They are not $n\bar n$ and $s\bar s$states. The $f_0(980)$ is rather 
determined to be the SU(3) singlet state, nearly mass
degenerated with the \etp . The two isosinglet states
\etp\ and $f_0(980)$ form a parity doublet. 

The full spectrum of scalar states calculated for the two
different Lorentz structures of the confinement potential
\cite{Koll:2000ke}\cite{Ricken:2000kf}
are compared to experiment in Table \ref{vergl}.
The octet state is found at 1470 MeV, it is readily identified with the 
$f_0(1500)$. From the $T$ matrices for 
\p\p\ra\etg\etg\ and \kkb\ra\kkb\ scattering, 
Minkowski and Ochs \cite{minkochs} concluded that
the $n\bar n$ and $s\bar s$ components of the $f_0(1500)$ must have opposite 
signs, an observation providing strong evidence against a
large glueball component in the $f_0(1500)$ wave function.
\par
The first radial excitations of the $f_0(980)$ and $f_0(1500)$ 
are calculated to have masses of 1776 and 2113 MeV, respectively.
The agreement with the experimental candidates discussed above
is excellent, much better than one could expect.
\par
The calculated masses of the two isodoublets states 
K$_0$ also agree very well with the experimental values, but 
there is a problem in the isovector meson masses. The calculated values
are 1320 and 1930 MeV, respectively. Experimentally, we
have one state at 980 and one at 1480 MeV. The first one
is suspected of being a \kkb\ molecule; the latter mass value
is consistently found by the Crystal Barrel Collaboration
in \p\etg\ \cite{spanier} , \p\etp\ \cite{resag} , and \kkb\ 
\cite{Abele:1998qd}. The Obelix Collaboration, however, finds
a mass of 1290 MeV \cite{masoni}. 
\par 
The agreement in the isovector sector is improved, the overall
agreement weakened if a different Lorentz structure is chosen.
Table \ref{vergl} shows a comparison of the scalar mesons with
calculated values using two different Lorentz structures
for the confinement potential. 
\par
The results of model B are in striking agreement with the
$K$ matrix poles of a coupled-channel analysis of various
reactions in which scalar states are produced \cite{anisov}.
(Note that the experimental entries in Table \ref{vergl}
correspond to poles in the scattering amplitude or
$T$-matrix.) In this analysis, 5 poles are introduced.
Four poles are enforced to make up two nonets 
with couplings which are fixed to guarantee SU(3)
symmetry in their decays. The mixing angle of the nonets
is a free parameter in the fit. 
One of the five poles is considered to be of exotic nature.
Its decay modes are compatible with full flavor symmetry. 
In the scattering amplitude, this pole is rather wide:
it has a mass in the range from $\sim 1200$ to 1600 MeV 
and a width of about $\sim 1000 $MeV. 
\par
Their starting
point is the observation that quark model calculations
yield bare meson masses which could be and are 
effected by their couplings to decay channels. 
This coupling leads to particularly strong shifts
in case of scalar mesons. The authors in \cite{anisov}
propose to identify the $K$-matrix poles with 
bare meson masses,  $T$-matrix poles with the observed
meson masses. This is an ansatz which is worth to be tested.
The authors do not, however, discuss the nature of their
5 poles. From the discussion of \p\p\ scattering, 
central production, radiative J$\psi$ data,
of \pbp\ra\piz\etg\etg\ in flight and D$_s$ decays I
believe 3 scalar isoscalar resonances to exist below
1.8 GeV. Hence I believe that 2 of the poles in \cite{anisov}
are used to describe $t$-channel exchange processes.
Of course, the presence of these poles has an impact
on the position of all other poles. Therefore, in my
view, this ansatz does not lead to a reliable and
interpretable result. 
In spite of the striking agreement between
the $K$-matrix analysis and model B, I am personally
convinced that the results of the $K$-matrix analysis
must be misleading. 
\par
\begin{table}
\begin{tiny}
\bc
\begin{tabular}{|ccc||ccc|ccc|}
\multicolumn{3}{c}{\normalsize Experiment}&
\multicolumn{3}{c}{\normalsize model A}&
\multicolumn{3}{c}{\normalsize model B}\\
\hline
             &           &           &        
             &           &           &        
             &           &                \\  
             &           &           &        
             &           &           &        
             &           &                \\  
             &           &           &        
             &           &           &        
             &           &                \\  
             &           &           &        
             &           &           &        
             &           &                \\  
             &\hspace*{-4mm}\small\bf $a_0(980)$ &\hspace*{-4mm}\small\bf $f_0(980)$\hspace*{-2mm}&
             &\hspace*{-4mm}\small\bf $a_0(1321)$ &\hspace*{-4mm}\small\bf $f_0(984)$\hspace*{-2mm}& 
             &\hspace*{-4mm}\small\bf $a_0(1057)$ &\hspace*{-4mm}\small\bf $f_0(665)$\hspace*{-2mm}      \\  
             &           &           &        
             &           &           &        
             &           &                \\  
             &           &                    
             &           &           &        
             &           &                \\  
             &           &           &        
             &           &           &        
             &           &                \\  
             &           &           &        
             &           &           &        
             &           &                \\  
\hspace*{-2mm}\small\bf $K_0^*(1430)$&\hspace*{-4mm}\small\bf $a_0(1490)$&\hspace*{-4mm}\small\bf $f_0(1500)$\hspace*{-2mm}&        
\hspace*{-2mm}\small\bf $K_0^*(1426)$&\hspace*{-4mm}\small\bf $a_0(1931)$&\hspace*{-4mm}\small\bf $f_0(1468)$\hspace*{-2mm}&        
\hspace*{-2mm}\small\bf $K_0^*(1187)$&\hspace*{-4mm}\small\bf $a_0(1665)$&\hspace*{-4mm}\small\bf $f_0(1262)$\hspace*{-2mm}     \\  
             &           &           &        
             &           &           &        
             &           &                \\  
             &           &\hspace*{-4mm}\small\bf $f_0(1740)$\hspace*{-2mm}&        
             &           &\hspace*{-4mm}\small\bf $f_0(1776)$\hspace*{-2mm}&        
             &           &\hspace*{-4mm}\small\bf $f_0(1554)$\hspace*{-2mm}     \\  
             &           &           &        
             &           &           &        
             &           &                \\  
\hspace*{-2mm}\small\bf $K_0^*(1950)$&           &           &        
\hspace*{-2mm}\small\bf $K_0^*(2058)$&           &           &        
\hspace*{-2mm}\small\bf $K_0^*(1788)$&           &                \\  
             &           &\hspace*{-4mm}\small\bf $f_0(2100)$\hspace*{-2mm}&        
             &           &\hspace*{-4mm}\small\bf $f_0(2113)$\hspace*{-2mm}&        
             &           &\hspace*{-4mm}\small\bf $f_0(1870)$\hspace*{-2mm}     \\  
\hline
\end{tabular}
\ec
\end{tiny}
\caption{The scalar mesons in a relativistic quark model
with an instanton-induced interaction. The two models
are different in the Lorentz structure of the (linear) 
confinement potential. 
(A): $\frac{1}{2}(1\otimes 1 - \gamma_0\otimes\gamma_0)$;
(B): $\frac{1}{2}(1\otimes 1 - \gamma_{\mu}\otimes\gamma^{\mu}
- \gamma_{5}\otimes\gamma_{5})$.
}
\label{vergl}
\end{table}

\subsection{What is wrong with the scalar glueball\,?}

We have identified two full nonets of scalar mesons, no meson is
left to play the role of the ground-state glueball. Also, the number
of states is just what we expect from the quark model: there is
no additional state, there is no evidence that the scalar
glueball has intruded the spectrum of scalar quarkonia and mixes
with them. Could the scalar glueball have escaped detection? 
\par
This is very unlikely. At least it is incompatible with experimental 
findings in radiative J/$\psi$ decays. The 2\p , \kkb\ and 4\p\ 
intensities 
is measured, the full intensity is ascribed to conventional
$q\bar q$ states. The scalar states are well separated; their mass difference
is large compared to their widths. Their is no room for the scalar
glueball to hide away. So the question arises how we can understand
the absence of the scalar glueball which is so firmly predicted
by QCD on the lattice. 
\par
The reason has to lay in the approximations made on the lattice.
QCD on the lattice neglects the coupling of the gluon field to 
$q\bar q$ pairs, to pions, to Goldstone bosons. This is called 
quenched approximation. Recent glueball mass calculations
on the lattice include couplings to fermion loops \cite{Bali:2000vr}
but pions
are still too heavy to represent the true chiral limit.

We may estimate the strength of this
coupling from the pseudoscalar mixing angle.
The \etg\ and \etp\ mesons are nearly octet and singlet
states, the $s\bar s - n\bar n$ mass difference leads to mixing with
a mixing angle of about $-18^{\circ}$ degrees or of 1/3 radian.
In a basis of $n\bar n$ and $s\bar s$eigenstates the matrix element which mixes
the two states is therefore large compared to the  
mass difference before mixing, large compared to 
$m_{s\bar s} - m_{n\bar n} \sim\ 300$\,MeV. 
The matrix element for the transition 
$n\bar n$\ra\ gluon fields \ra\ $s\bar s$ in the \etg\
or \etp\ wave function must therefore be of the order of a GeV. 
Using the arguments advocated for by 
Schwinger~\cite{Schwinger:1951nm}
the transition $n\bar n$\ra\ gluon fields \ra\ $s\bar s$
is a tunneling phenomenon through a potential barrier, 
with a probability which falls off exponentially with the generated
quark mass. The energy gap to $s\bar s$is of course much 
larger than that to $n\bar n$: 
gluon fields must have a very large coupling to pions, and
the scalar glueball may acquire a width of several GeV\,!
Lattice gauge calculation do not have these small pion
masses, they fail to observe the strong coupling of gluon
fields to nearly massless quarks. The neglect of the Goldstone
aspects of pseudoscalar mesons may lead to long-lived glueball
states. If the coupling of gluon fields to light quarks is
indeed governed by Schwinger's tunneling process, no 
resonant-like behavior of scalar gluon-gluon interactions 
are to be expected.  

\section{Hybrids}
\par
Hybrids, mesons with an intrinsic gluonic excitation, were first
predicted  shortly after the development of the
bag model \cite{Chodos:1974pn}. At that time, hybrids were
thought of as $q\bar q$ pair in color octet 
neutralized in color by a constituent gluon 
\cite{Horn:1978rq,Barnes:1983tx}.
Today we expect hybrids as excitations of the gluon fields
providing the binding forces between quark and antiquark, 
as excitations of the color flux
tube linking quark and antiquark \cite{Isgur:1985bm}.
\par
Hybrids are expected with a wide range of different quantum numbers.
Particularly 
exciting is the possibility of states with exotic $J^{PC}$ configurations 
like $J^{PC}=1^{-+}$, with quantum numbers which are not accessible to 
$q\bar q$ systems \cite{Close:1995hc}. 
They are expected at masses around 2 GeV and higher 
and to decay into two mesons with one of them
having one unit of orbital angular momentum \cite{Isgur:1985vy}. 

\subsection{Exotica with $I^G(J^{PC})=1^-(1^{-+})$} 
\subsubsection{The  $\pi_1(1400)$}

Indeed, an exotic meson 
has been seen to decay into a $p$-wave $\eta\pi$ system.
The quantum numbers in this partial wave are
$I^G(J^{PC})=1^-(1^{-+})$. These are not quantum numbers which
are accessible to the $q\bar q$ system, they are exotic. 
\par
A meson with quantum numbers $I^G(J^{PC})=1^-(0^{-+})$
is called a \p , one with $I^G(J^{PC})=1^-(2^{-+})$
is called $\pi_2$. These latter two mesons are well established
$q\bar q$ mesons. A meson with quantum numbers $I^G(J^{PC})=1^-(1^{-+})$
is thus called  $\pi_1$. Its mass is added to the name 
in the form  $\pi_1(1400)$ to identify 
the meson since  there could be and there are more than one resonance
in this partial wave.
\par
A meson with exotic quantum numbers like the  $\pi_1(1400)$ cannot
be a regular $q\bar q$ meson. It must have a more complex structure. 
It could be a hybrid but it might also be a four-quark
$qq\bar q\bar q$ resonance. The quantum numbers give no hint which
of the two possibilities is realized in nature. Before
we discuss arguments in favor of a  four-quark assignment
let us first have a look at the experimental findings.
\par
\begin{figure}[h!]
\bc
\epsfig{file=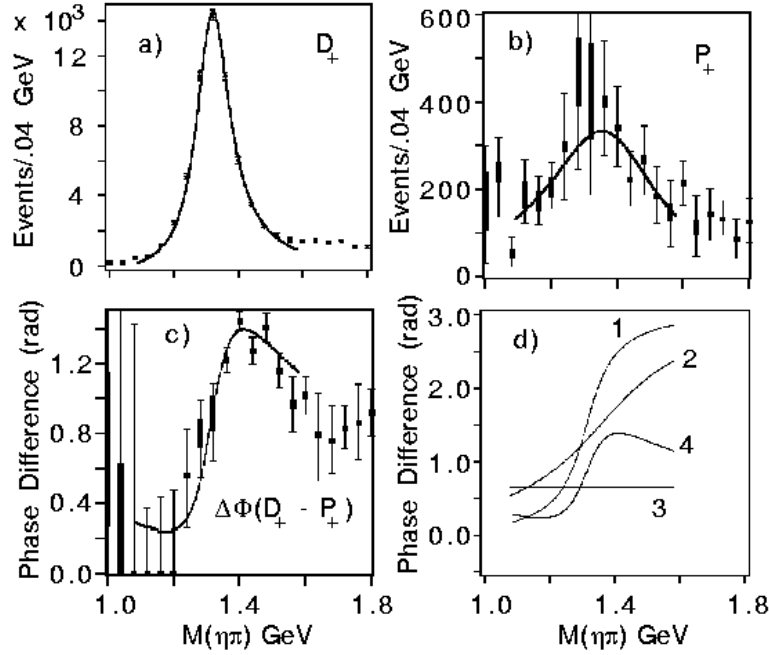,width=0.8\textwidth}
\ec
\caption{The squared scattering amplitude for the $D^+$ (a)
and $P^+$ (b) waves. The +sign indicates natural parity exchange.
amplitude. The relative phase between the two waves is shown
in (c). The lines
correspond to the expectation for two Breit-Wigner amplitudes.
In (d) the (fitted) phases for the D- (1) and P-wave (2) are
shown. The $P$- and $D$-production phases are free parameters
in the fits; their differences are plotted as line 3 and -
with a different scale - as 4.}
\label{bnl}
  \end{figure}

At BNL, the reaction
$$
\pim\ p \ra\ \pim\etg\ p
$$
was studied at 18 GeV/c \cite{Thompson:1997bs}\cite{Chung:1999we}. 
The data show a large asymmetry
in the angular distribution evidencing interference
between even and odd angular momentum contributions.
Fig.~\ref{bnl} shows data and the results of the partial wave
analysis. In a scattering process, the \p\etg\ system can be produced
in different partial waves ($S, P, D$ waves). In the $t$-channel
quantum numbers are exchanged corresponding to natural 
($0^{++},1^{--},2^{++}$) or unnatural ($0^{-+},1^{+-},2^{-+}$) 
parity. The naturality is a good quantum number for a given partial wave
and is added as suffix, + for natural, - for unnatural exchange.
\par
The data are fully compatible with the existence
of a resonance in the 
$I^G(J^{PC})=1^-(1^{-+})$ partial wave produced
via natural parity exchange. 
Mass and width are fitted to values given in Table \ref{exotics}.
\par
Since the spin in the 
final state is one, the exchanged particle cannot 
have scalar quantum numbers. The resonance is not observed
in the charge exchange reaction \cite{chungpr}
(with \piz\etg\ in the final state), hence the exchanged particle
cannot be the \rh . The particle which is exchanged
is the $f_2(1270)$ (or the tensor part of a Pomeron).
\par
\begin{figure}[thb]
\vskip 3mm
\begin{minipage}[t][4cm][t]{.55\textwidth}
\hspace*{-5mm}
\includegraphics[width=\textwidth]{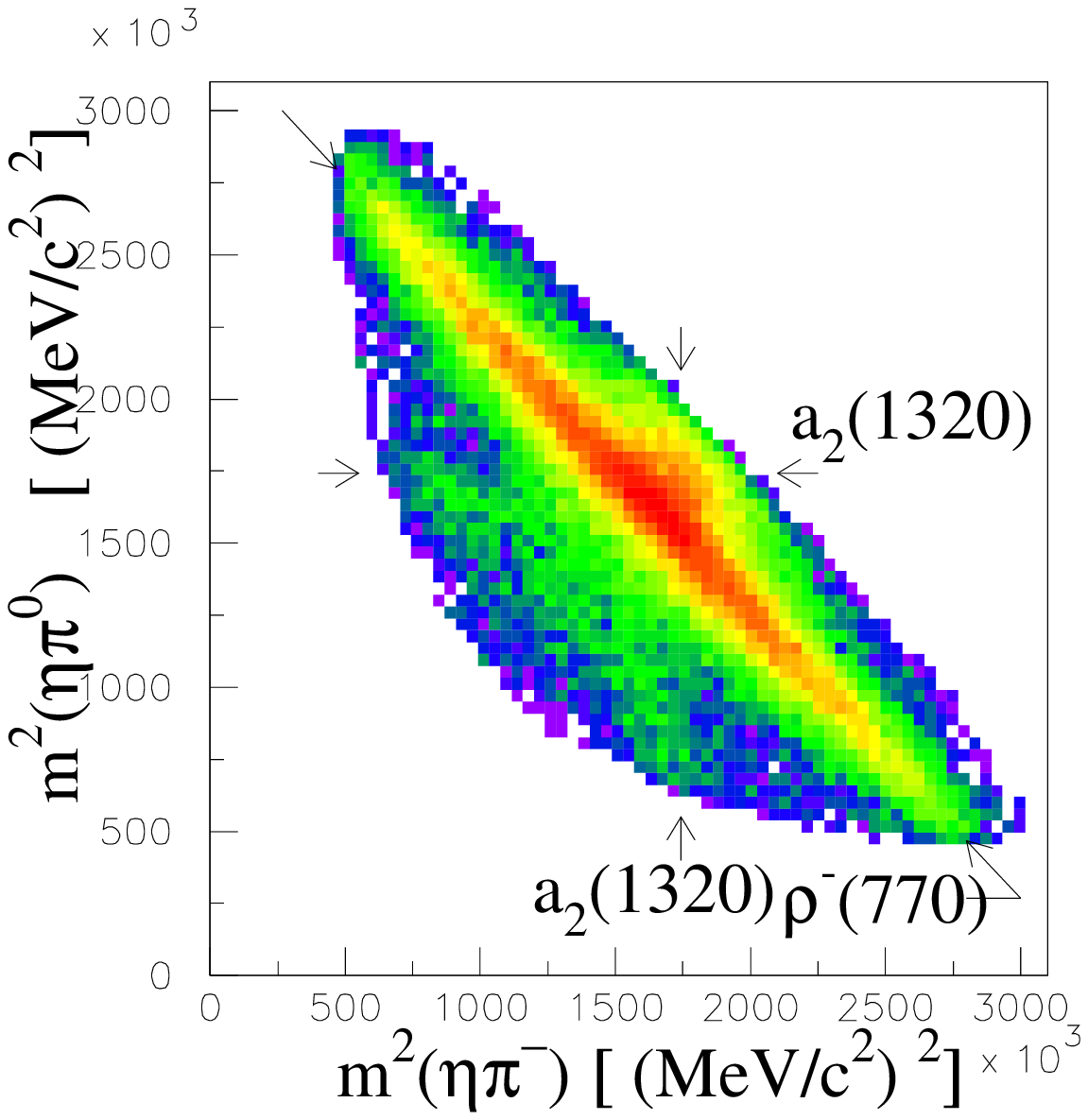}
\end{minipage}
\begin{minipage}[t][4cm][t]{.51\textwidth}
\vspace*{-73mm}
\includegraphics[width=\textwidth]{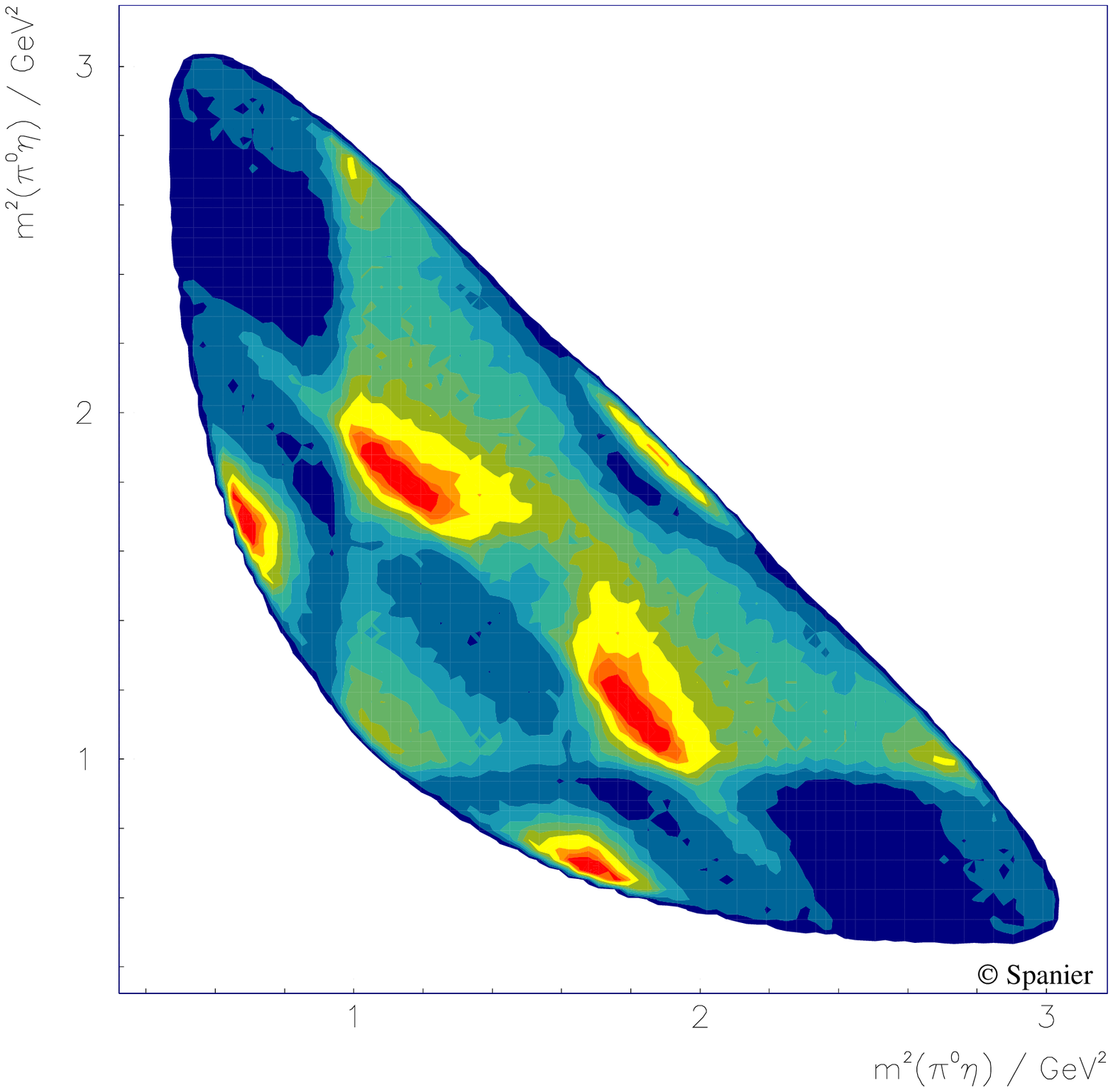}
\phantom{xxxxxx}\\
\end{minipage}
\vspace*{-40mm}
\caption{Dalitz plot for the reaction \pbn\ra
\pim\piz\etg\ for antiproton annihilation at rest
in liquid D$_2$. Annhilation on quasi-free neutrons is
enforced by a cut in the proton momentum 
($p_{proton}\leq 100$\,MeV/c. The data require
contributions from the 
$I^G(J^{PC})=1^-(1^{-+})$ partial wave in the
\etg\p\ system.}
\label{wd}
\end{figure}

The Crystal Barrel Collaboration studied the
reaction \pbn\ra\pim\piz\etg . Fig.~\ref{wd} shows
the \pim\piz\etg\ Dalitz plot. Clearly visible
are \rhm\etg , $a_2(1320)\pi$ with $a_2(1320)
\ra\eta\pi$ (in two charge modes) as intermediate states. 
\par
A fit with only conventional mesons gives 
a bad description only: the difference between data and
predicted Dalitz plot shows a pattern which is very
similar to the contributions expected from the interference
of the $\pi_1(1400)$ with the amplitudes for production of conventional
mesons \cite{Abele:1998gn}. 
\par
Introducing the exotic partial wave, the fit optimizes for
values listed in Table \ref{exotics}. Selection rules 
(and the PWA) attribute the production of the exotic partial wave
to the \pbp\ $^3S_1$ initial state.
\par
A similar analysis on the reaction \pbp\ra 2\piz\etg\ was
carried out. In this case the $\pi_1(1400)$ can only be
produced from the $^1S_0$ state; its production is much 
reduced in this
situation. The small contribution could only be unrevealed
when data taken by stopping antiprotons in liquid and gaseous
H$_2$ where analyzed. 
In these two data sets the fraction of annihilation contributions from
atomic S and P states is different (and their ratio known
from cascade models). Thus S and P wave contributions are 
constrained. It is only under these conditions that
positive evidence for the small contribution from the exotic 
partial wave could be found \cite{Abele:1999tf}.
\par
\begin{table}[htb]        
\begin{center}
\begin{tabular}[t]{c l l c c}
\hline
\hline
Experiment & mass (MeV/c$^2$) & width (MeV/c$^2$) &decay mode& reaction\\
\hline
BNL \cite{Thompson:1997bs}
& 1370 $\pm$ 16 $^{+\,\,\,\,\,\,50}_{-\,\,\,\,\,\,30}$& 385 $\pm$ 40 $^{+\,\,\,65}_{-105}$& $\eta \pi$&$\pi ^-
\textrm{p}\to \eta \pi ^- \textrm{p} $\\
BNL \cite{Chung:1999we} 
& 1359 $\,^{+\,\,\,\,\,16}_{-\,\,\,\,\,14}$ $\,^{+\,\,\,\,\,10}_{-\,\,\,\,\,24}$ & 314 $\,^{+31}_{-29}$ $\,\,\,\,^{+\,\,9}_{-66}$ & $\eta \pi$&$\pi ^-
\textrm{p}\to \eta \pi ^- \textrm{p} $\\
CBar \cite{Abele:1998gn}& 1400 $\pm$ 20 $\pm$ 20& 310 $\pm$ 50
$^{+50}_{-30}$ &$\eta \pi$&
$\bar{\textrm{p}}\textrm{n}\to \pi ^- \pi ^0\eta$\\
CBar \cite{Abele:1999tf}& 1360 $\pm$ 25 & 220 $\pm$ 90 &$\eta
\pi$&$\bar{\textrm{p}}\textrm{p}\to \pi ^0\pi ^0\eta$ \\
\hline
\hline
BNL \cite{Adams:1998ff}& 1593 $\pm$ 8 $^{+29}_{-47}$ & 168 $\pm$ 20
$^{+150}_{-\,\,\,12}$ & $\rho \pi$&$\pi ^- \textrm{p}\to \pi
^+\pi ^-\pi ^-\textrm{p}$\\
BNL \cite{Chung:1999mc}& 1596 $\pm$ 8 & 387 $\pm$ 23 &$\eta' \pi$&$\pi ^-
\textrm{p}\to 
\pi^-\eta'\textrm{p}$\\
VES \cite{Khokhlov:2000tk}& 1610 $\pm$ 20 & 290 $\pm$ 30 & $\rho \pi , \eta '\pi$&$\pi^-$N$\to\pi^-\eta'$N\\
\hline
\hline
\end{tabular}
\caption{Evidence for  J$^{\textrm{PC}}$ = $1^{-+}$ exotics}
\label{exotics}
\end{center}  
\end{table} 

\subsubsection{The $\pi_1(1600)$}

\begin{figure}[h]
\epsfig{file=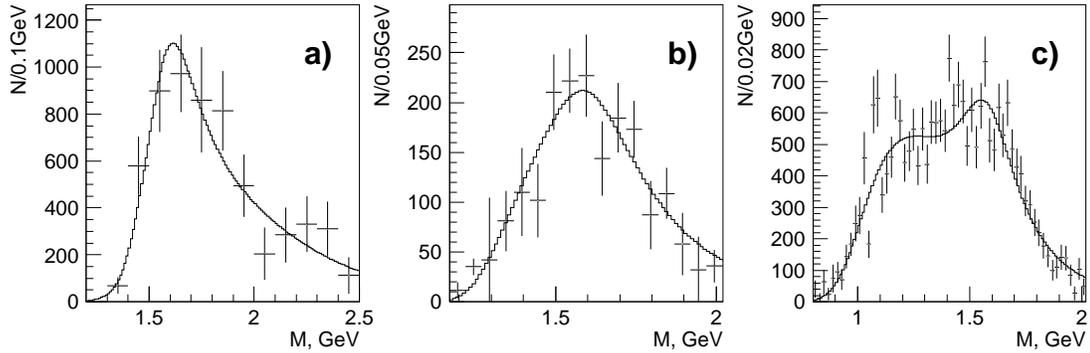}
\caption{Intensities in the $I^G(J^{PC})=1^-(1^{-+})$ 
partial wave for the (from left to right) $b_1\pi$,
\etg\p\ and \rh\p\ systems. The curves represent a fit to the 
data.}
\label{hyb1600}
  \end{figure}

The $\pi_1(1400)$ is not the only resonance observed in this partial wave.  
At Serpukhov, the \p\etp , \rh\p\ and the $b_1(1235)\pi$
systems  are studied in a 40\,GeV/c \pim\ beam. In all three systems
a resonant contribution in the exotic  $I^G(J^{PC})=1^-(1^{-+})$
partial wave is found. A combined fit finds a mass of
$\sim$\,1600 MeV and a width of $\sim$\,300 MeV \cite{Khokhlov:2000tk}.
Likely, these are three different 
decay modes of one resonances. Fig.~\ref{hyb1600} shows the VES data
with fit.
\par
At BNL, the $\eta^{\prime}\pi$ is also observed to 
exhibit a resonant behaviour \cite{Chung:1999mc} 
at about 1600 MeV. A partial wave
analysis of the $\rho\pi$ system \cite{Adams:1998ff}
reveals an exotic meson  with 
mass and width given in Table \ref{exotics}

\subsubsection{Higher-mass exotics}
There is evidence for further 
states from an analysis of $\pi^0f_1(1285)$ production 
\cite{Lee:1994bh}. The two mesons are produced in part with
zero orbital angular momentum between them, and this leads to
the same quantum numbers we have seen so often now, to the exotic  
$I^G(J^{PC})=1^-(1^{-+})$ partial wave. The phase motion in this partial wave
is not so well determined as in the other cases. It suggests
that even two resonances might exist, at 1800 and 2100 MeV.

\subsection{Four-quark or hybrid\,?}
Neither the  $\pi_1(1400)$ nor the $\pi_1(1600)$ have a mass as predicted for
hybrids; the  $\pi_1(1400)$ does not
have the predicted decay mode into a meson with one unit of
orbital angular momentum. Only the  $\pi_1(1600)$ decays into
$b_1(1235)\pi$ but this decay mode is not particularly dominant.
The  exotic partial wave in the   $\pi^0f_1(1285)$ system at
about 1800 MeV corresponds much more to what we should expect. 
Then the questions remains why so many resonances exist in this
one partial wave. The large number of states in one partial and 
in such a narrow mass interval is certainly surprising.
All have exotic quantum numbers, so they cannot 
possibly be $q\bar q$ states. 
\par 
We now 
discuss whether they are likely four-quark states or the searched-for
hybrid mesons.

\subsubsection{The Fock-space expansion}

The majority of established mesons can be interpreted as $q\bar q$
bound states. This can be an approximation only; the $\rho$-meson
e.g. with its large coupling to \p\p\ must have a four-quark
component and could as well have contributions from
gluonic excitations. The Fock space of the $\rho$
must be more complicated than just $q\bar q$. We may
write
\begin{equation}
\rho\ = \alpha q\bar q +  \beta_1b\bar qq\bar q + ... +  
\gamma_1q\bar qg + ... 
\label{fock}  
\end{equation}
where we have used $q\bar qg$ as short-name for a gluonic excitation. 
The orthogonal states may be shifted into the \p\p\ continuum.
Now we may ask: are the higher-order terms important and what is
the relative importance of the $\beta$ and $\gamma$ series\,?
\par
Possibly this question can be answered by truncating the
$\alpha$-term. Exotic mesons do not contain a $q\bar q$ component
and they are rare. Naively we may expect the production of
exotics in hadronic reactions to be suppressed by a factor 10 when 
one of the coefficients,
$\alpha_1$ or $\beta_1$, is of the order 0.3. 
We thus expect additional states 
having exotic quantum numbers, quantum numbers which are not accessible 
to the $q\bar q$ system. Their production rate should be
suppressed compared to those for regular $q\bar q$ mesons. In 
non-exotic waves the four-quark and hybrid configurations are
likely subsumed into the Fock expansion. If we can decide what
kind of exotic mesons we observe, four-quarks or hybrids or both, we 
can say what the most important contributions in (\ref{fock}) are.
\par
\subsubsection{SU(3) relations}
\begin{figure}[h]
\bc
\epsfig{file=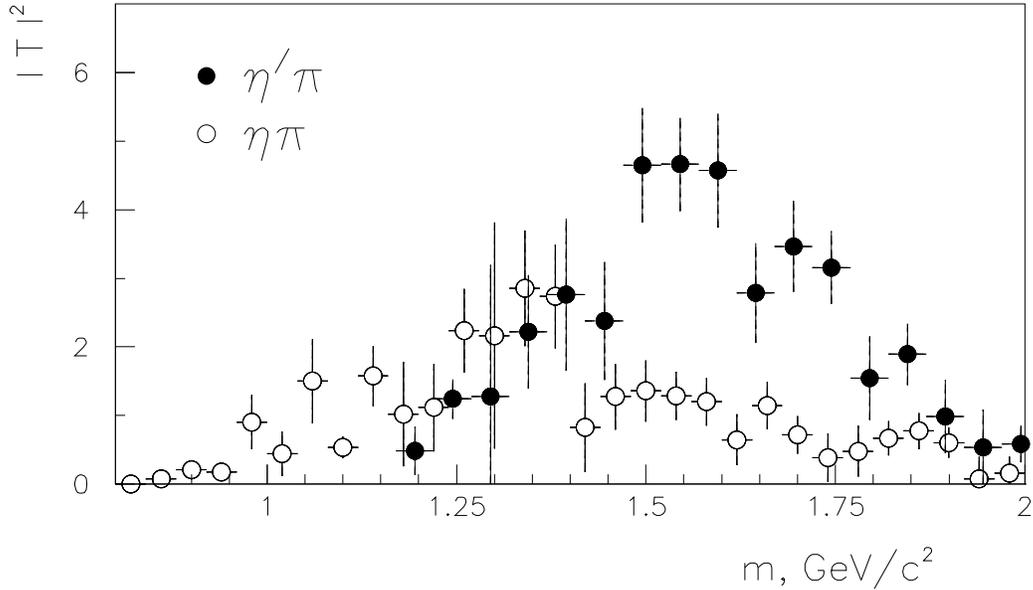,width=\textwidth}
\ec
\caption{The squared scattering amplitude in the 
$I^G(J^{PC})=1^-(1^{-+})$ partial wave for the
\etg\p\ and \etp\p\ systems; (from~\cite{Khokhlov:2000tk}).}
\label{etapietp}
  \end{figure}

The  $\pi_1(1400)$ decays strongly into \p\etg , the 
$\pi_1(1600)$ into \p\etp ; decays of the  $\pi_1(1400)$ into
\p\etp\ and of the $\pi_1(1600)$ into \p\etg\ were not observed
or reported. Fig.~\ref{etapietp} shows the exotic wave
for the \p\etg\ and \p\etp\ systems as a function of their mass:
the \p\etg\ intensity is concentrated around 1400 MeV, the
\p\etp\ intensity at 1600 MeV. A resonance decaying
into \p\etp\ should also decay into \p\etg , and a \p\etg\
resonance should also have a sizable coupling to \p\etp .
Why is there such  a strange decay pattern\,?
\par
We first consider the limit of flavor symmetry: the $\eta$ is
supposed to belong to the pseudoscalar octet, the
\etp\ is considered as pure singlet SU(3) state; mixing is neglected.
The $\pi_1$ states, having isospin one, cannot be isoscalar
states. Now I claim, that a meson belonging to an octet 
with exotic quantum numbers  $J^{PC}=1^{-+}$
cannot decay into two octet pseudoscalar mesons.
\par
The argument goes as follows: decays of particles belonging to an octet
of states into two other octet mesons, decays 
of the type 
$8 \ra 8 \otimes 8$ may have symmetric or antisymmetric
couplings. The two octets can be combined using
symmetric structure constants $d_{ijk}$ or antisymmetric
structure constants $f_{ijk}$. The decay $\pi_1(1400)$ into
two pseudoscalar mesons is governed by the symmetric
couplings. SU(3) demands the decay amplitude
for decays into two pseudoscalar 
mesons not to change sign when the two mesons are
exchanged. The orbital angular momentum $l=1$
between the two mesons requires the opposite: the
two mesons must be in a state \p\etg -\etg\p .
Both requirements cannot be fulfilled at the same time: the 
decay of a $\pi_1$ which belongs to an SU(3) octet
into two octet pseudoscalar mesons is forbidden. 
\par
There are immediate consequences: let us begin
with the $\pi_1(1600)$ and assume that it belongs to
an octet of states. Then it must decay
into \p\etp\ while the decay into \p\etg\ is forbidden.
This is precisely what we see. But what happens
in case of the $\pi_1(1400)$\,? It does decay into
\p\etg , why\,? As we have seen, it cannot belong to
an SU(3) octet, hence it must belong to a multiplet
of higher order. The easiest choice is a decuplet.
The  $\pi_1(1400)$ must be a SU(3)
decuplet state\,! As member of a decuplet, it cannot
decay into \p\etp , into an octet and a singlet meson,
and it 
cannot possibly be a hybrid: gluonic excitations do
not contribute to the flavor. Mesonic hybrids can only
be SU(3) singlets or octets. 
The strange phenomenon that the  $\pi_1(1600)$ does
not decay into \p\etg\ thus provides the clue for the 
interpretation of the  $\pi_1(1400)$ as decuplet state. 
\par
The above arguments hold in the limit of flavor symmetry. 
Due to  $\eta -\eta^{\prime}$ mixing, the exotic $\pi_1(1600)$ 
could decay into $\eta\pi$ via the small singlet component of the 
$\eta$. Also a small coupling of the  $\pi_1(1400)$ 
to $\eta^{\prime}\pi$ is possible. 
\par
\subsubsection{Four-quark states in SU(3)}
In the limit of SU(3) symmetry, the $\pi_1(1400)$ with its large \p\etg\ decay 
rate must belong to a decuplet and must hence be a four-quark state. 
The $\pi_1(1600)$ must belong to an octet of states and could thus be a hybrid.
There is no rigid argument against this conjecture. However, the mass difference 
between the $\pi_1(1400)$ and $\pi_1(1600)$ is typical for SU(3) multiplet
splittings. It has the same order of magnitude as the octet-decuplet splitting
in baryon spectroscopy, only the mass ordering is reversed. 
\par
Now let us dicuss how we can construct a decuplet of states
from two quarks and two antiquarks. Two quarks in flavor 3
combine to $3 \otimes 3 = \bar 3 + 6$, two antiquarks to
$3 + \bar 6$. Now we construct
$$
(\bar 3 + 6) \otimes (3 + \bar 6) = \bar 3 \otimes 3 +
\bar 3 \otimes \bar 6 +  6 \otimes 3  + 6 \otimes \bar 6 
$$
$$
= 1 + 8 + 8 + 10 + 8 + \bar{10} + 1 + 8 + 27
$$
Hence we can construct $10+\bar{10}$ and $10-\bar{10}$ multiplets and four
different octets. Hence a large 
number of different states with the same quantum numbers should be expected
from four-quark states. 
\subsubsection{Exotic(s) summary}
Several exotic mesons are observed, all in one partial wave  
$I^G(J^{PC})=1^-(1^{-+})$. The
decay pattern of the two resonances at 1400 and 1600 MeV
suggests that the $\pi_1(1400)$ should be a four-quark
resonance belonging to a decuplet of states. Then further
resonances with quantum numbers $I^G(J^{PC})=1^-(1^{-+})$
are to be expected and may have been found.
\par
Even though there is no argument against the hypothesis
that one of the observed resonances could be a hybrid,
there is also at present no experimental support for
this hypothesis. Once Pandora's box of four-quark
states being open, it is very hard to establish experimentally
that mesons with gluonic excitations be found
in an experiment.
\section{Conclusions}
This review challenges the wide-held believe
that gluons may act as constituent parts of hadronic
matter. Gluons exist as we know, e.g. from
3-jet events in $e^+e^-$ annihilation. Gluons interact;
this we know from the jet distribution in 
4-jet events in $e^+e^-$ annihilation. Gluons
are confined since they carry color. Do these
facts imply that glueballs must exist\,? I do
not believe so. Let us recall what a resonance
is: it may be defined, of course, in many different
ways but it needs to have specific decay modes, and
a phase variation must be associated with it. If the
gluon-gluon interactions lead to such a fast coupling
to final state particles that the phase advance during its
life time is, let's say 1 degree, you cannot identify
a resonant phase motion and the statement 'a glueball
exists' looses its meaning. 
\par
What is the experimental basis for this interpretation\,?
\par
First we discuss glueballs. 
Scalar isoscalar interactions play an important role
in many reactions. In purely hadronic reactions its
is difficult to decide if a resonance or better a
pole in the complex energy plane originates from a true
$s$-channel resonance or if it originates from $t$-channel
exchange processes. I believe that both types of reactions
can lead to observable poles and that one has to identify
those which come from $s$-channel resonances. 
The dependence on the momentum transfer in the production
discriminates long-range  $t$-channel
exchange processes from $s$-channel resonances. Also,
$s$-channel resonances have defined couplings to its
different decay modes while e.g. \rh\ exchange 
scatters \p\p\ dominantly to \p\p , Pomeron-Pomeron to
\p\p\ or to 
\rh\rh , \kkb\ to \kkb , and so on. 
\par
In the glueball-$q\bar q$-scalar-meson mixing scenarios 
no attempt was made to exclude poles originating from
$t$-channel exchange processes. One of the poles is however - as
discussed at length in this review - likely due to such processes.
This observation reduces the number of $s$-channel resonances
which has of course significant impact on the  mixing scenarios.
With the $f_0(980)$ included in the list of $q\bar q$ mesons,
the  $f_0(1500)$ and $f_0(1740)$ belong to different nonets,
to the 1$^3P_0$ ground state nonet and to the first radial 
excitations. 
\par
The broad 'background' component in scalar isoscalar
interactions depends 
on the momentum transfer in a way which shows that   
its physical origin is a long-range process. So it cannot
be the compact ground-state scalar glueball expected 
from lattice QCD. 
\par
There are processes which are free (or at least less
influenced) by $t$-channel exchanges. These are 
Z$^0$ or J/$\psi$ decays, radiative J/$\psi$ decays,
D$_s$ decays. Even in these processes $t$-channel
exchange may contribute: the data on Fig. 14c
on the left side seem to suggest that the two gluons
in radiative J/$\psi$ decays may interact via
pion exchange to create a two-pion pair and that
this process interferes with $f_0(1500)$ production,
the characteristic feature being the dip at 1500
MeV in the \p\p\ mass distribution. 
\par
Unfortunately, data from experiments with a
cleaner environment are statistically poor or
- in case of Z$^0$ decays - difficult to analyze. 
From BES we hope for a significant increase in 
statistics in radiative or non-radiative 
J/$\psi$ decays, the Babar experiment will 
deliver high-statistics data on D$_s$ decays
into three pions.
In this review large credibility is laid on these
reactions with scarce statistics. This is certainly
a weak point. So you should take the article
as contribution to an ongoing discussion. 
\par
We now turn to the discussion of exotic mesons.
First we have to make the point that at least two exotic
mesons, mesons with quantum numbers $I^G(J^{PC})$ = 
$1^-(1^{-+})$ which cannot be reached within the 
$q\bar q$ scheme,  now have been seen in different experiments
which come to (nearly) consistent results for masses 
and widths for these states. This is great progress\,!
These mesons can be four-quark states or hybrids. Here
I argue that the decay pattern of the \p$_1(1400)$ and 
\p$_1(1600)$ suggests that these two states should be
four-quark states. In the case of the two $\pi_1$
states, 
a decision on their
hybrid or four-quark nature can be based on 
selection rules. In other cases, in particular in those
in which non-exotic quantum numbers,
the decision has to be based on dynamical arguments.
Obviously, it is then very hard 
to argue that resonances should be interpreted as 
hybrids and not as four-quark states. As the
first exotic states observed experimentally
can be interpreted as four-quark states,
production of four-quark states seems to be more likely
than production of hybrids -- if the latter exist
at all.
\par
I would like to draw the following consequences
for our understanding of low-energy QCD: 
\par
Fundamental predictions of lattice gauge theory
are challenged by data and their interpretation
offered here. It is possible that lattice gauge
theories are still too far away from the chiral
limit. The assumption that static potentials QCD
potentials can be calculated 
by simulating QCD on a lattice seems to be
unjustified. Lattice QCD does - at present - 
not work sufficiently close to the chiral limit, it does not
reproduce the Goldstone nature of pions and thus
leads to wrong predictions concerning the existence
of new types of hadronic matter, of glueballs and hybrids.
We have to revise our understanding of low-energy QCD.  
\subsubsection*{Acknowledgments}
I would like to thank the Organizers of the
Zuoz Summer School for the invitation to this beautiful
place and the opportunity to talk
about glueballs, hybrids and $q\bar q$ mesons.
I appreciate the help of many colleagues in numerous discussions
and would like to express my gratitude to those who have given
advice, comments and help when I prepared and wrote this review.
They often do not share my view; whenever 
you disagree with interpretations offered here you should
not held them responsible.  
A few of them need to be mentioned here: 
V.V. Anisovich,  F.E. Barnes, S.U. Chung, A. Kirk, 
J. K\"orner, V. Markushin, B. Metsch, P. Minkowski, 
W. Ochs, M. Pennington, H. Petry, R. Ricken, 
A. Sarantsev, U. Thoma, H. Willutzki and A. Zaitsev. 

\end{document}